\documentclass[galaxies,review,accept,oneauthor,pdftex,10pt,a4paper]{mdpi}

\def\lsim{~\rlap{$<$}{\lower 1.0ex\hbox{$\sim$}}}
\def\bsim{~\rlap{$>$}{\lower 1.0ex\hbox{$\sim$}}}

\def\dd{{\rm d}}
\def\ln{{\rm ln}}

\def\vk{\mathrm{\bf k}}

\def\vq{\mathrm{\bf q}}

\def\vx{\mathrm{\bf x}}
\def\vy{\mathrm{\bf y}}

\newcommand{\esp}{\textnormal{\small \textsc{esp}}}

\usepackage{ amssymb }
\usepackage{ mathrsfs }
\usepackage{tabularx}
\usepackage{enumitem}
\usepackage{graphicx}
\usepackage{color}
\usepackage{multirow}
\usepackage{booktabs}
\usepackage{tabu}
\usepackage{float}

\def\d{{\rm d}}
\def\bx{{{\bf x}}}
\def\bk{{{\bf k}}}
\def\bq{{{\bf q}}}

\def\fnl{f_{\rm NL}}
\def\pk{{\rm pk}}
\newcommand{\ggam}{\gamma_1}
\newcommand{\ggamum}{\gamma_{J_1 \mu}}
\newcommand{\ggamnm}{\gamma_{\nu\mu}}
\firstpage{1} 
\pubvolume{7}
\issuenum{3}
\articlenumber{5}
\pubyear{2019}
\copyrightyear{2019}
\history{Received: 31 May 2019; Accepted: 7 August 2019; Published: 8 August 2019}

\Title{The Hunt for Primordial Interactions in the Large Scale Structures of the Universe}

\Author{Matteo Biagetti $^{\dagger}$\orcidA{}}

\AuthorNames{Matteo Biagetti}

\address{%
$^{\dagger}$ \quad Institute for Theoretical Physics Amsterdam, University of Amsterdam, Science Park 904, 1098 XH Amsterdam, The Netherlands; m.biagetti@uva.nl}

\abstract{
The understanding of the primordial mechanism that seeded the cosmic structures we observe today in the sky is one of the major goals in cosmology. The leading paradigm for such a mechanism is provided by the inflationary scenario, a period of violent accelerated expansion in the very early stages of evolution of the Universe. While our current knowledge of the physics of inflation is limited to phenomenological models which fit observations, an exquisite understanding of the particle content and interactions taking place during inflation would provide breakthroughs in our understanding of fundamental physics at high energies. In this review, we summarize recent theoretical progress in the modelling of the imprint of primordial interactions in the large scale structures of the Universe. We focus specifically on the effects of such interactions on the statistical distribution of dark matter halos, providing a consistent treatment of the steps required to connect the correlations generated among fields during inflation all the way to the late-time correlations of halos.
}

\keyword{Inflation; Large-Scale Structure; Primordial non-Gaussianity; Dark Matter Halos.}

\begin{document}


\tableofcontents
 \newpage
\section*{Introduction}

Cosmological observations reveal a Universe filled with structures over a wide range of scales. The last three decades of research in the field of cosmology have seen a huge development in our understanding of how these cosmic structures were formed throughout the history of the Universe. The current standard cosmological model is able to make a consistent timeline of the dynamical evolution of the Universe over the course of $13$ billion years. 
During the earliest known stage of cosmic evolution, an accelerated expansion phase known as \emph{cosmic inflation} \cite{Guth:1980zm,Starobinsky:1980te,Linde:1981mu,Albrecht:1982wi}, primordial perturbations are believed to have formed , providing the seed for the formation of all structures at later times.
In the inflationary scenario, these perturbations come from quantum fluctuations of scalar fields in an expanding background \cite{Mukhanov:1981xt}. As they are produced, they are stretched to very large scales, outside of the causal horizon, where they remain frozen. In subsequent stages of the cosmic evolution, these perturbations reenter the horizon, providing small inhomogeneities over the whole universe. The inhomogeneities grow due to gravitational instability and form the cosmic structures that we observe today.

The characteristics of the primordial perturbations have been best constrained by the statistical analysis of the temperature anisotropies in the Cosmic Microwave Background (CMB), the relic light that decoupled from all interactions in the moment in which electrons and protons combined to form neutral hydrogen atoms, 380,000 years after inflation. 
Observations of the CMB temperature anisotropies require that i) super-horizon, ii) nearly scale invariant, iii) very close to Gaussian and iv) adiabatic perturbations are produced in the early Universe \cite{Akrami:2019izv}. These features are strong hints that an inflationary mechanism indeed took place\footnote{We should not forget however that alternative models to inflation are also able to pass these tests (see for instance \cite{Ijjas:2015hcc} for a recent comparison of ekpyrotic models \cite{Khoury:2001wf} with Planck data).}.  
Despite the great success of observing such features, it is somewhat underwhelming to realize that the constraints from the CMB power spectrum are far from restrictive on inflationary models. Indeed, hundreds of models of inflation exist which are able to satisfy these constraints. 

A deeper understanding of the physics of inflation would fill our knowledge of the Universe up to $10^{-30}$ seconds after the Big Bang. The theoretical and observational challenges pertaining such a search are somewhat hard to tackle.  Other observational windows into the primordial Universe, such as primordial nucleosynthesis (happened around $3$ minutes after the Big Bang) and recombination ($\sim 380,000$ years), are governed by the laws of nuclear and atomic physics which have been established and extensively tested on Earth over the last century. On the other hand, the processes involved during inflation are still modeled using mostly simplified phenomenological mechanisms, because the typical energy scale at which inflation takes place is far above the TeV scale, thus inaccessible to experiments on Earth. Moreover, the inflationary environment most probably  does not involve fields and interactions of our standard model of particle physics, the elementary particles we are made of being generated at a later stage of cosmic evolution. The bright side is that any new discovery is a clear window on new physics, being able to explore energies as high as $10^{14}$ GeV, just a couple of orders of magnitude away from the Planck energy scale, thus providing us the best hope of experimentally probing quantum gravity.

A strong prediction which is common to all inflationary constructions is the production of primordial gravitational waves. Unfortunately, we have not been able to observe them yet. The Planck data, combined with the BICEP/Keck array measurements, constrain primordial gravitational waves perturbations to have amplitude $6.4\%$ smaller than scalar ones \cite{Akrami:2018odb}. While a minimal amount of gravitational waves is necessarily produced by any model of inflation, lower bounds can be unobservably low. For instance, if the amplitude of tensor fluctuations scales as the fourth root of the energy scale of inflation \cite{Lyth:1996im}, as expected for a large class of models of inflation, primordial gravitational waves might be as far as $\sim 40$ orders of magnitude away from current limits\footnote{This limit is calculated by imposing the minimal reheating temperature which realizes successful big bang nucleosynthesis.}.

Beside the existence of primordial gravitational waves, a great deal of information is still hidden in the statistics of scalar perturbations. Indeed, the characteristics listed above concern only the power spectrum, which tells us about the underlying free-field theory of the inflaton. On the other hand, attempts at building inflationary models from string theory and even particle physics perspective are usually characterized by a very rich particle content and interactions (see \cite{Lyth:1998xn,Baumann:2014nda} for reviews). 
Current limits on higher-order statistics impose interactions among fields to be rather weak, at least on the inflationary direction in field space \cite{Akrami:2019izv}. Nevertheless, there is still plenty of room to explore possibilities. Moreover, even in the simplest model of single-field inflation a minimal coupling to gravity is present.  This is usually called \emph{gravitational floor} of interactions and it is a ``must observe'' feature of all inflationary models. Although the detection of such an imprint is still at least two orders of magnitudes away from the sensitivity of current experiments, it does represent a guaranteed discovery which should be a main focus in cosmological searches.


Given the importance of such a search, there has been a growing effort in finding new observables that can constrain inflation. A promising observational probe of inflation is the study of structure formation at large scales and late times. Primordial perturbations provide the initial conditions with which matter overdensities grew under the effect of  gravitational instability and formed all the structures in the Universe. It is therefore natural to hope to extract information about inflation by studying how matter is distributed in the Universe. This task is complicated by the fact that gravitational instability is a non-linear process and most of the information on initial conditions is washed out once away from the linear regime. This is the reason why CMB searches for primordial non-Gaussianities have dominated the efforts of the last two decades: at the last scattering surface, perturbations are still mostly linear. However, recent developments in large-scale structure (LSS) theory and observations have demonstrated that LSS can provide better constraints than the CMB in the near future. A recent analysis showed the first such example: using Baryon Acoustic Oscillations data from the BOSS collaboration \cite{Beutler:2016ixs}, the authors of \cite{Beutler:2019ojk} have put constraints on primordial features that are stronger than CMB ones \cite{Akrami:2019izv}. For primordial non-Gaussianity, results from eBOSS collaboration have given the most recent constraints on local type primordial non-Gaussianity \cite{Castorina:2019wmr} and are expected to reach CMB sensitivity including the full set of data in the next few months. In the close future, the recently funded SPHEREx mission is one of the most promising examples of this great potential \cite{Dore:2014cca}, along with Euclid \cite{Amendola:2012ys}, LSST \cite{Abate:2012za} and SKA \cite{Bacon:2018dui}.  A strong feature of LSS observations is the fact that they explore a tridimensional volume, as opposed to the bidimensional photon last scattering surface probed by the CMB. The number of modes available to constrain the statistical distribution of perturbations is therefore greatly enhanced. Moreover, a variety of different probes can be exploited: the statistical distribution of galaxies can be mapped through spectroscopic and photometric surveys, while weak gravitational lensing can trace the dark matter distribution directly. In addition, galaxy intrinsic alignments can also provide constraints on primordial scalar and tensor perturbations \cite{Chisari:2013dda,Chisari:2016xki,Kogai:2018nse}. Higher-redshift probes include the 21-cm neutral hydrogen line \cite{Joudaki:2011sv,Chongchitnan:2012we,Chongchitnan:2013oxa,Camera:2013kpa,Sekiguchi:2018kqe,Witzemann:2018cdx} and intensity mapping with other emission lines  \cite{MoradinezhadDizgah:2018zrs,MoradinezhadDizgah:2018lac}. While experimental efforts on these directions are still not as developed as for lower-redshift ones, their potential is huge given that they explore various ranges in red-shift and therefore observed volume.


{\bf Plan of the Review.} In this review, we want to summarize recent theoretical progress in the modelling of the imprint of primordial interactions, taking place during inflation, in the clustering of dark matter halos. We follow primordial perturbations in chronological order from the very early Universe to the present day, considering their impact in the statistics of rare objects and of the lowest order clustering statistics. Particular attention is devoted to reviewing the latest theoretical developments in characterizing interactions taking place during, or right after, inflation as a source of primordial non-Gaussianity. We will therefore start by providing a quick overview of the inflationary mechanism with a focus on the generation of primordial perturbations in Section \S\ref{sec:power}. The following Section \S\ref{sec:milan} investigates interactions taking place during inflation and their observable signature in the three-point function of the primordial curvature perturbation. Section \S\ref{sec:transfer} is a bridge connecting the primordial curvature perturbation to perturbations in the dark matter density field. We then proceed with the two main sections of the review: the imprint of primordial interactions in one- and two-point statistics of dark matter halos, Sections \S\ref{sec:onepoint} and \S\ref{sec:two-point} respectively. We conclude by providing an overview of observational prospects in Section \S\ref{sec:final}.

{\bf Not in this Review.} Before we begin, we find it useful to briefly stress what this review will not be about. On the inflationary side, this review will deal with interactions among fields, or self-interactions, which generate a non-zero three-point correlation function of primordial perturbations, neglecting all higher-order correlations which are also generated. This choice is partially justified by the fact that we expect most models to respect a perturbative expansion, and therefore higher-order correlations to be more and more suppressed. On the other hand, there are various examples in which the study of the four-point correlation function is interesting and can lead to observable imprints. Another limitation on the inflationary side is that, even restricting to interactions generating three-point correlations only, the list of models investigated here is not exhaustive. On the LSS side, the main limitation of ths review is that it does not deal with the real observable, i.e. galaxies and clusters of galaxies. While luminous galaxies are the main observable tracers of the dark matter statistical distribution, dark matter halos provide the building blocks for their formation. Their study is therefore a crucial step for connecting theoretical predictions from inflation to observable imprints in LSS. It is clearly not the last, nor the only step: indeed, this review will not deal with a number of interesting issues, such as, for instance, the way in which galaxies populate halos. Even within the modeling of dark matter halos, we will restrict mostly to the analytic treatment of the evolution of perturbations, from inflation all the way to the present Universe, although short paragraphs will be devoted to recent numerical progress in the study of structure formation. A further limitation is that we will not make any computation in red-shift space, which is where galaxies are observed. Finally, we will not investigate the three-point correlation function of halos (nor galaxies), which is a natural observable for primordial three-point functions.

\section{Inflation and primordial perturbations}\label{sec:power}
It is a remarkable achievement of the generic inflationary scenario to provide a mechanism to generate primordial perturbations, which source the formation of structure in the Universe, considering that it was not designed for it, but to solve the well-known hot big bang problems \cite{Guth:1980zm,Starobinsky:1980te,Linde:1981mu}. 
There are countless good introductions to inflation and the production of primordial perturbations, including textbook material \cite{Kolb:1990vq, Liddle:2000cg, Dodelson:2003ft, Linde:2005ht,Mukhanov:2005sc,Weinberg:2008zzc} and reviews \cite{Lyth:1998xn, Malik:2008im,Baumann:2009ds,Riotto:2018pcx}. The reader is encouraged to look at these references for a detailed analysis. Here we will give the main physical intuition without enter any technical detail. 

\subsection{Background evolution}
 A good simple model to start with is a single scalar field, called generically the \emph{inflaton}, coupled to gravity through the metric $g_{\mu\nu}$ and slowly rolling down its potential. Its vacuum energy drives the accelerated expansion of the Universe. The corresponding action reads

\begin{equation}\label{eq:singleaction}
S= \frac{M_P}{2}\int {\rm d}^4 x \sqrt{-g} R + \int {\rm d} ^4 x \sqrt{-g} \left[-\frac 12 g^{\mu\nu}\partial_\mu \phi\partial_\nu\phi - V(\phi)\right],
\end{equation}

where $M_P$ is the reduced Planck mass, the first term is the Einstein-Hilbert action and $V(\phi)$ is the slow-roll potential. The background evolution is studied by assuming a flat Friedmann-Lema\^ itre-Robertson-Walker (FLRW) metric

\begin{equation}
{\rm d}s^2 = -{\rm d}t^2+a^2(t) {\rm d}{\bf x}^2,
\end{equation}

where $a(t)$ is the scale factor, for which the equations of motion are

\begin{align}
&H^2 = \frac{1}{3 M^2_P} \left(\frac 12 \dot\phi_0^2+V(\phi_0)\right),\\
&\dot H = -\frac{\dot\phi_0^2}{2 M_P^2},\\
&\ddot\phi_0+3H\dot\phi_0+V'=0.
\end{align}

Here the prime on the potential $V$ indicates derivation with respect to the background field $\phi_0$, while the dot denotes the time derivative with respect to cosmic time $t$. The expansion rate $H = \dot a/a$, known as the Hubble parameter, is determined by the first equation, while the evolution of the background field $\phi_0$ is determined by the third equation and the two are connected by the second equation. Inflation can be achieved when the potential $V$ dominates over the kinetic energy of the inflation. As a consequence, $\dot H\approx 0$ and the scale factor grows exponentially. Observations of the CMB require a minimum duration for inflation in order to solve the horizon and flatness problems. This is usually quantified in number e-folds $N$

\begin{equation}
N(\phi) \equiv \int_a^{a_{e}} {\rm d} \ln\, a = \int_t^{t_e} H\, {\rm d} t
\end{equation}

here measured from a time $t$ during inflation until the end of inflation, $t_e$. In this parametrization, inflation is required by CMB to run for $O(60)$ e-folds \cite{Akrami:2019izv} and as a consequence the Hubble parameter has to stay almost constant within a typical Hubble time $H^{-1}$. The slow-roll conditions therefore are

\begin{equation}\label{eq:slowrollp}
\epsilon \equiv -\frac{\dot H}{H^2} \ll 1, \qquad \eta \equiv  \frac{\dot\epsilon}{\epsilon H}<1
\end{equation}

and have to be satisfied during inflation. Using Friedmann equations, we can impose the slow-roll conditions on the shape of the potential as well

\begin{equation}\label{eq:slowpot}
\epsilon_V \equiv \frac{M_P^2}{2}\left( \frac{V'}{V}\right)^2<1, \qquad \eta_V \equiv M_P^2\left| \frac{V''}{V}\right|<1.
\end{equation}

Within the approximation of perfectly constant rate of expansion, the scale factor grows exponentially in time $a \propto e^{Ht}$ and the space time is called \emph{De-Sitter} (DS) space-time. 

\subsection{Quantum fluctuations of the inflaton}
Let us now turn to perturbations.  The inflaton field sets a ``clock'' for the amount of the expansion that the universe goes through during this phase and this amount is dictated, as a lower limit, by observations. A quantum-mechanical clock can not be infinitely precise though, as a consequence of Heisenberg's uncertainty principle, instead it has a variance. The inflaton is therefore subject to spatially varying fluctuations, such that

\begin{equation}
\phi({\bf x},t) = \phi_0(t) + \delta \phi({\bf x},t).
\end{equation}

These spatial fluctuations determine small differences in the time at which inflation ends, so that the Universe inflates by different amounts in different regions. This physical phenomenon is at the basis of the generation of the primordial perturbations throughout the density distribution in the Universe.

Because of the quantum nature of these primordial perturbations, the full computation would involve the quantization of the coupled fluctuations of the inflaton and the metric.  For the purpose of this review, it is sufficient to only show the case of a perturbed scalar field in DS without coupling to gravity. Counter-intuitively, this treatment captures most of the crucial points of the full coupled system and it is is somewhat technically simpler. We also ignore terms suppressed by the slow-roll parameters for now, so that, for example, the inflaton is effectively massless, being the second derivative of the potential constrained by Eq. \eqref{eq:slowpot}. The action for the perturbed massless free field at second order reads

\begin{equation}
S=\frac 12 \int {\rm d} \tau {\rm d}^3 x\, a^2\,\left[  (\partial_\tau{\delta\phi})^2- (\partial_i \delta\phi)^2\right],
\end{equation}

where we have defined conformal time $\tau$ as ${\rm d} t \equiv a {\rm d}\tau$. Note that $\tau \in (-\infty,0)$. The equations of motion in Fourier space for this action read

\begin{equation}
\delta\phi''({\bf k},\tau) + 2 \mathcal{H} \delta\phi'({\bf k},\tau) + k^2\delta\phi({\bf k},\tau) = 0,
\end{equation}

where $\mathcal H$ is the Hubble parameter in conformal time. The generic solution to this differential equation is\footnote{Here we used the approximation of perfectly DS spacetime, which implies $\tau = -1/Ha$. Corrections are proportional to slow-roll parameters, which indeed we are neglecting at this stage.}

\begin{equation}\label{eq:diffsol}
u({\bf k},\tau) = c_1 \frac{H}{\sqrt{2 k^3}}(1+i k \tau) e^{-ik\tau} + c_2 \frac{H}{\sqrt{2 k^3}} (1-i k \tau) e^{i k \tau}.
\end{equation}

To quantize the perturbations, we promote the field $\delta\phi$ and its conjugate momentum $\delta\pi \equiv \partial L/\partial \delta\phi'$ to operators and write canonical commutation relations

\begin{align}
[\delta\hat\phi({\bf x},\tau), \delta\hat\pi({\bf y},\tau)] &= i\delta({\bf x} - {\bf y}),\\
[\delta\hat\phi({\bf x},\tau), \delta\hat\phi({\bf y},\tau)] &=0, \quad [\delta\hat\pi({\bf x},\tau), \delta\hat\pi({\bf y},\tau)] = 0.
\end{align}

It is useful to decompose the field defining time-independent creation and annihilation operators in Fourier space

\begin{align}\label{eq:oper}
\delta\hat\phi({\bf k},\tau) &= u({\bf k},\tau) a_{{\rm k}} +u^{*}(-{\bf k},\tau)a^\dagger_{-{\rm k}},\\
\delta\hat\pi({\bf k},\tau) &= a^2u'({\bf k},\tau) a_{{\rm k}} +a^2 u'^{*}(-{\bf k},\tau)a^\dagger_{-{\rm k}},
\end{align}

with commutation relations

\begin{align}
[a_{\bf p}, a^\dagger_{-{\bf q}}] &= (2\pi)^3 \delta^{(3)}({\bf p} + {\bf q}),\\
[a_{\bf p}, a_{-{\bf q}}] &=0, \quad [a^\dagger_{\bf p}, a^\dagger_{-{\bf q}}]= 0,
\end{align}

which follow from the fact that $a^2 u({\bf k},\tau) u'^{*}({\bf k},\tau) - \mbox{ c.c.} =t$ -independent constant. We would like now to fix the initial conditions $c_1$ and $c_2$. The first constraint comes from the normalization conditions

\begin{equation}
|c_1|^2 - |c_2|^2=1
\end{equation}

and the second from choosing the vacuum state. To do this, we notice that on sub-horizon scales, that is for $k\tau\gg 1$, the system changes on time scales much shorter than the typical time scale of expansion, $H^{-1}$. In this limit, the first term of the solution Eq. \eqref{eq:diffsol} approaches the vacuum mode of Minkowski space-time and hence defines a preferable set of mode functions and a unique physical vacuum, usually referred to as Bunch-Davies state \cite{Bunch:1978yq}. With this choice of initial conditions, the solution reads

\begin{equation}\label{eq:diffsolBD}
u({\bf k},\tau) = \frac{H}{\sqrt{2 k^3}} (1+ik\tau) e^{-ik\tau}.
\end{equation}

While well within the horizon, $k\tau\gg 1$, this solution is highly oscillatory, in the opposite limit, on super horizon scales, the amplitude asymptotes to a constant. This is the essential feature of models of inflation, because it states that patches of the size of the horizon or bigger evolve classically with spatially modulated amplitude given by the different values of $\delta\phi$ at horizon exit for that patch. As a consequence, the ``inflation clock'' stops at slightly different times in different patches of the Universe, or, in other words, the inflationary e-folding has space-dependent variations of size

\begin{equation}\label{eq:zeta}
\zeta \approx H\delta t \approx H\frac{\delta\phi}{\dot\phi_0},
\end{equation}

where the approximate equality indicates again corrections suppressed by slow-roll parameters. An analysis extended to the coupled inflaton-metric fluctuations would show that $\zeta$ can be defined as a Gauge-invariant quantity, called \emph{comoving curvature perturbation}, which is conserved on superhorizon scales and it is directly related to the  temperature anisotropies we observe in the CMB, $\zeta \approx -5 \Delta T/T$. Correlation functions of this quantity therefore characterize perturbations produced during inflation. The  power spectrum $ P_\zeta$ is defined as

\begin{equation}
\langle 0 | \zeta( {\bf k}_1, 0) \zeta({\bf k}_2,0)|0\rangle = (2\pi)^3 \delta^{(3)}({\bf k}_1+{\bf k}_2) P_\zeta(k_1).
\end{equation}

Combining Eqs. \eqref{eq:zeta}, \eqref{eq:oper} and \eqref{eq:diffsolBD} we get

\begin{equation}\label{eq:powless}
\mathcal P_\zeta = \frac{H^2_*}{8\pi^2 M^2_P \epsilon_*} \left(\frac{k}{k_*}\right)^{n_s-1},
\end{equation}

where it is customary to define a dimensionless power spectrum as $\mathcal P_\zeta (k) \equiv k^3 /2 \pi^2 P_\zeta(k)$. Here $*$ indicates that the quantities are computed at horizon crossing, $k\tau_*=1$, and we reinstated slow-roll suppressed parameters in the last factor, giving the power spectrum a spectral index

\begin{equation}
n_s-1\equiv \frac{{\rm d} \ln P_\zeta}{{\rm d} \ln k} = -2\epsilon-\eta.
\end{equation}

The amplitude $\mathcal A_s =  H^2_*/ 8\pi^2 M^2_P \epsilon_*$ and spectral index $n_s$ of the power spectrum have been measured by Planck \cite{Akrami:2018odb} to be

\begin{align}
\mathcal A_s &= (3.044\pm 0.014)\times 10^{-10},\\
n_s &= 0.9649\pm 0.0042.
\end{align}

It is remarkable to notice that the deviation of $n_s$ from $1$ is statistically significant, indicating the first direct measurement of time dependence in the inflationary evolution. On the other hand, since there is no direct constraint on the slow-roll parameter $\epsilon$, the value of the Hubble parameter during inflation, which is related to the energy scale of inflation, is still unknown. A measurement of the primordial gravitational waves power spectrum would break the degeneracy.


\section{Interactions from inflationary models}\label{sec:milan}

In the previous section, we have shown how a simple phenomenological model of inflation with a single scalar field driving the accelerated expansion is able to produce primordial perturbations with characteristics that match CMB observations. Constructing an inflationary setting featuring these predictions is not as hard as it would seem: indeed a wide range of viable inflation models are available in the market. For this reason, it becomes crucial to extract information about higher-order primordial correlators, which would indicate that the statistics of $\zeta$ is not Gaussian and that nonlinear interactions are taking place during or after inflation. As we briefly show in this section, the nature of these nonlinear couplings is encoded in the higher-order correlators. 

{\bf Plan of the section.} We start by briefly introducing the basic points of the in-in formalism and connect it to the deviation from Gaussian statistics of the primordial perturbation $\zeta$ in Sec. \S\ref{sec:interactionsasnonG}, we then review different types of interactions as produced by various models, or classes of models, of inflation in Sec. \S\ref{sec:interactions} and we conclude with remarks about the current status and future prospects of detection of primordial non-Gaussianity, Sec. \S\ref{sec:remarksinf}.

\subsection{Interactions as non-Gaussianities}\label{sec:interactionsasnonG}


The computation of correlation functions in inflation is somewhat different than the usual quantum field theory methods applied to particle physics. In the latter case, scattering amplitudes are considered as non-interacting at some very early and very late times, far enough from the interaction region. In this way, the boundary conditions are taken to be vacuum states of the free theory on these limits, respectively called \emph{in} and \emph{out} states. 
On the other hand, during inflation the universe undergoes accelerating expansion and correlation functions are to be computed at \emph{fixed time}. As we have seen in the previous section, only modes with wavelengths much smaller than the horizon can be approximated as living in a flat Minkowski space. 
This is the limit where the  Bunch-Davies vacuum has been defined. Boundary conditions are therefore defined only at very early times, when most of the wavelengths are well within the horizon. The formalism to compute correlation functions in cosmological settings is called the \emph{in-in} formalism \cite{Schwinger:1960qe,Bakshi:1962dv,Keldysh:1964ud,Calzetta:1986ey,Weinberg:2005vy}. 
A generic form for an in-in correlator is

\begin{equation}\label{eq:corrQ}
\langle \hat Q(\tau)\rangle = \langle \Omega | \hat Q(\tau)|\Omega\rangle,
\end{equation}

 where $Q(\tau)$ represents the operator of the product of $n$-curvature perturbation $\zeta$s so that $\hat Q = \zeta_{{\bf k}_1} \zeta_{{\bf k}_2}... \zeta_{{\bf k}_n}$ and $\Omega$ is the \emph{in} state, the  vacuum of the interacting theory. Note that, more in general, there could be combinations of both $\zeta$ and $\gamma$, the tensor perturbation, in $\hat Q$. In this review we only focus on $\zeta$, but tensor perturbations are equally important in constraining the inflationary scenario \cite{Arutyunov:1999nw,Maldacena:2002vr,Maldacena:2011nz,Creminelli:2012ed,Hinterbichler:2013dpa}.
 
 The time $\tau$ at which the correlator is computed is usually either at horizon crossing for the modes of interest or at the end of inflation. The strategy to compute correlators such as the one in Eq. \eqref{eq:corrQ} is to evolve $Q(\tau)$ back to initial time $\tau_i$ where the vacuum state is defined. In order to do that, the \emph{interaction} picture is used, in which the background time dependence is determined by the quadratic Hamiltonian $H_0$, while interactions arise as corrections to $H_0$ through the interaction Hamiltonian $H_{\rm int}$. Eq. \eqref{eq:corrQ} is therefore evaluated formally as
 
 \begin{equation}\label{eq:inin}
 \langle \hat Q(\tau)\rangle = \langle 0 |\, \bar T\, e^{i \int_{-\infty(1-i\epsilon)}^\tau \hat H_{\rm int}^I(\tau'){\rm d}\tau'}\,\hat Q^I(\tau) \,T \,e^{i \int_{-\infty(1+i\epsilon)}^\tau \hat H_{\rm int}^I(\tau''){\rm d}\tau''}|0\rangle,
 \end{equation}
 where $T$ and $\bar T$ are time and reversed time-ordering symbols and $\hat Q^I$ and $\hat H^I_{\rm int}$ are evaluated using interaction picture operators. The $i \epsilon$ prescription allows to project the interacting vacuum state, $\Omega$ to the free one. The above expression can be expanded as a power series in $H_{\rm int}$. Interactions are then organized as usual using Feynman diagrams. As an example, let us take the three-point correlator of $\zeta$ expanding Eq. \eqref{eq:inin} to first order
 
 \begin{equation}\label{eq:threezinin}
 \langle \zeta_{{\bf k}_1}\zeta_{{\bf k}_2}\zeta_{{\bf k}_3}\rangle \approx -i \int_{-\infty}^0 {\rm d} \tau \langle 0 |\,[  \hat\zeta_{{\bf k}_1}\hat\zeta_{{\bf k}_2}\hat\zeta_{{\bf k}_3}(0), \hat H_{\rm int}(\tau)]\, |0\rangle,
 \end{equation}
 
 where $H_{\rm int} = -L_3 + \mathcal O(\zeta^4)$ and $L_3$ is the perturbed Lagrangian up to cubic order in $\zeta$ and we have taken the superhorizon limit $k\tau\rightarrow0$. This is how interactions connect to non-Gaussianities through the perturbed action: they generate higher-order correlators, such as the bispectrum in Eq. \eqref{eq:threezinin}. 
 
In this review, we exclusively consider non-Gaussian signatures which come from a non-zero three-point correlation function, or more frequently used, its equivalent in Fourier space, the bispectrum
 
\begin{equation}\label{eq:bispe}
B_\zeta( {\bf k}_1, {\bf k}_2, {\bf k}_3) \equiv  \langle \zeta_{{\bf k}_1}\zeta_{{\bf k}_2}\zeta_{{\bf k}_3}\rangle.
\end{equation}


It is customary to decompose $B_\zeta$ as

\begin{equation}
B_\zeta( {\bf k}_1, {\bf k}_2, {\bf k}_3) = (2\pi)^3 \delta^{(3)}({\bf k}_1 + {\bf k}_2+ {\bf k}_3)\, \frac{(2\pi^2)^2}{(k_1 k_2 k_3)^2 } \mathcal P^2_\zeta(k_*) \mathcal S(k_1,k_2,k_3),
\end{equation}

where we assumed statistical homogeneity and isotropy and we evaluate the dimensionless power spectrum, $\mathcal P_\zeta$ at pivot value, neglecting the small scale dependence $(k/k_p)^{n_s-1}$ from now on. All the momentum dependence goes therefore in the function $\mathcal S$, which encodes all the crucial information about the bispectrum. This information can be categorized into three features: 

\begin{itemize}[leftmargin=*,labelsep=5.8mm]

\item the \emph{shape} of the bispectrum, which is usually expressed through the dependence of $\mathcal S$ on the ratio of the momenta, for instance $k_2/k_1$ and $k_3/k_1$.

\item the \emph{running} of the bispectrum, which refers to the dependence of $\mathcal S$ on the sum of the amplitude of the wave numbers  $K = k_1+k_2+k_3$.

\item the \emph{amplitude} of the bispectrum, usually denoted as $f_{\rm NL}$ and defined as

\begin{equation}\label{eq:fnl}
\mathcal S(k_1,k_2,k_3)\stackrel{k_1=k_2=k_3}{ \xrightarrow {\hspace*{1.2cm}}}\frac{18}{5} f_{\rm NL}(K),
\end{equation}

where, as we can see, $f_{\rm NL}$ can generically depend on $K$.

\end{itemize}

 
\subsection{Interactions in models of inflation}\label{sec:interactions}
Having made the connection between interactions and non-Gaussianities, we now classify models of inflation by the interactions, and consequently non-Gaussianities, they produce during inflation. All these models satisfy the minimal current constraints on inflation that we outlined in the previous 
section \S\ref{sec:power}. Therefore, the only way to discriminate among them is to observe some degree of non-Gaussianity in the CMB temperature anisotropies or in the LSS.


In the phenomenological example we outlined in \S\ref{sec:power}, inflation is run by a single, scalar field. Even in this simple case, non-Gaussianities through various types of interactions are produced,  leading to a rich phenomenology in correlation functions of the curvature perturbation $\zeta$. 
On the other hand, it is reasonable to think that inflation was populated by many fields. The case of more than one field during inflation is clearly even richer, as it includes all the features of single-field models plus the possible interactions among the fields. 
These fields might have very diverse functions: i) contribute to the background, some/all of which ii) generate primordial perturbations, or some might simply iii) spectate, i.e. they do not give a substantial contribution to neither the accelerated expansion nor the perturbations. 
Here we are interested in the effects that a multi-field scenario has on the correlation functions of the curvature perturbation $\zeta$, and consequently on structure formation, thus we will restrict to the study of perturbations. We distinguish between two cases: first, the case of massive particles present during inflation, which decay right after horizon crossing and therefore their signature is left in the curvature perturbation during inflation. Second, the case in which perturbations from massless particles survive even on superhorizon scales and imprint non-Gaussianities after inflation.

\subsubsection{Interactions in single-field models}\label{sec:singleint}

In this section we review different ways to produce non-Gaussianities in single-field models. They range from very small (though never zero), i.e. of order of the slow-roll parameters $\epsilon$ and $\eta$, to possibly larger amplitudes when relaxing the assumption of minimal interaction of the inflaton with gravity. The list that follows is not complete, but rather it gives a schematic idea of the class of models that can generate non-Gaussianities in single-field inflation.

\paragraph{ {\bf Gravitational floor}}

The example in Eq. \eqref{eq:threezinin} was not selected casually: it is the case of nonlinearities produced by the lowest order interaction of the inflaton with gravity. This is why it is called \emph{gravitational floor}: every model of inflation is expected to have this minimal amount of interaction and non-Gaussianity produced. The integral in Eq. \eqref{eq:threezinin} was first performed in the seminal paper by Maldacena \cite{Maldacena:2002vr} and we redirect directly to it for details on the computation. The dimensionless bispectrum that results from the computation reads

\begin{align}\label{eq:malda}
\mathcal S_{\rm GF}(k_1,k_2,k_3) = &\frac{\epsilon}{2} \left[ -\left(\frac{k_1^2}{k_2 k_3} + 2 \mbox{ perms. } \right) + \left(\frac{k_1}{k_2} + 5 \mbox{ perms. }\right) + \frac{8}{K} \left(\frac{k_1 k_2}{k_3} + 2 \mbox{ perms. } \right)\right]\nonumber\\
&+ \frac{\eta}{2} \left( \frac{k_1^2}{k_2 k_3} + 2 \mbox{ perms. }\right),
\end{align}

where $K=k_1+k_2+k_3$. By taking the limit where all the momenta are equal and using Eq. \eqref{eq:fnl} we extract the amplitude  $\fnl^{\rm GF} = 55/9\, \epsilon + 15/9\, \eta$. The running of this shape is typically small, given that it is of order the running of the slow-roll parameters.

The \emph{squeezed limit} of the bispectrum, i.e. the limit where one of the momenta is soft, provides the \emph{single-field consistency relation}

\begin{equation}\label{eq:maldasqueezed}
\lim_{k_3\rightarrow 0} B_\zeta(k_1,k_2,k_3) = (1-n_s) P_\zeta (k_1)P_\zeta(k_3).
\end{equation}

Extensions and generalizations of this statement have been made in several subsequent efforts \cite{Maldacena:2002vr,Creminelli:2004yq,Cheung:2007sv,Creminelli:2011rh,Creminelli:2012ed,Assassi:2012zq}. This relation is valid for any single-field model of inflation, not necessarily slow-roll, since it can be demonstrated on the sole assumption that the curvature perturbation $\zeta$ is constant on super-horizon scale, i.e. that it is adiabatic. It follows from the fact that a long-wavelength perturbation, if adiabatic, corresponds to a local rescaling of the background experienced by short-wavelength ones. The direct consequence of Eq. \eqref{eq:maldasqueezed} is that any detection of non-Gaussianity for a bispectrum computed in the squeezed limit indicates that perturbations were generated by more than one field during or after inflation. 

The consistency relation can be extended to finite long-wavelengths by an expansion in $k_{\rm long} /k_{\rm short}$

\begin{equation}
\lim_{k_3\ll k_1} B_\zeta(k_1,k_2,k_3) = P_\zeta (k_1)P_\zeta(k_3) \sum_{n=0}^{\infty} b_n \left(\frac{k_3}{k_1}\right)^n,
\end{equation}

where $b_0 = 1-n_s$ and $b_1$ is also fixed by symmetries, corresponding to a local constant gradient rescaling of short modes. Owing to the consistency relation, all powers of the expansion are integer. Interestingly, the presence of additional fields introduces non-analytic scalings, as we will show below for the case of quasi-single field inflation. 

\paragraph{ {\bf Higher-Derivative kinetic terms} }

A first extension of the model presented in the previous section, Eq. \eqref{eq:singleaction}  is done by writing the most general Lorentz-invariant Lagrangian as a function of the inflaton $\phi$ and its derivative \cite{Garriga:1999vw}

\begin{equation}
S= \frac{M_P}{2} \int \d^4 x \sqrt{-g} + \int \d^4 x \sqrt{-g} P(X,\phi),
\end{equation}

where $X\equiv -1/2 g^{\mu\nu} \partial_\mu\phi\partial_\nu\phi$. In the context of an effective field theory of inflation, the form of $P(X,\phi)$ is constrained by the following considerations: 

\begin{itemize}[leftmargin=*,labelsep=5.8mm]

\item Non-derivative operators, such as $\phi^n / \Lambda^{n-4}$, being $\Lambda$ the largest energy scale where the effective description holds, contribute directly to the inflaton potential and are therefore strongly constrained by the background \cite{Copeland:1994vg}. 

\item Derivative operators of the form $X^n / \Lambda^{4n-4}$ do not suffer this limitation. However, a simple estimation of the amplitude of the leading correction to the slow-roll Lagrangian, $X^2/\Lambda^4$, gives

\begin{equation}
\fnl \approx \frac{\dot \phi_0^2}{\Lambda^4}.
\end{equation}

Non-Gaussianities of order unity are therefore generated when $\dot \phi_0 \approx \Lambda^2$, which is the regime where the effective description brakes down. In other words, the effective Lagrangian becomes unstable to radiative corrections. Non-gaussianities of order unity of this type  therefore represent a particularly well motivated target for observational searches, because $\dot \phi_0$ is already a relevant energy scale for the dynamics of the inflationary background, being the scale related to the breaking of exact DS background evolution.

\end{itemize}

Following these considerations, the only way to escape the break down of the effective description and to have sizeable non-Gaussianities is to write down a UV-complete model. One such example is provided by the \emph{Dirac-Born-Infeld} (DBI) model \cite{Silverstein:2003hf,Alishahiha:2004eh}, for which

\begin{equation}\label{eq:pxdbi}
P(X,\phi) = -\frac{\sqrt{1-2 f(\phi) X}}{f(\phi)} + \frac{1}{f(\phi)} - V(\phi),
\end{equation}

where $f(\phi)$ is the (squared) warp factor of the AdS-like throat related to these models\footnote{ One should bear in mind that there is a degree of fine-tuning needed also for this model \cite{Chen:2008hz}.}.
The dimensionless bispectrum for a generic $P(X,\phi)$ model can be written in a rather model-independent way \cite{Seery:2005wm}

\begin{align}\label{eq:shapepx}
\mathcal S_{\rm HD}(k_1,k_2,k_3) = S_\lambda + S_c + \mathcal O(\epsilon, \eta) ,
\end{align}

where the subscript HD stands for ``Higher-Derivative'' terms and 

\begin{align}
S_\lambda &= \left(\frac{1}{c_s^2} -1 - \frac{2\lambda}{\Sigma}\right) \frac{6\, k_1 k_2 k_3}{ K^3}\\
S_c & =\left(\frac{1}{c_s^2}-1\right) \left[ - \frac{4}{K} \left(\frac{k_1 k_2}{k_3} + 2 \mbox{ perms. }\right)+ \frac{2}{K^2} \left( \frac{k_1 k_2^2}{k_3} + 5 \mbox{ perms. }\right) + \frac 12 \left(\frac{k_1^2}{k_2 k_3} + 2 \mbox{ perms. }\right)\right] 
\end{align}

and we have defined the following quantities \cite{Garriga:1999vw}

\begin{align}\label{eq:pxparams}
c^2_s &= \frac{P_X}{P_X+2XP_{XX}},\\
\Sigma &= X P_X+2X^2P_{XX} = \frac{H^2 \epsilon}{c^2_s},\\
\lambda &= X^2 P_{XX}+\frac 23 X^3 P_{XXX},
\end{align}

where $P_X$ denotes the derivative of $P$ with respect to $X$ and we stopped at the third derivative because we limit to the bispectrum and we quoted only the leading order in slow-roll parameters. In $P(X)$ models, $c_s$ is the speed of propagation of the scalar perturbations and can be typically different then unity, thus leading to sizeable bispectra. The bispectrum amplitude for these models indeed is given by

\begin{align}
\label{eq:fnlambda}
\fnl^{\lambda} &=  \frac{5}{3} \left(\frac{1}{c_s^2} -1 - \frac{2\lambda}{\Sigma}\right)\\
\label{eq:fnlc}
\fnl^c & = -\frac{5}{12} \left(\frac{1}{c_s^2}-1\right)
\end{align}

As it is clear from Eq. \eqref{eq:shapepx}, models with large non-Gaussianity, as the DBI model, need to have $c_s^2 \ll 1$ and/or $\lambda/\Sigma \gg 1$. In the specific case of the DBI model, by combining Eq. \eqref{eq:pxdbi} into Eq. \eqref{eq:pxparams} and \eqref{eq:shapepx}, Eq. \eqref{eq:fnlambda} vanishes so that only $S_c$ contributes with a sizable non-Gaussianity.

This bispectrum peaks in the \emph{equilateral} triangle configuration, where all the momenta are similar. The physical intuition for this fact is straightforward:  modes that are much longer than the others, once out of the horizon, cannot interact with those within the horizon. Large interactions can occur only when all momenta are similar and therefore exit the horizon at the same time. The running of this bispectrum is small.

%
%


%
%
%
%
%

\paragraph{ {\bf Features during inflation}}

The presence of features in the primordial power spectrum is frequently linked to a sizeble bispectrum. From the observational point of view, a feature represents a breaking of scale-invariance of the primordial spectrum within a range of scales. From the point of view of building inflationary models based on quantum gravity, or string theory, constructions, scale invariance is often a result of various mechanisms, while features can appear rather naturally \cite{Chen:2010xka,Chluba:2015bqa}. These features can be broadly classified into two categories: sharp features or periodic oscillations. Sharp features in the potential or in the internal field space can arise from a variety of models \cite{Adams:2001vc,Chen:2006xjb,Bean:2008na,Achucarro:2010da,Miranda:2012rm} and they can also be studied in the context of the effective field theory of inflation \cite{Bartolo:2013exa}. Such features typically show up primarily in the power spectrum and constrains can be put with observations both of the CMB \cite{Akrami:2018odb} and LSS \cite{Beutler:2019ojk}, so non-Gaussianities can be used as a cross-check. 

{\it Sharp feature. }The feature causes the inflaton to momentarily exit the attractor phase and consequently the slow-roll parameters to vary over a few e-folds. If we characterize the feature by its relative height $\mu \sim \Delta V/V$ and width $\sigma$, it can be easily shown on general grounds that observations of the power spectrum from the CMB constrain the ratio $\mu/\epsilon \lesssim 1$ \cite{Akrami:2018odb}. The slow-roll parameter $\eta$ on the other hand can change a lot if the change in $\epsilon$ occurs within a short time. If we take $\Delta \epsilon \sim \Delta (\dot\phi_0^2)/H^2 \sim \mu$ and $\Delta t \sim \Delta\phi / \dot\phi_0 \sim \sigma / \sqrt{V(c+\epsilon)}$ we get

\begin{equation}
\Delta\eta \sim \frac{\Delta\epsilon}{H\epsilon \Delta t} \sim \frac{\mu \sqrt{\mu+\epsilon}}{\sigma\epsilon}.
\end{equation}

For most models with a sharp feature, the calculation of the bispectrum requires to use numerical solutions. Here we quote an approximated shape \cite{Chen:2010xka}

\begin{equation}\label{eq:sharp}
\mathcal S_{\rm sharp}(k_1,k_2,k_3) \sim \frac{\mu \sqrt{\mu+\epsilon}}{\sigma\epsilon} \sin \left(\frac{K}{k_*}+\varphi_0\right) \left(\frac{K}{k_*}\right)^n e^{-\frac{K}{mk_*}},
\end{equation}

where $1/k_*$ is the oscillatory frequency in Fourier space corresponding  to the feature. The exponential cuts off long-wavelength modes which are much longer than $k_*$ and feature is smoothed with a power $n$ to be fit with numerical results, along with $m$. The most important property of this type of non-Gaussianity is the running of the bispectrum, which is explicit in the dependence on $K$.

{\it Resonant running.} 

A different type of features might be generated by an oscillatory component in the background evolution, as predicted, for instance, by axion-monodromy inspired models of inflation \cite{McAllister:2008hb,Flauger:2009ab}. In these models, the inflaton potential is characterized by an oscillatory term added to the usual slow-roll one

\begin{equation}
V(\phi)=V_0(\phi)+ \Lambda^4\cos\left(\frac{\phi}{f}\right),
\end{equation}

where here $\Lambda$ is some high energy scale and $f$ is the axion decay constant. As we have already seen, each mode oscillates during inflation with decreasing frequency as it is stretched by the accelerated expansion, until it reaches $H$ where it becomes frozen. Therefore, for any oscillatory feature with frequency $\omega > H$, we might expect a resonance between couplings and modes which sources non-Gaussianities \cite{Chen:2008wn,Flauger:2010ja}. This type of non-Gaussianity, differently from the other cases analyzed before, is generated on sub-horizon scales. It can be shown that the parameter space for this type of resonance can be large, since

\begin{equation}\label{eq:axionmono}
\frac{\omega}{H}>1 \Rightarrow \frac{\sqrt{2\epsilon}}{f M_P}>1,
\end{equation}

being $\epsilon$ the slow roll parameter and $f\ll M_P$ is predicted in string theory constructions with sub-Planckian decay constants \cite{Banks:2003sx,Svrcek:2006yi}. The corresponding dimensionless bispectrum for this type of models reads

\begin{align}\label{eq:resrun}
\mathcal S_{\rm res} = \sin\left[\alpha\, \ln\left(\frac{K}{k_*}\right)\right] + \frac{1}{\alpha}\sum_{i\neq j}\frac{k_i}{k_j}\cos\left[\alpha\, \ln\left(\frac{K}{k_*}\right)\right]+ \mathcal O \left(\frac{1}{\alpha^2}\right),
\end{align}

where $\alpha = \sqrt{2\epsilon}/f M_P$ is constrained to be large by Eq. \eqref{eq:axionmono} and $k_*$ is the pivot scale at which the amplitude of the dimensionless power spectrum is defined. Similarly to the sharp feature case, this type of non-Gaussianity has sizeble running.

\paragraph{ {\bf Non Bunch-Davies vacua}}

The choice of the vacuum state during inflation is not unique. To address the ambiguity from basic principles one would need to know the full theory at the highest energies, where we expect the free-field approximation to break down, as well as the physics preceding inflation. 
Notwithstanding this, the issue can be addressed phenomenologically: any deviation from the attractor solution of the inflaton, such as the ones related to sharp features studied above, lead generically to a deviation from the standard Bunch-Davies vacuum.  This is because the Bunch-Davies vacuum is chosen out the asymptotic limit of $k\tau\gg 1$ of the attractor solution. Non-Gaussianities produced by choosing different prescriptions for the vacua are rather model-dependent and have been explored in a number of papers (see \cite{Chen:2008wn} for a summary and list of references). 
A common feature of non-Gaussianities resulting from non Bunch-Davies vacua is the fact that they are enhanced in the \emph{folded} triangle limit, i.e. for $k_1 + k_2 - k_3=0$. The intuitive explanation can be understood looking at Eq. \eqref{eq:diffsol}: in the Bunch-Davies case, only positive frequencies are considered and therefore the second term proportional to $c_2$ is neglected. Negative frequencies are instead produced when deviating from the attractor solution, and the leading order deviation in the in-in calculation of the bispectrum of $\zeta$ has therefore at least one contribution from them. Effectively, this translates into sending one of the three momenta from $k\rightarrow -k$. 


\paragraph{{\bf Solid inflation}}

Solids can be described in the context of field theory \cite{Dubovsky:2005xd}  by introducing three scalar fields $\phi^I$ whose background values are identified with spatial coordinates

\begin{equation}
\langle\phi^I(x)\rangle = x^I
\end{equation} 

and they are time-independent. Using this framework, \cite{Endlich:2012pz} showed that inflation can be driven by a particular type of solid which has approximate dilation symmetry and exact rotational and translational internal symmetries. These symmetries allow to consistently build a solid that stretches during the accelerated expansion by many orders of magnitude. Even though the treatment of this model is done in the context of an effective field theory, the time-independence of the background fields implies that there is no breaking of time-translational invariance as in conventional effective field theory approaches of inflation. The computation of cosmological perturbations also shows peculiarities: for instance, adiabatic perturbations during inflation are absent \cite{Endlich:2012pz}.

 Most interestingly for this review, the three-point function of scalar perturbations drastically violates \cite{Endlich:2013jia} the standard single-field consistency relation of Eq. \eqref{eq:maldasqueezed}. The dimensionless bispectrum is calculated to be \cite{Endlich:2012pz}
 
 \begin{equation}\label{eq:solid}
 \mathcal S_{\rm SI}(k_1,k_2,k_3) = \frac {5}{8\pi^4} \, \frac{F_Y}{F}\, \frac{1}{\epsilon\, c_L^2}\, \left(\frac{\tau_e}{\tau_c}\right)^{-4\epsilon c^2_T/3} \frac{Q(k_1,k_2,k_3)\, U(k_1,k_2,k_3)}{k_1 k_2 k_3},
 \end{equation}
 
where $F$ and $F_Y$ are free parameters of the solid Lagrangian, $\epsilon$ plays the role of the slow-roll parameter, $c_L$ and $c_T$ are the speeds of longitudinal and transverse phonons of the solid, respectively, $\tau_c$ is the conformal time at which the longest modes of observational relevance today exit the horizon, while $\tau_e$ is the conformal time at the end of inflation. The functions $Q$ and $U$ are given in Eq. (7.4) and (7.13) of \cite{Endlich:2012pz}, respectively. It is important to notice that Eq. \eqref{eq:solid} is computed for \emph{not-too-squeezed} momenta, that is, for $k_{\rm long}/k_{\rm short} > \sqrt{\epsilon}$. The squeezed limit was investigated in detail in \cite{Endlich:2013jia}, at leading order in slow-roll expansion it can be written in a much more compact form as

\begin{equation}\label{eq:solidsqueezed}
\lim_{k_3\rightarrow 0} B_\zeta(k_1,k_2,k_3) \simeq -\frac{20}{9} \, \frac{F_Y}{F}\, \frac{1}{\epsilon\, c^2_L}\, (1-3\cos^2\theta) \,P_\zeta (k_1)P_\zeta(k_3),
\end{equation}

which should be compared to the consistency relation of Eq. \eqref{eq:maldasqueezed}. Here $\theta$ is the angle between $\bq$ and $\bk$. The fact that the angular dependence is that of a quadrupole and the overall amplitude not being constrained to be small are clear signals of the breaking of the consistency relation, despite the fact that solid inflation propagates only a single scalar mode.

\subsubsection{Multi-field interactions during inflation}\label{sec:massivep}

In this section, we provide a schematic summary of the interactions between massive particles and the inflaton taking place during inflation, which lead to non-Gaussian signatures in the correlation functions of $\zeta$. 

Massive particles might be spontaneously created in the expanding space-time during inflation through non-perturbative effects \cite{Parker:1968mv,Parker:1969au,Parker:1971pt}. 
Their production is particularly interesting as a probe to ultraviolet completions of inflation motivated by string theory constructions \cite{Baumann:2014nda}. We must specify that they are \emph{massive} since their mass cannot be arbitrarily light: to avoid back-reaction on the inflationary background, their typical mass must be of order $H$ or higher. 
On the other hand, the production rate is exponentially suppressed as a function of mass in De-Sitter space-time, roughly as $e^{-m/H}$. Therefore very massive particles, $m\gg H$, decay exponentially fast after horizon crossing and their effect  can be \emph{integrated out} from the dynamics of $\zeta$. 
In this case, inflation is effectively single-field, so that, for instance, the inflationary consistency relation of Eq. \eqref{eq:maldasqueezed} has to be satisfied. Particles with mass of order $H$ instead produce characteristic \emph{non-local} signatures in the correlation functions of $\zeta$ and will generically violate the single-field consistency relations. 
Here we want to stress that we are not only restricting to scalar fields: indeed a tower of high-spin states can arise in string theory constructions \cite{Rindani:1985pi,Aragone:1988yx}. While the effect of massive scalar fields have been thoroughly investigated in the context of quasi-single field inflation \cite{Chen:2009zp,Baumann:2011nk,Assassi:2012zq,Noumi:2012vr,Gong:2013sma,Baumann:2014nda,Arkani-Hamed:2015bza,Lee:2016vti,Bordin:2016ruc,Flauger:2016idt,Baumann:2017jvh,McAneny:2019epy}, higher-spin particles studies arose more recently \cite{Arkani-Hamed:2015bza,Lee:2016vti,Kehagias:2017rpe,Kehagias:2017cym,Baumann:2017jvh,Franciolini:2017ktv,Biagetti:2017viz,Dimastrogiovanni:2018uqy}. 

\paragraph{ Massive particles in De-Sitter}

In the De-Sitter background space-time of inflation, particles can be classified as unitary irreducible representations of the De-Sitter group SO(1,4). Particles are characterized by their spin and mass, and the condition of unitarity imposes three allowed categories \cite{Wigner:1939cj,Bargmann:1948ck} for particles with spin $s\geq 1$

\vspace{2mm}
\begin{center}
\begin{tabular}{ccc}
{\rm principal series} & {\rm complementary series} & {\rm discrete series} \\
$\qquad \frac{m^2}{H^2} \geq \left(s-\frac 12\right)^2\qquad $ & $\qquad s(s-1)<\frac{m^2}{H^2}< \left(s-\frac 12\right)^2\qquad$ & $\qquad\frac{m^2}{H^2} = s(s-1) - t(t+1)\qquad$
\end{tabular}
\end{center}
\vspace{2mm}

for $t=0,1,2, ..., s-1$. Similarly, scalar particles with mass $m\leq 3/2 H$ belong to the principal series, while lightest particles belong to the complementary series. Massless scalar particles are conformally invariant in DS. Notice that particles with spin $s\geq 1$ are required to have a minimal mass, $m^2\geq s(s-1)H^2$, unless they belong to the discrete series\footnote{Light particles with spin are allowed in inflationary set ups which breake DS isometries, as shown in \cite{Bordin:2018pca}}. For these values, the system acquires an additional gauge invariance and the corresponding fields are called \emph{partially massless} fields \cite{Deser:2001pe}. Partially massless fields produce sizeable tensor-scalar-scalar bispectra and the scalar trispectra \cite{Kehagias:2017rpe,Baumann:2017jvh,Franciolini:2017ktv}. In the case of study here, i.e. the scalar bispectrum, partially massless fields do not contribute, as their lack of a longitudinal degree of freedom kinematically prevents them to oscillate into a single scalar field.

As we already anticipated, massive particles decay at late times. The DS group acts as conformal group on the three-dimensional Euclidean space for the super-Hubble fluctuations, so that the asymptotic scaling at late times of a scalar particles is

\begin{equation}
\lim_{\eta\rightarrow 0} \sigma(\eta, \bx) = \sigma^+ (\bx) \eta^{\Delta^+} + \sigma^- (\bx) \eta^{\Delta^-},
\end{equation}

where $\Delta=3/2\pm i \mu$ and 

\begin{equation}
\mu=\sqrt{\frac{m^2}{H^2}-\frac 94},
\end{equation}

where the massless case $m=0$ corresponds to a conformally coupled field that does not decay at late times. We will investigate this case in the next section. 
Similarly, for a spin-$s$ field we get

\begin{equation}
\lim_{\eta\rightarrow 0} \sigma_{i_1 ... i_s} (\eta, \bx) = \sigma^+_{i_1 ... i_s} (\bx) \eta^{\Delta_s^+-s} + \sigma^-_{i_1 ... i_s} (\bx) \eta^{\Delta_s^--s},
\end{equation}

where $\eta$ is conformal time and  $\Delta_s^\pm = 3/2 \pm i \mu_s$ is the conformal weight of the field and

\begin{equation}
 \mu_s \equiv \sqrt{\frac{m^2}{H^2}-\left(s-\frac 12\right)^2}
\end{equation}

will be a crucial parameter in the non-Gaussian shapes, as we will show shortly.  For finite mass fields, the decay scales as the conformal weight and can be distinguished in real values of $\mu_s$, for which the wavefunction oscillates logarithmically in conformal time, and imaginary $\mu_s$, for which particles belong to the complementary series and survive longer at late-times.

\paragraph{Effective approach to interactions of massive particles}

\begin{figure}
\centering
\includegraphics[width=\linewidth]{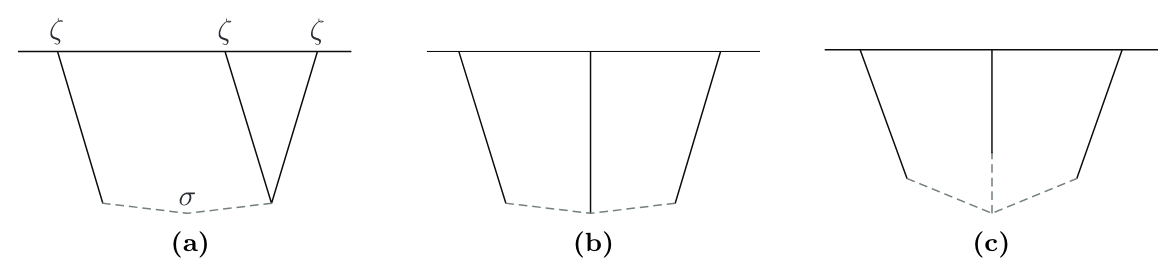}
\caption{ Tree-level diagrams contributing to the three-point correlator $\langle \zeta\zeta\zeta\rangle$. The diagrams represent the exchange of spin-$s$ particles $\sigma$ (dashed lines) with the curvature perturbation $\zeta$ (solid lines). Figure from \cite{Lee:2016vti}.}
\label{fig:diag}
\end{figure}

The interactions of the massive particles described above in the context of inflation has been investigated using the effective field theory of inflation in \cite{Lee:2016vti}. This extends earlier works focusing only on scalar fields. At tree-level, \cite{Lee:2016vti} distinguished three diagrams, illustrated in Fig. \ref{fig:diag}, which represent three different ways in which spin-$s$ fields can be exchanged in the three-point correlator $\langle \zeta\zeta\zeta\rangle$.

The corresponding dimensionless bispectrum for these three diagrams can be written in the following compact form

\begin{equation}\label{eq:massbis}
\mathcal S_{\rm MP} (k_1,k_2,k_3) = \alpha_s^{(\kappa)}\, \mathcal P_\zeta(k_*)^{-1} \, \mathcal L_s^{(\kappa)}(\hat \bk_1,\hat \bk_2,\hat \bk_3) \, \mathcal I_s^{(\kappa)}(\mu_s,c_\pi,k_1,k_2,k_3) + 5\mbox{ perms.},
\end{equation}

where $(\kappa) = (a), (b)$ and $(c)$ represents the three diagrams, $\alpha^{(\kappa)}_s$ are dimensionless parameters, $L_s^{(\kappa)}$ are functions of the angles between momenta (typically Legendre polynomials) and $\mathcal I^{(\kappa)}_s$ are complicated integrals whose expressions can be found in \cite{Lee:2016vti}\footnote{ Note that there is a conversion factor of $(k_1 k_2 k_3)^2$ between $\mathcal I_s^{(\kappa)}$ and those defined in \cite{Lee:2016vti}. For instance, $\mathcal I^{(a)}_s= (k_1 k_2 k_3)^2\mathcal I^{(s)}$.} and $c_\pi$ is the sound speed of the Goldstone boson in the effective field theory of inflation. Although in a compact form, the importance of Eq. \eqref{eq:massbis} is manifest: $\mathcal S_{\rm MP}$ depends on the mass and spin of the particle which mediates the exchange, so it provides a promising way to detect new particles at the high energies at which inflation takes place \cite{Arkani-Hamed:2015bza}. 

To get a more explicit idea of this type of bispectra, let us restrict to the squeezed limit of the single-exchange diagram $(a)$. Looking at the squeezed limit is relevant because of the single-field consistency relation, Eq. \eqref{eq:maldasqueezed}, which implies that in this limit non-Gaussianity of order $\mathcal O(\fnl) \gtrsim 0.1$ is only sourced by the presence of multi-field scenarios.  In this limit, the bispectrum splits into an analytic part and a non-analytic part. The analytic part reflects local effects of massive particles with a scaling similar to the case of single-field models, Eq. \eqref{eq:malda} and it does not contain information on the spin and mass of the particle at leading order,

\begin{equation}\label{eq:shapean}
\lim_{k_1\ll k_3} \mathcal S^{\rm an}_{\rm MP}(k_1,k_2,k_3) \propto \frac{k_1}{k_3}.
\end{equation}

The non-analytic part for scalar particles has been derived in the context of quasi-single field inflation \cite{Chen:2009zp}. It has to be distinguished into two cases: for massive particles belonging to the principal series, $\mu \geq 0$ it reads

\begin{equation}
\lim_{k_1\ll k_3} \mathcal S^{\rm non-an}_{\rm QSF}(k_1,k_2,k_3) \propto \left(\frac{k_1}{k_3}\right)^{1/2} \cos \left[ \mu\, \ln \left(\frac{k_1}{k_3}\right)\right],
\end{equation}

while for particles belonging to the complementary series for which $\mu$ becomes imaginary, the scaling changes to 

\begin{equation}
\lim_{k_1\ll k_3} \mathcal S^{\rm non-an}_{\rm QSF}(k_1,k_2,k_3) \propto \left(\frac{k_1}{k_3}\right)^{1/2-\nu},
\end{equation}

where $\nu = -i\mu$ is real. The case of higher-spin particles proceeds analogously: massive particles belonging to the principal series, $\mu_s \geq 0$, have a squeezed limit of the form

\begin{equation}
\lim_{k_1\ll k_3} \mathcal S^{\rm non-an}_{\rm HS}(k_1,k_2,k_3) \propto \left(\frac{k_1}{k_3}\right)^{1/2} \mathcal L_s(\hat \bk_1,\hat \bk_3) \cos \left[ \mu_s\, \ln \left(\frac{k_1}{k_3}\right)+\phi_s\right],
\end{equation}

for even spins, where the $\mathcal L_s$ is the Legendre polynomial of order $s$ and $\phi_s$ is a phase that is uniquely fixed in terms of $\mu_s$ and $c_\pi$ \cite{Lee:2016vti}. This oscillatory scaling is the distinctive feature of this type of interactions and it tells information about the mass and spin of the particle involved in the exchange. 

For particles belonging to the complementary series for which $\mu_s$ becomes imaginary, the scaling changes to 

\begin{equation}
\lim_{k_1\ll k_3} \mathcal S^{\rm non-an}_{\rm HS}(k_1,k_2,k_3) \propto \left(\frac{k_1}{k_3}\right)^{1/2-\nu_s} \mathcal L_s(\hat \bk_1,\hat \bk_3),
\end{equation}

where $\nu_s = -i\mu_s$ is real. In the case of odd spins, for symmetry reasons the squeezed limit of the non-analytic part is suppressed by at least an additional power of $k_1/k_3$, therefore the leading piece is always the analytic one of Eq. \eqref{eq:shapean}. Both in the case of scalar and higher-spin massive particles, the least suppressed scaling in powers of the squeezed ratio is the limiting case of $m=s(s-1)$ and corresponds to a constant shape $\mathcal S \propto (k_1/k_3)^0$. 


\subsubsection{Multi-field interactions after inflation}\label{sec:after}

Massive particles necessarily decay after horizon crossing during inflation, as seen in the previous section. This is not the case for massless particles: they survive at late times and typically have a non-linear evolution in multi-field space on superhorizon scales, which generate isocurvature perturbations (see \cite{Riotto:2018pcx} for recent lectures on the topic). Observations of the CMB constrain the amount of isocurvature perturbations to be very small \cite{Akrami:2018odb}. Nevertheless, there are several mechanisms with which isocurvature perturbations can be converted to curvature ones after inflation and evade CMB constraints \cite{Salopek:1990jq,Bartolo:2001cw,Bernardeau:2002jy,Bernardeau:2002jf,Rigopoulos:2005xx,Rigopoulos:2005ae,Seery:2005gb,Vernizzi:2006ve}. We will not go into details of specific realizations of this conversion mechanism. Instead, we will summarize the results obtained in the framework of the $\delta N$-formalism \cite{Starobinsky:1986fxa,Sasaki:1995aw,Lyth:2005fi}, which is rather model-independent. It will quickly become clear that the  shape of this type of non-Gaussianities is generic and goes under the name of \emph{local} non-Gaussianity. Local-type non-Gaussianities are the most studied in the literature (see \cite{Salopek:1990jq,Gangui:1993tt,Verde:1999ij,Komatsu:2001rj} for the earliest studies) and can be realized by a wide range of models. The name \emph{local} comes from the fact that it can be expressed as a simple Taylor expansion around a Gaussian field at position $\bx$,

\begin{equation}
\zeta(\bx) = \zeta_{\rm G}(\bx) + \frac 35 \fnl^{\rm loc} \left(\zeta^2_{\rm G}(\bx) - \langle \zeta^2_{\rm G}\rangle\right) + \mathcal O\left( \zeta^3_{\rm G} \right).
\end{equation}

The corresponding dimensionless bispectrum in Fourier space is of the form

\begin{equation}\label{eq:slocal}
\mathcal S_{\rm loc}(k_1,k_2,k_3) = \frac{6}{5} \fnl^{\rm loc} \left( \frac{k_1^2}{k_2 k_3} +  \frac{k_2^2}{k_1 k_3} +  \frac{k_3^2}{k_1 k_2}\right)
\end{equation}

and it has the nice feature of being already factorizable and simple to implement as optimal estimator in observed quantities. Moreover, it is a distinctive signature of multi-field models, since it peaks in the squeezed limit, where single-field models are necessarily producing very small non-Gaussianity as implied by the consistency relation Eq. \eqref{eq:maldasqueezed}. 

\paragraph{The $\delta N$-formalism}

In the single-field model we introduced in \S\ref{sec:power}, the comoving curvature perturbation $\zeta$ is constant on superhorizon scales. Let us consider the gauge in which the inflaton perturbations are zero and only fluctuations on the metric are present. In this gauge, the spatial metric is simply given by $a^2 e^{2\zeta}$. 
Each comoving superhorizon patch in the universe will have its own frozen value of $\zeta$ which determines the local evolution in space, independently of the other disconnected patches, through the scale-dependent scale factor. It can be shown that this is equivalent to say that each patch evolves as a \emph{separate universe} each with slightly different number of e-folds, $\delta N$, at different positions. This interpretation is dubbed $\delta N$-formalism.
The generalization of this picture to the case of multi-field inflation is well summarized in \cite{Chen:2010xka}, here we broadly follow its steps. 

Let us consider a set of scalars $\phi_i  = \phi_{0i}+\delta\phi_i$ during inflation.  We are interested in the behavior of the perturbations of these fields on superhorizon scales, looking at different causally disconnected patches. We choose an initial spatially flat slice where scalar metric fluctuations are zero. Note that modes we are interested in are already superhorizon on this slice, so we will evolve them classically. Moreover, we assume their statistics to be Gaussian. We now select uniform density final slices, that is, slices with the same energy density at each point of them. We define $N_0(\phi_{0i})$ as the number of e-folds between the initial and final slices for the unperturbed field $\phi_{0i}$, while the one for the perturbed fields  will be $N( \phi_{0i}+\delta\phi_i(\bx))$. The curvature perturbation $\zeta$ can be expressed as a Taylor expansion of the variation of $N$ around the initial values $\phi_{0i}$ as

\begin{align}\label{eq:deltan}
\zeta=\delta N& =N( \phi_{0i}+\delta\phi_i(\bx))-N_0(\phi_{0i})\\
&= N_i \delta\phi_i + \frac 12 N_{ij} \delta\phi_i\delta\phi_j+ ...,
\end{align}

where the subscripts on $N$ denote partial derivatives with respect to $\phi_i$ and indices are summed using the Einstein convention, with $i=1,..., N$, being $N$ the number of fields. Correlation functions of $\zeta$ are then related to correlation functions of $\delta\phi_i$, which we assumed to be Gaussian distributed massless fields. We have computed the mode functions of each of these fields in Eq. \eqref{eq:diffsolBD}, their power spectrum is

\begin{equation}
\langle \delta\phi_i(\bk_1)\delta\phi(\bk_2)\rangle' = \frac{H_*^2}{2 k_1^3} \delta_{ij},
\end{equation}

where the prime indicates that we are leaving the usual delta function implicit and $H_*$ is the Hubble parameter at horizon exit. Consequently, using Eq. \eqref{eq:deltan} we can express the bispectrum of the curvature perturbation as

\begin{equation}
\langle\zeta(\bk_1)\zeta(\bk_2)\zeta(\bk_3)\rangle' = N_{ij} N_i N_j \frac{H^4_*}{4} \left(\frac{1}{k_1^3 k_2^3}+\frac{1}{k_1^3 k_3^3}+\frac{1}{k_2^3 k_3^3}\right)
\end{equation}

and it can be easily shown that the corresponding dimensionless bispectrum $\mathcal S$ is the local one of Eq. \eqref{eq:slocal} and 

\begin{equation}
\fnl^{\rm loc} = \frac 56 \frac{N_{ij} N_i N_j}{(N_l^2)^2}.
\end{equation}

The fact that this shape is local should not come as a surprise: this bispectrum is sourced by local interactions of fields on superhorizon scales. A number of multi-field models can be described using the  $\delta N$-formalism, such as the \emph{curvaton} model, the modulated reheating and preheating  scenarios (see \cite{Bartolo:2004if} for a review).

\subsection{Final remarks of this section}\label{sec:remarksinf}

\begin{figure}[!htb]
\centering
\includegraphics[width=\linewidth]{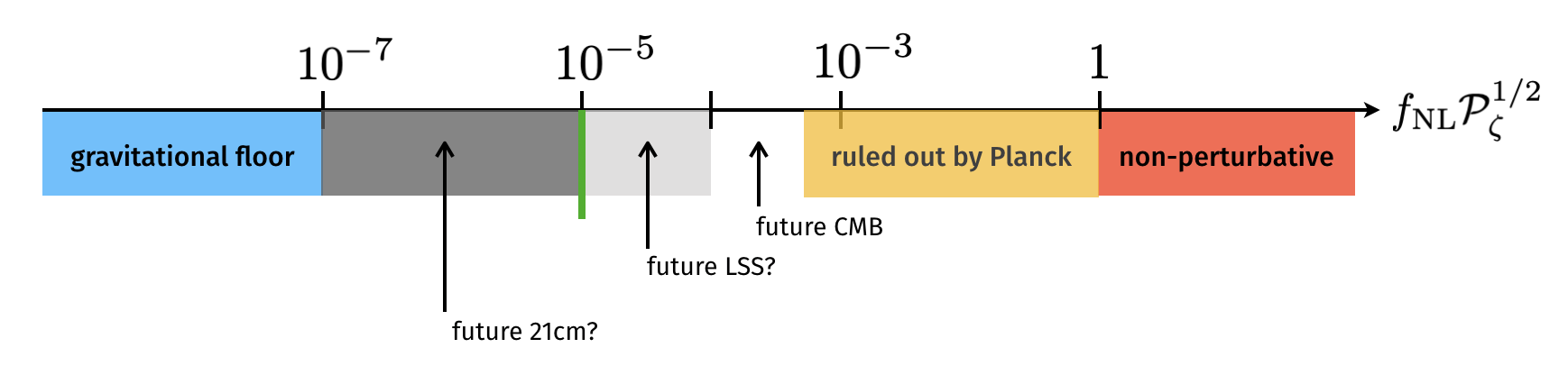}
\caption{ A schematic representation of current and future limits on the primordial non-Gaussianity amplitude parametrized by $\fnl$.  Credits to \cite{Lee:2016vti}.}
\label{fig:fnltime}
\end{figure}

We have identified three main features that we can extract from the bispectrum of the curvature perturbation $\zeta$: shape, running and amplitude. Among these three, the amplitude, parametrized via a single parameter $\fnl$, is the easiest to constrain from data and it tells us already a lot about what is the source of non-Gaussianity, as shown schematically in Fig. \ref{fig:fnltime}. 

The limit of $\fnl \sim \mathcal O(1)$ is particularly important for interactions generated by non-trivial higher-order kinetic terms in single-field inflation: it would signal that interactions come at an energy scale which coincides with $\dot\phi_0$, which is the scale at which the exact DS background evolution is broken. These are non-Gaussianities typically peaking on equilateral shapes of the bispectrum (i.e. with $k_1\approx k_2\approx k_3$). Furthermore, a positive detection of an order unity $\fnl$ for squeezed bispectra (i.e. with $k_1\ll k_2\approx k_3$) would rule out single-field models of inflation in one shot. The ultimate goal is to measure $\fnl$ down to the level of slow-roll parameters, $\fnl \lesssim \mathcal O(0.01)$. At these limits, inflation predicts that non-Gaussianity has to be there no matter the inflationary model. Experimental sensitivity seems to be still far from this goal.

Current best limits on non-Gaussianity are set by observations of the bispectrum of CMB temperature anisotropy from the Planck satellite \cite{Akrami:2019izv}

\begin{equation}
\fnl^{\rm loc} = -0.9\pm 5.1,\qquad \fnl^{\rm equil}= -26 \pm 47,\qquad \fnl^{\rm orth} = -38\pm 24.
\end{equation}

at $68\%$ confidence level. While these limits refer directly to Eq. \eqref{eq:slocal} for the local shape, the equilateral and orthogonal constraints come from \emph{templates}, rather than one of the shapes presented above. This is because the analysis from data require optimal estimators of the bispectrum which need to sum over all modes available in the survey. This implies that only factorizable shapes, such as the local one of Eq. \eqref{eq:slocal}, are usable in practice \cite{Creminelli:2003iq,Fergusson:2009nv}. For this reason, templates need to be used for CMB constraints in place of the realistic predictions we have outlined above and the two can be related using so-called ``fudge-factors''. This has been applied also to LSS \cite{Wagner:2010me,Fergusson:2010ia,Wagner:2011wx}. We will only briefly mention this issue in Sec. \S\ref{sec:nbodymulti} in the context of N-body simulations.

In this section, we have shown that interactions during  and after inflation necessarily take place even in the minimal scenario of single-field models through minimal coupling with gravity. This implies that there is a \emph{gravitational floor} for detectable non-Gaussianities in the CMB and LSS and therefore a guaranteed signal. In non-minimal scenarios, consistency relations and constraints from current observations allow to classify interactions from realistic models of the early Universe and in some cases even open the possibility to use inflation as a cosmological particle collider \cite{Arkani-Hamed:2015bza} and explore new physics at energies as high as $10^{14}$ GeV. Such powerful predictions are rare in cosmology, and should be a primary target for experimental searches. 

\section{From primordial interactions to matter overdensities}\label{sec:transfer}

In this section, we briefly review how non-Gaussianities are transferred from the primordial curvature perturbation to the distribution of dark matter and its correlation functions.  We will show how non-Gaussian initial conditions affect directly the mass density probability function and also generate additional terms in the correlation functions of dark matter at different positions.  We will argue that these effects are typically very small and their search is complicated by the fact that we do not have direct access to dark matter correlation functions\footnote{Weak lensing \cite{Bartelmann:1999yn} and temperature anisotropies in the 21cm background from the pre-reionization epoch \cite{Loeb:2003ya,Cooray:2004kt,Pillepich:2006fj} can be used to trace DM directly, although the latter might also be biased \cite{Pritchard:2006sq}}.

The initial conditions for structure formation are set by connecting the primordial curvature perturbation $\zeta$ to the Newtonian potential $\Phi$ via a transfer function, $T(k)$. At the linear level and using the Poisson equation, we can write the useful relation

\begin{equation}\label{eq:transf}
\delta_{\rm L}({\bf k},z) = \mathcal M(k,z) \zeta({\bf k})\quad \mbox{where}\quad \mathcal M(k,z) = \frac 25 \frac{k^2 T(k) D(z)}{\Omega_{m,0} H^2_0},
\end{equation}

which defines the linear (L) matter overdensity. Here $D(z)$ is the linear growth factor normalised to unity at present day and $\Omega_{m,0}$ and $H$ are the matter density and hubble parameter today, respectively. The linear relation is reliable as long as the matter overdensity $\delta$ is much smaller than unity. In this regime, one can  solve the (Newtonian) dynamical equations perturbatively (see \cite{Bernardeau:2001qr} for a review and \cite{Baumann:2010tm,Carrasco:2012cv,Pajer:2013jj,Carrasco:2013sva,Mercolli:2013bsa,Carrasco:2013mua,Carroll:2013oxa,Porto:2013qua,Senatore:2014via,Baldauf:2015tla,Assassi:2015jqa,Baldauf:2015aha,Baldauf:2015zga,Lewandowski:2017kes,Senatore:2017pbn} for an approach using effective field theory). Even pushing perturbation theories to their limits, theoretical predictions describing the evolution of matter are valid only in the weakly nonlinear regime,  that is for $k \lesssim 0.2$ h/Mpc at redshift $0$, even in the case of Gaussian initial conditions. Consequently, for all practical purposes, it is convenient to smooth small scale perturbations with a window function $W_R$, being $R$ the smoothing scale, and we indicate the smoothed field as $\delta_R$. Commonly used window functions in LSS are spherically symmetric functions, such as top-hat filters in real space or Gaussian windows. Consequently, the linear density power spectrum smoothed on a scale $R$ is given by

\begin{equation}\label{eq:mtrps}
P_R(k,z) = \mathcal M^2(k,z) W_R^2(k) P_\zeta(k) 
\end{equation}

and its variance

\begin{equation}
\sigma^2_R(z) = \int \frac{\d^3 k}{(2\pi)^3}   \mathcal M^2(k,z) W_R^2(k)  P_\zeta(k) .
\end{equation}

To avoid clutter, we will condense the transfer function and the smoothing window into $\mathcal M_R(k,z)\equiv \mathcal M (k,z) W_R(k)$ and drop the redshift dependence unless needed.



\subsection{The density probability distribution}\label{subsec:pdf}

The probability distribution function (PDF) of the smoothed density field at early times is usually assumed to be Gaussian at each fixed point in space ${\bf x}$,

\begin{equation}
\mathcal P(\delta_R)\d \delta_R = \frac{1}{\sqrt{2\pi}\sigma_R}\exp \left(-\frac{\delta_R^2}{2\sigma_R^2}\right),
\end{equation}

i.e. the value of the smoothed density field $\delta_R$ at each ${\bf x}$ is drawn from a Gaussian distribution.
We know however from the previous section that small initial non-Gaussianities are always present, even in the minimal case in which only a single-field and gravity are present during inflation. 
These non-Gaussianities source higher-than-two reduced smoothed cumulants, defined as

\begin{equation}
S_J (R) \equiv \frac{\langle \delta_R^J \rangle_c}{\langle \delta_R^2 \rangle_c^{J-1}},
\end{equation}

where $\langle \delta_R^N \rangle_c$ is the connected n-point moment. Note that, in the linear regime, $\sigma_R \propto D(z)$ and therefore $S_N \propto 1/ D^{N-2}$, which means that for $N\geq 3$, higher order cumulants are increasingly suppressed by powers of the linear growth factor $D$. As a working example, let us compute the lowest order cumulant which is non-zero in the presence of non-Gaussianity, the \emph{skewness}

\begin{equation}
\sigma^4_R S_3 = \langle \delta_R^3 \rangle_c =\int \frac{\d^3 k_1}{(2\pi)^3} \frac{\d^3 k_2}{(2\pi)^3} \frac{\d^3 k_3}{(2\pi)^3} B_R (\bk_1,\bk_2,\bk_3)
\end{equation}

where

\begin{equation}
B_R (\bk_1,\bk_2,\bk_3) = \mathcal M_R(k_1) \mathcal M_R(k_2) \mathcal M_R(k_3) \langle \zeta(\bk_1)\zeta(\bk_2)\zeta(\bk_3) \rangle.
\end{equation}

It is easy to recognize the inflationary three-point function as a source of the skewness. The minimal amount of skewness produced in the initial conditions can be computed by making use of Eq. \eqref{eq:malda} and gives $\sigma_R S_{3,GF} = A(R) \epsilon + B(R) \eta$, being $A$ and $B$ mild functions of the scale $R$ and with amplitude $A,B \propto \mathcal O(10^{-4})$ for $R\approx 3 Mpc/h$. Similar values apply also for the local type non-Gaussianities of Eq. \eqref{eq:slocal}.
To get the physical intuition of how the PDF changes, it is enough to take the simplest approach and perform an Edgeworth expansion on the smoothed field $\delta_R$, or more commonly the so-called peak height $\nu = \delta_R / \sigma_R$, to get

\begin{equation}\label{eq:ploverde}
\mathcal P(\nu)\d\nu = \frac{e^{-\frac{\nu^2}{2}}}{\sqrt{2\pi}} \left[ 1+ \frac{\sigma_R S_3 (R)}{6} H_3 (\nu) + \frac{ \sigma^2_R S_4(R)}{24} H_4(\nu) + \frac{(\sigma_R S_3)^2}{72} H_6(\nu) + ... \right] \d\nu,
\end{equation}

where $H_N$ are Hermite polynomials. Note that, as a consequence of how the reduced cumulants scale with the linear growth factor, the combinations $\sigma S_3$ and $\sigma^2 S_4$ are redshift independent in the linear regime. More details on these expansion, and its refinements, can be found in several analyses dating back to the 90s \cite{Juszkiewicz:1993vk,Fry:1993xy,Chodorowski:1995qj,Gaztanaga:1997nz,Durrer:2000gi}.

The generic effect of a positive (negative) skewness on the PDF is to produce more overdense (underdense) regions in the matter distribution. As perturbations grow from the initial time to later times, gravitational instabilities generate a positive skewness in the PDF which
eventually dominates over primordial contributions. The effect of the skewness from gravitational evolution can be easily computed within perturbation theory to be $S_3 \approx 34/7$ at lowest order  \cite{Bernardeau:2001qr}, which is much larger than the one sourced by primordial bispectra respecting the current constraints from the CMB.

Beside studies in N-body simulations, the PDF of the matter density field is not a direct observable. Nevertheless, a PDF with non-Gaussian initial conditions has a direct impact on the abundance of clustered objects and on the clustering statistics. This is what we discuss in the following section \S\ref{sec:onepoint}.

\subsection{Dark matter correlation functions}

Primordial perturbations are transferred to the matter density field via Eq. \eqref{eq:transf} and correlation functions are necessarily affected. The matter $N$-point correlation functions reads, at the linear level,

\begin{equation}\label{eq:mtrcorr}
\langle \delta_{\rm L} ({\bf k}_1,z) \cdots \delta_{\rm L} ({\bf k}_N,z) \rangle = \left(\prod_{i=1}^{N} \mathcal M({\bf k},z) \right) \langle \zeta({\bf k}_1) \cdots \zeta({\bf k}_N)\rangle.
\end{equation}

It is immediately clear from Eq. \eqref{eq:mtrcorr} that a bispectrum, or higher-order correlations, of $\zeta$ source at least the corresponding matter field correlator. As remarked above, gravitational evolution also sources secondary non-Gaussianities though, and these eventually dominate on the primordial ones. 
In the next section, we will show that the most promising observables for disentangling secondary non-Gaussianities from primary ones involve  \emph{biased} tracers of the matter field, rather than directly probing the dark matter field.
Nevertheless, let us review a few essential points about dark matter correlation functions with non-Gaussian initial conditions, while we redirect the interested reader to the latest analyses in the context of the effective field theory for a wide range of non-Gaussianities \cite{Assassi:2015jqa}.

In standard Eulerian perturbation theory (EPT) (see \cite{Bernardeau:2001qr} for a review), one follows the evolution in time of the matter density contrast $\delta$ at position ${\bf x}$. In the weakly non-linear regime, the Fourier mode of the density field reads 

\begin{equation}
\delta(\bk,z) = \delta_{\rm L}(\bk,z) + \int \frac{\d^3 q}{(2\pi)^3} F_2(\bq, \bk-\bq) \delta_{\rm L}(\bq,z)\delta_{\rm L}(\bk-\bq,z) +\mathcal O (\delta_{\rm L}^3),
\end{equation}

where we simply denote $\delta$ as the unsmoothed, non-linear matter field and $F_2$ is the PT kernel

\begin{equation}
F_2(\bq_1,\bq_2) = \frac 57 + \mu\left(\frac{q_1}{q_2}+\frac{q_2}{q_1}\right) + \frac 27 \mu^2
\end{equation}

and $\mu$ is the cosine between $\bq_1$ and $\bq_2$. For the present analysis, it is not essential to smooth the matter field, so we work only with the unsmoothed $\delta$. A contribution from a non-zero primordial bispectrum therefore arises already on the matter power spectrum at the $1$-loop order and reads

\begin{equation}\label{eq:mmng}
P^{\rm NG} (k,z) = \int\frac{\d^3 q}{(2\pi)^3} 2 F_2(\bq, \bk-\bq) \mathcal M(k,z) \mathcal M(q,z) \mathcal M(|\bk-\bq|,z)B_\zeta(-\bk,\bq,\bk-\bq),
\end{equation}

where notice that here we have used the statistical homogeneity and isotropy to write the momentum dependence of $B_\zeta$.  The full 1-loop power spectrum is defined as

\begin{equation}
P_{1-loop}(k,z) = P^{\rm G}_{\rm L}(k,z) + P^{\rm G}_{22}(k,z) + P^{\rm G}_{13}(k,z) + P^{\rm NG} (k,z),
\end{equation}

where $P_{22}$ and $P_{13}$ are the $1$-loop contributions to the Gaussian case \cite{Bernardeau:2001qr}. It is interesting to notice that this term scales as $D(z)^3$, while $1$-loop terms from purely Gaussian initial conditions scale as $D(z)^4$. The non-Gaussian correction is suppressed at large scales, since the kernel $F_2$ vanishes in that limit. On the other hand, at small scales late-time non-linearities become quickly important as midly non-linear scales $k\sim 0.1$ h/Mpc are approached. Moreover, in \cite{Assassi:2015jqa} it was shown that the $2$-loop Gaussian terms are comparable to the non-Gaussian ones at $1$-loop even in the mildy non-linear regime. Therefore, deviations from the Gaussian case are hardly reaching the percent level for non-Gaussianities within current constraints. 

The matter bispectrum is sensitive at tree-level to primordial corrections as well as to gravitational non-linearities,

\begin{equation}\label{eq:bistreedm}
B^{\rm tree}(\bk_1,\bk_2,\bk_3,z) = \mathcal M(k_1,z) \mathcal M(k_2,z) \mathcal M(k_3,z)B_\zeta(\bk_1,\bk_2,\bk_3) + \left[ 2F_2(\bk_1,\bk_2) P_{\rm L}(k_1,z) P_{\rm L}(k_2,z) + 2 \mbox{ perms. } \right].
\end{equation}

Extensions to higher loops have been computed in the context of perturbation theory (see \cite{Assassi:2015jqa,Welling:2016dng} for a treatment in the effective field theory of LSS). It is customary to define a \emph{reduced} bispectrum

\begin{equation}
Q_3 (\bk_1,\bk_2,\bk_3,z) = \frac{B (\bk_1,\bk_2,\bk_3,z)}{\left[P(k_1,z) P(k_2,z) + 2 \mbox{ perms. } \right]},
\end{equation}

which is time- and scale-independent in the Gaussian case at tree-level in perturbation theory for equilateral configurations, so it is the appropriate quantity to study different types of non-Gaussianities. Several studies have tested theoretical predictions against N-body simulations \cite{Sefusatti:2010ee,Figueroa:2012ws,Lazanu:2015bqo}. The prospects for observing the imprint of non-Gaussianities in the matter bispectrum are also rather weak for two main reasons: first, the dark matter field is not directly observable, with the exception of the case of weak leansing bispectrum measurements, for which, however, it was shown that the sensitivity to primordial non-Gaussianity is around two orders of magnitude away from current limits \cite{Schaefer:2011kb}. Secondly, the amplitude of late-time non-linearities largely surpasses the primordial contributions on all scales of interests for future observations. Moreover, similarly to the case of the one-loop power-spectrum, the contribution from the two-loop gaussian bispectrum is larger than the one-loop non-Gaussian one \cite{Welling:2016dng}.

\section{Imprints of primordial interactions on one-point halo statistics}\label{sec:onepoint}

As anticipated in the previous section \S\ref{subsec:pdf}, any higher-order correlation function generated during or right after inflation sources non-Gaussian terms on the PDF of the dark matter field. At early times, this is the only source of non-Gaussianity on the matter distribution and it implies a deviation from an equally probable abundance of overdense and underdense regions. Even as gravitational evolution generates secondary non-Gaussianities, the trace of the initial conditions can be disentangled to a certain degree by looking at very massive halos, whose formation is highly sensitive to the tails of the initial PDF. Massive halos are also more likely to be associated to peaks of the early time density field as shown in multiple studies in N-body simulations \cite{Ludlow:2010xd,Elia:2011ds,Ludlow:2011jx}, confirming that they are sensitive to the initial PDF. 
The pioneers in the analytic treatment of non-Gaussianities in the mass density PDF and in the abundance of DM tracers date back to the $1980$s  \cite{Adler1981,Matarrese:1986et,Lucchin:1987yv,Catelan:1988zb, 1989A&A...215...17B,Matarrese:2000iz,Scoccimarro:2003wn}. Almost as early, numerical methods have been used to test predictions and provide fits to data \cite{Moscardini:1990zh,Weinberg:1991qe,Matarrese:1991sj,Park:1991mh,Gooding:1991ys,Borgani:1993nz}. Since then, the growing interest in primordial non-Gaussianity has pushed for multiple developments in both directions. 

{\bf Plan of the section.} This section has two main goals: i) broadly summarize the theoretical advances by providing a background with the earliest attempts and then focusing on a few of the most recent progresses (Section \S\ref{sec:onepointanalytic}) and ii) presenting the most recent numerical attempts at testing current predictions and providing with semi-analytic or fully fitted phenomenological models in Section \S\ref{sec:onepointnumeric}. 
We then conclude with a few remarks in Section \S\ref{sec:onepointremarks}. This plan is far from being exhaustive and will inevitably refer the reader to the available literature for many details. 

\subsection{Analytic approaches}\label{sec:onepointanalytic}
In order to extract most efficiently information about the early PDF of the matter field, which would give constraints on primordial non-Gaussianity, a consistent theoretical framework of halo formation is needed. In practice, however, what we really measure is number counts of biased tracers, such as galaxies and clusters of galaxies.
It is therefore sufficient to have  a working model for the number density of halos of a certain mass per unite volume, known as the \emph{halo mass function}. Since the earliest attempts of modeling the abundance of bound objects in the presence of non-Gaussian initial  conditions, this search has progressed side to side with models of the simpler Gaussian case: for instance, several attempts tried to extend the Press and Schechter (PS) mass function \cite{Press:1973iz} to local-type non-Gaussianities in the initial conditions \cite{LoVerde:2007ri,Jimenez:2009us,Desjacques:2009jb,Lam:2008rk,Lam:2009nb,Lam:2009nd}. 
This extension was also studied for higher order primordial non-Gaussianities \cite{Desjacques:2009jb,Chongchitnan:2010xz,LoVerde:2011iz} and a range of other bispectrum shapes \cite{Barnaby:2011pe} ; the excursion set approach \cite{Epstein01111983,Bond:1990iw,Lacey:1993iv,Sheth:1998ew,Porciani1998}, which was introduced to solve problems suffered by the PS model, was thoroughtly also investigated with non-Gaussianity in the initial conditions  \cite{Inoue:2001fa,Avelino:1999hs,Lam:2008rk,Lam:2009nb,Lam:2009nd,Lam:2009rz,Maggiore:2009rx,DeSimone:2010mu,Musso:2011ck,DAmico:2010dwy,DAmico:2010ywu,Paranjape:2011uk,Achitouv:2011sq,Achitouv:2012mk,Achitouv:2012ux}.
Lastly, the peak model \cite{Bardeen:1985tr}, which was recently combined with the excursion set approach in the Excursion Set Peaks (ESP) model \cite{Paranjape:2012ks,Paranjape:2012jt}, has been applied to non-Gaussian peak statistics  \cite{Gay:2011wz,Codis:2013exa,Desjacques:2013qx,Biagetti:2015exa,Lazeyras:2015giz,Uhlemann:2017tex}.  Recently, methods based on spherically averaging cosmic densities have been shown to successfully disentangle primordial non-Gaussianities with late time ones \cite{Uhlemann:2017tex}.

\subsubsection{The Press-Schechter mass function}

The Press-Schechter (PS) model \cite{Press:1973iz} is a good framework to make simple analytic computations, hence we use it here to show the main physical intuition on the problem and then briefly mention extensions which improve it. It is based on the fundamental assumption that collapse is spherical \cite{Gunn:1972sv}. Let us first take the case of Gaussian initial conditions. The PDF for the density field $\delta_R$ smoothed on a scale $R \propto M^{1/3}$, where $M$ is the halo mass, is a simple Gaussian with zero mean and variance $\sigma^2_R$. In the PS approach the halo mass function reads

\begin{equation}\label{eq:psmf}
\frac{\d n}{\d M} \equiv \bar n(M) = -\frac{\bar \rho}{M} \frac{\d F_{>\delta_{sc}}(M)}{\d M} = \sqrt{\frac{2}{\pi}} \frac{\bar \rho}{M^2} \nu e^{-\nu^2/2} \frac{\d\ln \nu}{\d\ln M},
\end{equation}

where $\bar\rho$ is the mean matter density, $F_{>\delta_{sc}}$ is the level excursion probability that the density field smoothed on a scale $M$ has overcome the threshold for spherical collapse, $\delta_{sc}\approx 1.687$\footnote{ This value corresponds to a Universe completely dominated by matter. Small corrections need to be taken into account when including the late time domination of dark energy.}, and $\nu(M,z) = \delta_{sc} / \sigma_R(z)$ is the peak height\footnote{ The smoothing scale $R$ is directly related to the mass of the halo $M$, we will use them interchangeably.}. Here we have corrected for a factor of $2$ which accounts for the \emph{cloud-in-cloud} problem,  that is, the fact that the PS mass function does not include the possibility that overdense regions can be contained in bigger, underdense, ones. Excursion set models solve this problem \cite{Epstein01111983,Bond:1990iw,Lacey:1993iv,Sheth:1998ew,Porciani1998}. The halo mass function of Eq. \eqref{eq:psmf} represents the differential number density of halos per unit mass and volume and it is an exponentially decreasing function of $\nu$, i.e. more massive halos are more rare. Despite the fact that the linear theory value of the threshold for spherical collapse is used, $\delta_{sc}$, the calculation is fully non-linear \cite{Gunn:1972sv}.  A feature of Eq. \eqref{eq:psmf}, and common to all \emph{universal} mass functions\footnote{As we discuss in Sec. \S\ref{sec:universality}, the assumption of universality of the mass function, although well tested in N-body simulations with Gaussian initial conditions, is cause of concern in the presence of primordial non-Gaussianity \cite{Biagetti:2016ywx,Desjacques:2017msa}. }, is that it is entirely specified by a function of $\nu$ only, 

\begin{equation}\label{eq:fnu}
 \bar n(M) = \frac{\bar \rho}{M^2} \nu f(\nu) \frac{\d\ln \nu}{\d\ln M},
\end{equation}

where $f(\nu)$ is called \emph{multiplicity} function and we have dropped a subscript $\nu_R$ to avoid clutter, but the smoothing on a scale $R$, corresponding to a halo of mass $M$, is understood. In the PS case we have

\begin{equation}
f_{\rm PS}(\nu) = \sqrt{\frac{2}{\pi}} e^{-\nu^2/2}.
\end{equation} 

The PS mass function of Eq. \eqref{eq:psmf} provides a decent fit to data in the intermediate halo mass regime, but poorly predicts the high mass regime. Nevertheless, an estimation of the non-Gaussian mass function was first given in \cite{Matarrese:2000iz} by performing a saddle-point approximation on an Edgeworth expansion on $F_{>\delta_{sc}}$ and computing the ratio of the non-Gaussian-to-Gaussian multiplicity functions

\begin{equation}\label{eq:mvj}
\mathcal R_{\rm MVJ}(\nu, \fnl) \approx \exp\left(\frac{\nu^3}{6}\sigma S_3\right)\left[-\frac{\sigma\nu^2}{6\nu_*}\frac{\d S_3}{\d\ln \nu_*} + \frac{\nu_*}{\nu}\right],
\end{equation}

where $\nu_*=\delta_{sc}\sqrt{1-S_3\delta_{sc}/3}$ is introduced to enforce the normalisation of the halo mass function.
The non-Gaussian mass function was then obtained by multiplying Eq, \eqref{eq:mvj} to the Gaussian PS mass function.
Alternatively, \cite{LoVerde:2007ri} expanded directly the PDF,  Eq. \eqref{eq:ploverde}, to calculate $F_{>\delta_{sc}}$ and then the ratio

\begin{equation}\label{eq:lv}
\mathcal R_{\rm LV}(\nu, \fnl) \approx 1 + \frac 16 \sigma S_3(\nu^3-3\nu) -\frac 16 \frac{\d(\sigma S_3)}{\d \ln\nu}\left(\nu - \frac{1}{\nu}\right),
\end{equation}

where here we only considered the skewness. Combining the two predictions to better match numerical simulations, \cite{Desjacques:2009jb} got

\begin{equation}\label{eq:ds}
\mathcal R_{\rm DS}(\nu, \fnl) \approx \exp\left(\frac{\nu^3}{6}\sigma S_3\right)\left[1-\frac{\nu}{2}\sigma S_3-\frac{\nu}{6}\frac{\d S_3}{\d\ln \nu}\right],
\end{equation}

where $\nu_*$ from Eq. \eqref{eq:mvj} was expanded to first order in $\fnl$. Earliest checks of these predictions against N-body simulations showed that they tend to overestimate the effect of non-Gaussianity at increasing mass and redshift \cite{Dalal:2007cu,Pillepich:2008ka,Grossi:2009an}.

These formulae can be extended in case of primordial non-Gaussianity from trispectra of $\zeta$. In this case, the kurtosis $S_4$ needs to be added \cite{Desjacques:2009jb,Chongchitnan:2010xz,LoVerde:2011iz}. 

\subsubsection{The excursion set approach}\label{sec:excu}
Press-Schechter  models are known to suffer the cloud-in-cloud problem. The excursion set approach solves this issue and provides an elegant method to count regions above a certain threshold.
The goal of this approach is precisely to find regions that are sufficiently overdense on a given smoothing scale, but not on larger ones. To perform this check, one needs to consider the density field $\delta$ at any given point as a function of the smoothing scale. This function looks similar to a random walk, the starting point being the limit of infinitely large smoothing scales, where the overdensity is zero. In this case, the critical density for collapse defines another curve, typically independent of the point of space, namely a``barrier'' for the random walk. In the spherical collapse model, this barrier is constant and flat (in $R$), $B(R)=\delta_{sc}$, but one may consider more involved, and more realistic, models.
The cloud-in-cloud problem is solved by identifying the largest smoothing scale on which the walk first crosses the barrier. The excursion set ansatz therefore relates the abundance of haloes of mass $M$ to the fraction of random walks that first cross a barrier $B(R)$ on the scale $R$, the halo mass being the one contained inside a sphere of radius $R$,

\begin{equation}
M=\frac{4\pi}{3} R^3 \bar\rho.
\end{equation} 

With a similar meaning to the multiplicity function, the first crossing fraction $f (s)\d s$ is defined as  

\begin{equation}
\frac{M}{\bar\rho}\frac{\d n(M)}{\d M} \d M = f (s)\d s
\end{equation}

where it is customary in this framework to use the variance $s = \sigma^2_0(R)$ as the reference variable for the random walk. As we said, the excursion set ansatz consists in requiring that $\delta < B(s)$ for all $s < S$. This is an infinite set of conditions and, in a generic case, calculating the first crossing distribution can be very complicated. Let us assume that, when the walk is strongly correlated, we can instead impose the condition on the one preceding step, that is, that $\delta < B(S-\Delta s)$ for $\Delta s\rightarrow 0$. We can then Taylor expand in $\delta$ and $B$ and get the following condition:

\begin{equation}\label{eq:barrexc}
B(S) \leq \delta \leq B(S) + \Delta s \left (\frac{\d \delta}{\d s} - \frac{\d B}{\d s}\right)
\end{equation}

with $\d \delta/\d s \geq \d B/\d s$ at $s=S$, which is a condition on the `velocity' of $\delta$ being greater than the tilt of the barrier \cite{appel1990mass,Paranjape:2011wa,Paranjape:2012ks}. This implies that now we have to deal with a bivariate distribution on $\delta$ and its velocity $\delta'\equiv \d \delta/\d s$, $P(\delta,\delta')$, but with the simplification of having only one condition to impose, rather than an infinite number of them. The first crossing distribution reads
\begin{eqnarray}
f(s)\d s &\simeq & \int_{B'}^{+\infty} \d \delta' \int_{B(s)}^{B(s) + \Delta s (\delta' - B')} \d \delta\, P(\delta, \delta'),\label{eq:excgeneral}\\
&\simeq & \Delta s\,\int_{B'}^{+\infty} \d \delta' p(\delta',B)(\delta'-B').
\end{eqnarray}
If we now take the limit $\Delta s \rightarrow 0$ and change the integration variable $\delta' \rightarrow v=\delta' - B'$ we get the general formula
\begin{equation}\label{eq:fofs}
f(s) = \int_0^{+\infty} \d v\, v\, p(v+B',B).
\end{equation}
As an example, we can compute Eq. \eqref{eq:fofs} for Gaussian initial conditions and  a constant barrier, $B(s) = \delta_{sc}$. In this case, it is convenient to work with the following change of variables
\begin{equation}
\delta_{sc} \rightarrow \nu\equiv \delta_{sc} / \sqrt{s} \qquad v\rightarrow x\equiv 2\gamma\sqrt{s} v
\end{equation}
 and we have defined
\begin{equation}\label{eq:gamma}
\gamma^2 \equiv \frac{\langle \delta \delta' \rangle^2}{\langle \delta'^2 \rangle \langle \delta^2\rangle}.
\end{equation}
We obtain therefore the following first crossing distribution, as a function of $\nu$,
\begin{equation}\label{eq:fnues}
f(\nu) = \frac{e^{-\nu^2/2}}{\sqrt{2\pi}}\frac{1}{\gamma \nu } \int_0^{+\infty}\d x\, x\, p_{\rm G}(x-\gamma\nu;1-\gamma^2)
\end{equation}
where the subscript ${\rm G}$ indicates the Gaussian distribution and we have used the conditional probability $p_{\rm G}(x,\delta_{sc}) = p_{\rm G} (\delta_{sc}) p_{\rm G}(x|\delta_{sc})$.

There are several efforts that extend the excursion set to non-Gaussian initial conditions \cite{Inoue:2001fa,Avelino:1999hs,Lam:2008rk,Lam:2009nb,Lam:2009nd,Lam:2009rz,Maggiore:2009rx,DeSimone:2010mu,Musso:2011ck,DAmico:2010ywu,Paranjape:2011uk,Achitouv:2011sq,Achitouv:2012mk,Achitouv:2012ux}, some of which also relax the assumption of spherical collapse and consider, for instance, ellipsoidal barriers motivated by the fact that the tidal shear of the local density field contributes to the threshold for collapse. Here we will review the main results of \cite{Maggiore:2009rx,DeSimone:2010mu,DeSimone:2011dn}, which build up on results obtained for the Gaussian case \cite{Maggiore:2009rv,Maggiore:2009rw,Ma:2010ep}

\begin{enumerate}

\item They formulated the excursion set using top-hat filtering in real space, which is the preferred filter to compare with data and simulations \cite{Maggiore:2009rx}. Such a choice of filter introduces non-Markovianity of the random walks, that is, walks are correlated not only with their direct predecessor, but with the whole preceding path (see also \cite{Musso:2012qk,Musso:2014jda} for another approach). 

\item They gave a physical explanation of the correction to the spherical threshold for collapse $\delta_{sc}\rightarrow \sqrt{q} \delta_{sc}$ which is commonly used to improve fits to simulations (see \cite{Sheth:1998ew,Sheth:1999mn,Sheth:1999su,Sheth:2001dp} for the earliest applications ) and showed how it is expected that the factor depends on the halo finder used \cite{Maggiore:2009rw}.

\item Non-Gaussianities also introduce non-Markovian corrections, they were able to include them in their framework using the results from the Gaussian case with non-Markovianity \cite{Maggiore:2009rx}.

\item They extended the formalism to a generic moving barrier $B(\sigma)$ in \cite{DeSimone:2010mu}.

\end{enumerate}

These findings are summarized in the following prediction for the non-Gaussian multiplicity function

\begin{equation}\label{eq:mr}
f_{\rm MR}(\nu) = (1-\sqrt{q}\, \kappa) \sqrt{q}\nu f_{\rm PS}(\sqrt{q}\nu) \left[1+\frac 16 h_{\rm NG}(\sqrt{q} \nu)\right] + \frac{\kappa\,\sqrt{q} \nu}{\sqrt{2\pi}}\Gamma\left(0,\frac{q\nu^2}{2}\right)
\end{equation}

where $q = 1/(1+D_B)$ being $D_B$ the diffusion coefficient of the barrier, which advocates 2. of the previous list, 

\begin{equation}
\kappa(R) = \lim_{R'\rightarrow \infty} \frac{\langle\delta_{R'}\delta_R\rangle}{\langle \delta^2_{R'}\rangle}
\end{equation}

is the correction due to the use of top-hat filtering (addressing 1.), $\Gamma(0,x)$ is the incomplete Gamma function and 

\begin{equation}
h_{\rm NG}(\nu) =  \nu^3 \sigma S_3 - \nu \left(3 \sigma S_3+\sigma \mathcal U_3+\frac{\d (\sigma S_3)}{\d\ln \nu}\right)+\frac{1}{\nu}\left(\sigma \mathcal V_3 + \frac{\d (\sigma S_3)}{\d\ln \nu}- \frac{\d (\sigma \mathcal U_3)}{\d\ln \nu}\right) +\mathcal O\left(\frac{1}{\nu^3}\right)
\end{equation}

is the correction due to non-Gaussianity (addressing 3.), being $\mathcal U$ and $\mathcal V$ first and second derivatives of the skewness $S_3$, respectively. 

Eq. \eqref{eq:mr} is not generalized to account for moving barriers ( improvement 4. in the list above ), such a generalization can be found in \cite{DeSimone:2010mu}, where they also used an improved method based on the saddle-point approximation proposed by \cite{DAmico:2010ywu}. Another series of papers analyzes drifting diffusing barriers in the context of the excursion set approach \cite{Achitouv:2011sq}, including also an extension to a class of primordial trispectra \cite{Achitouv:2012ux}. In general, the introduction of a  consistent barrier for collapse is a non-trivial task because it requires a detailed modeling of halo formation, which at present day is not understood, being a fully non-linear process.

\subsubsection{The excursion set peaks model}\label{sec:espeaks}

It is rather natural to think that, to some degree of approximation, virialized haloes at late time may have formed out of  maxima of the initial density field \cite{Bardeen:1985tr}. This amounts to saying that, if we trace back halos of a certain mass $M$ to the initial conditions, they sit on a peak of the initial density field. The corresponding Lagrangian patch is called ``proto-halo'' and it defines the halo Lagrangian scale $R \propto (M/\bar\rho)^{1/3}$. This assumption has been extensively analyzed in numerical N-body simulations and shown to work rather well for a wide range of masses \cite{Ludlow:2010xd,Ludlow:2011jx,Elia:2011ds,Borzyszkowski:2014xua}. 
Several authors have developed the technical tools to deal with local density maxima in the presence of non-Gaussian initial conditions \cite{Bardeen:1985tr,Catelan:1988zb,Gay:2011wz,Codis:2013exa,Lazeyras:2015giz,Uhlemann:2017tex}.  Here we want to collect all their progresses and include them in the effort of combining the peak model and the excursion set approach. Only recently it has been realized that the two methods can be unified into one Excursion Set Peaks (ESP) model by imposing that peaks of the density field at a given smoothing scale satisfy a first crossing condition, in such a way as to have a first crossing distribution for peaks \cite{Paranjape:2012ks}. A full review and detailed description of the peak model and its extension to the ESP is nicely presented in \cite{Desjacques:2017msa}. Here we highlight the main results.

\paragraph{The Gaussian case}

We can generally define the number density of points that are local maxima of the Gaussian field $\delta(\vx)$ as a point process,
\begin{equation}\label{eq:npk}
n_{\rm pk}(\vx) = \sum_{\vx_{\rm pk}} \delta_{\rm D} (\vx - \vx_{\rm pk}),
\end{equation}
where $\delta_{\rm D}$ is the 3-dimensional delta function. The continuous field $\delta(\vx)$ can be Taylor expanded around a peak position $\vx_{\rm pk}$,
\begin{equation}\label{eq:delpk}
\delta(\vx) \approx \delta(\vx_\pk) + \frac 12 \sum_{i,j} [\nabla_i\nabla_j \delta(\vx_\pk)](\vx - \vx_\pk)_i (\vx - \vx_\pk)_j,
\end{equation}
where we imposed the peak condition on the first derivative, $\nabla_i\delta(\vx_\pk) = 0$. In a similar manner, we can Taylor expand $\nabla_i\delta(\vx)$ around the peak as
\begin{equation}\label{eq:etapk}
\nabla_i\delta(\vx) \approx  [\nabla_i\nabla_j \delta(\vx_\pk)] (\vx - \vx_\pk)_j.
\end{equation}
Using Eq. \eqref{eq:delpk} and Eq. \eqref{eq:etapk}, we can express the delta functions of Eq. \eqref{eq:npk} as
\begin{equation}
\delta_{\rm D}(\vx-\vx_\pk) \approx | \mbox{det } [\nabla_i\nabla_j \delta(\vx)] | \delta_{\rm D}(\boldsymbol{\nabla}\delta(\vx)),
\end{equation}
provided that the hessian $\nabla_i\nabla_j\delta$ is definite negative, to ensure to be looking at maxima, and non singular in $\vx_\pk$.

The identification of proto-halos inserts a scale into the problem, as we want to model the clustering of halo centres, and not their internal substructure. We therefore need to smooth the DM field on the Lagrangian size $R_s$. Moreover, we may impose a high enough threshold on the value of the density at the peak, that is, a \emph{peak height}, to make sure that we are dealing with regions sufficiently dense as to favour virialization, in the spirit of \cite{Kaiser:1984sw}. 

As a result, the number density of peaks of height $\nu'$ at position $x$  can be expressed in terms of the smoothed field $\delta_s$ and its
derivatives as \cite{Bardeen:1985tr}

\begin{equation}\label{eq:npkbbks}
n_\pk (\nu, R_s, \vx) = \frac{3^{3/2}}{R^3_*}\, | {\rm det} \zeta(\vx) |\, \delta_{\rm D} (\boldsymbol{\eta}(\vx)) \,\theta_{\rm H} (\lambda_3 (\vx))\, \delta_{\rm D} (\nu (\vx) - \nu'),
\end{equation}

where $R_*=\sqrt{3} \sigma_{1s} / \sigma_{2s}$ is the characteristic radius of the peak and we have conveniently defined the normalized peak density and its derivatives as

\begin{align}
\label{eq:normpeak1}
\nu(\vx) &= \frac{1}{\sigma_{0s}} \delta_s (\vx)\\
\label{eq:normpeak2}
\eta_i(\vx) &= \frac{1}{\sigma_{1s}} \nabla_i \delta_s (\vx)\\
\label{eq:normpeak3}
\zeta_{ij}(\vx) &=  \frac{1}{\sigma_{2s}} \nabla_i \nabla_j\delta_s (\vx)
\end{align} 

and we will adopt the following notation for the variance of the smoothed density field,
(linearly extrapolated to present-day) and its derivatives, 

\begin{equation}\label{eq:sigmaj}
\sigma^2_{j} = \frac{1}{2\pi^2}\int_0^\infty\!\! dk\, P(k) 
k^{2(j+1)} W_R^2(k).
\end{equation}

 We have imposed explicitly that the stress tensor $-\zeta_{ij}$ is definite negative through the positivity of $\lambda_3$, its lowest eigenvalue.

The number density $n_{\rm pk}$ can be used, in principle, to compute any N-point correlation functions among the peaks of the density field by ensemble averaging products of  $n_{\rm pk}$,
\begin{equation}
\rho^{(N)}_\pk(\nu,R_s,\mathbf{x}_1, ..., \mathbf{x}_N) = \langle n_{\pk}(\nu, R_s, \vx_1)\times ... \times n_{\pk}(\nu, R_s, \vx_N) \rangle,
\end{equation}
and in the Gaussian initial conditions considered here, multivariate normal distributions are assumed to perform the ensemble average. The case of $N=1$ is the averaged peak number density
\begin{equation}\label{eq:barnpk}
\bar n_{\rm pk} = \langle n_{\pk}(\nu, R_s, \vx) \rangle = \int \d \nu\,\d^3 \eta\, \d^6 \zeta\,   n_{\pk}(\nu, R_s, \vx) P_1(\nu, \boldsymbol{\eta}, \zeta_A)
\end{equation}
being $A= 1, ...,6$ since $\zeta$ is symmetric and the joint probability distribution $P_1$ is given by
\begin{equation}\label{eq:p1}
P_1 (\mathbf{y}) \d \mathbf{y} = \frac{1}{(2\pi)^5 |{\rm det} M|^{\frac{1}{2}}} e^{-Q_1(\mathbf{y})}  \d \mathbf{y},
\end{equation}

where $M$ is the corresponding covariance matrix. Owing to rotational invariance, we have regrouped the set of variables into the following vector

\begin{equation}\label{eq:peakvec}
{\bf y} \equiv \{ \nu, J_1, 3\eta^2, 5 J_2, J_3\},
\end{equation}

where $J_1=-{\rm tr}(\zeta_{ij})$ is the peak curvature, $J_2=\frac 32 {\rm tr}(\bar\zeta^2_{ij})$ and $J_3=\frac 92 {\rm tr} (\bar \zeta^3_{ij})$, being $\bar\zeta_{ij}= \zeta_{ij} + \delta_{ij}\, J_1 /3$ the components of the traceless part of the Hessian. We can then factorize the exponential into the following form

\begin{equation}
Q_1 =  \frac{\nu^2+J_1^2-2\gamma_1 \nu J_1}{2(1-\gamma_1^2)} + \frac 32 \eta^2 + \frac 52 J_2^2
\end{equation}

 and we define the correlation
 
\begin{equation}\label{eq:gamma1}
\gamma_1^2 = \frac{\langle \delta J_1 \rangle^2}{\langle \delta^2 \rangle \langle J_1^2 \rangle} = \frac{\sigma_{1s}^4}{\sigma_{0s}^{2}\,\sigma_{2s}^{2}}.
\end{equation}

This decomposition allows to write Eq. \eqref{eq:p1} in a more compact way as a product of a bivariate Gaussian $\mathcal N$ in the variables $\nu$ and $J_1$ and 3- and 5- degrees of freedom $\chi$-squared distributions in $3\eta^2$ and $5J_2$ respectively,
\begin{equation}\label{eq:p1peak}
P_1 (\mathbf{y}) \d \mathbf{y} = \mathcal{N}(\nu,J_1)\,\d \nu\, \d J_1\, \chi^2_3(3\eta^2)\, \d (3\eta^2)\,\chi^2_5(5J_2)\,\d (5J_2) \d J_3\,P(\boldsymbol{\Omega}),
\end{equation}
and $P(\boldsymbol{\Omega})$ represents the probability distribution of the five remaining d.o.f.. Since
they are all angular variables, they do not generate bias factors because the peak (and halo) overabundance can only depend on scalar quantities (see e.g. \cite{Catelan:1998pi, McDonald:2009dh}). The variable $J_3$ is uniformly distributed and constrained to satisfy $J_3^2 \leq J_2^3$ by the fact that $\zeta_{ij}$ is symmetric.

We now want to apply the excursion set ansatz to the statistics of peaks of the density field. This amounts to impose to select only those peaks which are found at a certain scale $R$ which have a smaller height on the next larger smoothing scale, that is,
\begin{equation}
\frac{B(R_s)}{\sigma_{0s}} \leq \nu \leq \frac{B(R_s)}{\sigma_{0s}} + \frac{\Delta R_s}{\sigma_{0s}} \left (\frac{\d B}{\d R_s} - \frac{\d \delta}{\d R_s}\right),
\end{equation}
where we just adapted Eq. \eqref{eq:barrexc} to the variable $R_s$, with the condition that $\delta' - B' \leq 0$, where now the prime indicates derivative with respect to $R_s$. Hence, the corresponding ESP discrete number density can be written as
\begin{equation}
n_\esp(\boldsymbol{\omega}) = -\frac{\mu}{\sigma'_{0s} \nu}\, \theta_{\rm H}( \mu)\, n_\pk(\mathbf{y}).
\end{equation}
where we have defined the variable $\mu =  B'-\delta'$ and extended the vector of variables to $\boldsymbol{\omega} = (\mathbf{y},\mu)$. Combining Eq. \eqref{eq:barnpk} and Eq. \eqref{eq:excgeneral} we get \cite{Desjacques:2013qx}

\begin{align}
\bar n_\esp (\nu, R_s) \Delta R_s &= \frac{3^{3/2}}{R_*} \int \d^6\zeta\, \int \d^3\eta\,\int^{B'}_{-\infty}\d \delta' \,\int_{B}^{B-\Delta R_s(\delta'-B')} \frac{\d\delta}{\sigma_{0s}}\, |{\rm det} \zeta| \delta_{\rm  D} (\eta) \theta_{\rm H} (\lambda_3) P_1(\boldsymbol{\omega})\nonumber\\
&= \int \d^6\zeta\,\int \d^3\eta\,\int^{\infty}_{0} \d \mu \, \frac{\mu}{\sigma_{0s}}\,n_\pk(\mathbf{y}) P_1(\boldsymbol{\omega})
\end{align}

and using a similar approach to Eq. \eqref{eq:npkxnu}, we can calculate the ESP multiplicity function to be

\begin{align}
\label{eq:fespnesp}
f_{\rm ESP}(\nu) &=  \frac{M}{\rho }\, \bar n_\esp\, \frac{\d R_s}{\d\nu} \\
&= -\frac{e^{-\nu^2/2}}{\sqrt{2\pi}}\frac{1}{\sigma'_{0s} \nu } \frac{V}{V_*}\mathcal{G}^{(1)}_1 (\gamma_1,\gamma_1\nu)
\end{align}

where $V=m/\bar{\rho}$, $\sigma'_{0s} = \d \sigma_{0s}/\d R_s$ and $\mathcal{G}^{(1)}_1(x)$ is defined in appendix \ref{app:bigg}. 

The multiplicity function of Eq. \eqref{eq:fespnesp} is a theoretical prediction of the halo mass function which does not depend on any free parameter. Its limits of applicability are directly related to the assumptions it is based on, namely

\begin{itemize}[leftmargin=*,labelsep=5.8mm]
    \item Halos form out of peaks of the Lagrangian density field,
    \item The barrier for collapse is described as a sole function of the density field.
    \item Walks are correlated enough on large scales such that we can impose the up-crossing condition on the one preceding step only, rather than the full walk.
\end{itemize}

The most important extension to be worked out on this model is the inclusion of the effect of tidal shear on the gravitational collapse of halos. The formalism has been wrote down in \cite{Desjacques:2017msa} for a generic collapse barrier which depends on the tidal field shear field. The next step would be to understand the particular form which this barrier should have, in order for the model to provide a prediction on the halo mass function. As argued in \cite{Desjacques:2017msa} and in Sec. \S\ref{sec:haloclustering}, this extension is particularly relevant for primordial non-Gaussianity.

\paragraph{The non-Gaussian extension}

The goal is now to embed a non-Gaussian initial PDF for peaks and therefore a non-Gaussian ESP mass function. 
%
%
%
%
%
The important point here is that in order to compute a non-Gaussian probability through an expansion similar to Eq. \eqref{eq:ploverde}, we need to expand all the variables in $\boldsymbol{\omega} = \{ \nu, J_1, \mu, 3\eta^2, 5 J_2, J_3\}$. This expansion takes the name of Gram-Charlier expansion and it has been developed in a number of papers \cite{Pogosyan:2009rg,Gay:2011wz,Gong:2011gx,Lazeyras:2015giz}. 




For generic non-Gaussian initial conditions we define $P_{\rm NG}=P_{\rm G}[1+\delta P_{\rm NG}]$ and the Gram-Charlier\footnote{For cumulants up to fifth order it can be demonstrated that the Gram-Charlier expansion coincides with the Edgeworth expansion. It is worth noticing that, especially for large non-Gaussianities, the Gram-Charlier expansion might not converge, see for example \cite{Blinnikov:1997jq}.} expansion up to third order in $\delta$ gives

\begin{align}
\delta P_{\rm NG} =  &\,\frac 16 \langle \nu^3 \rangle_{\rm GC} H_{300}(\nu, J_1, \mu)+\frac 16 \langle J_1^3 \rangle_{\rm GC} H_{030}(\nu, J_1, \mu)+\frac 16 \langle \mu^3 \rangle_{\rm GC} H_{003}(\nu, J_1, \mu)\nonumber\\
&+\frac 12 \langle \nu^2 J_1 \rangle_{\rm GC} H_{210}(\nu, J_1, \mu)+\frac 12 \langle \nu J_1^2 \rangle_{\rm GC} H_{120}(\nu, J_1, \mu) +\frac 12 \langle \nu^2 \mu \rangle_{\rm GC} H_{201}(\nu, J_1, \mu)\nonumber\\
&+\frac 12 \langle \nu \mu^2 \rangle_{\rm GC} H_{102}(\nu, J_1, \mu)+\frac 12 \langle  J_1^2 \mu \rangle_{\rm GC} H_{021}(\nu, J_1, \mu) +\frac 12 \langle J_1 \mu^2 \rangle_{\rm GC} H_{012}(\nu, J_1, \mu)\nonumber\\
&+ \langle \nu J_1 \mu \rangle_{\rm GC} H_{111}(\nu, J_1, \mu)- \langle \nu \eta^2 \rangle_{\rm GC} H_{100}(\nu, J_1, \mu) L_1^{(1/2)}\left(\frac 32 \eta^2\right)\nonumber\\
&-\langle J_1 \eta^2 \rangle_{\rm GC} H_{010}(\nu, J_1, \mu) L_1^{(1/2)}\left(\frac 32 \eta^2\right)-\langle \mu \eta^2 \rangle_{\rm GC} H_{001}(\nu, J_1, \mu) L_1^{(1/2)}\left(\frac 32 \eta^2\right)\nonumber\\
&-\langle \nu \zeta^2 \rangle_{\rm GC} H_{100}(\nu, J_1, \mu) L_1^{(3/2)}\left(\frac 52 \zeta^2\right)-\langle J_1 \zeta^2 \rangle_{\rm GC} H_{010}(\nu, J_1, \mu) L_1^{(3/2)}\left(\frac 52 \zeta^2\right)\nonumber\\
&-\langle \mu \zeta^2 \rangle_{\rm GC} H_{001}(\nu, J_1, \mu) L_1^{(3/2)}\left(\frac 52 \zeta^2\right)+\frac{25}{21}\langle J_3\rangle_{\rm GC} J_3
\end{align}

where we define the Hermite polynomials

\begin{equation}\label{eq:htransf}
H_{ijk}(\nu,u,\mu)= \frac{1}{\mathcal{N}(\nu,u,\mu)}\left(-\frac{\partial}{\partial \nu}\right)^i\left(-\frac{\partial}{\partial u}\right)^j\left(-\frac{\partial}{\partial \mu}\right)^k \mathcal{N}(\nu,u,\mu).
\end{equation}

 the generalized Laguerre polynomials

\begin{equation}
L_n^{(\alpha)}(x) = \frac{x^{-\alpha} e^x}{n!} \frac{\d^n}{\d x^n} (e^{-x} x^{n+\alpha})
\end{equation}

and the correlations $\langle \cdot \rangle_{\rm GC}$ as

\begin{equation}
\langle \nu^i z^j t^k \eta^{2q}J_2^{l} J_3^m \rangle_{\rm GC} = \frac{(-1)^{q+l} q! \,l!}{(3/2)^q(5/2)^l}\frac{1}{\bar n_{\rm ESP}}\left\langle n_{\rm ESP}(\vy)H_{ijk}(\nu, J_1, \mu) L_q^{(1/2)}\left(\frac 32 \eta^2\right) L_l^{(1/2)}\left(\frac 52 J_2\right)J_3^m\right\rangle,
\end{equation}

%

 where we have assumed $m\leq 1$ and the average is weighted by the ESP mass function to ensure that only peaks of the density field are selected.
%
%
The non-Gaussian correction to the mass function is

\begin{align}\label{eq:deltangESP}
\delta\bar n^{\rm NG}_{\rm ESP} \equiv \int \dd \boldsymbol{\omega}\, n_{\rm ESP}(\boldsymbol{\omega}) \delta P_{\rm NG}(\boldsymbol{\omega}) = \int \dd \boldsymbol{\omega}\, n_{\rm ESP}(\boldsymbol{\omega})  P_{\rm G}(\boldsymbol{\omega})\Biggr[\frac 16 \langle \nu^3 \rangle_{\rm GC} H_{300}(\nu, J_1, \mu)+...+\frac{25}{21}\langle J_3\rangle_{\rm GC} J_3\Biggr].
\end{align}

Here we notice an important feature of the expansion: each Hermite, or Laguerre, polynomial that multiplies the Gaussian mass function and PDF, when integrated over $\boldsymbol{\omega}$ defines itself a bias parameter for the Gaussian case, for instance

\begin{equation}
\sigma_0^3 b_{300} = \int \d \boldsymbol{\omega}\, n_{\rm ESP}(\boldsymbol{\omega}) H_{300}(\nu) P_{\rm G}(\boldsymbol{\omega}).
\end{equation} 






Hence we have found that each term in Eq. \eqref{eq:deltangESP} defines a third order bias multiplied by the corresponding ``skewness'' of the type $\langle X^3 \rangle_{\rm GC}$. If we were to include terms involving the kurtosis, they would generate fourth order bias \emph{Gaussian} parameters and so on. We refer to the Appendix \ref{app:dnesp} for the full expression for $\delta n^{\rm NG}_{\rm ESP}$, which is rather lengthy, but straightforward to calculate. Here we quote only the first three terms

\begin{equation}
\delta \bar n_{\rm ESP}^{\rm NG}\supset \frac{1}{\sqrt{6}} \Biggr\{b_{300} S_3^{(\nu^3)}  +b_{030} S_3^{(u^3)} + b_{003}  \sqrt{6} S_3^{(\mu^3)}+...,
\end{equation}

where we define the skewnesses as

\begin{equation}
S_3^{(X)}=2f_{\rm NL}\frac{g(\sigma_0,\sigma_1,\sigma_2)}{\sigma_0^4}\int \frac{\dd^3 k_1}{(2\pi)^3}\, \frac{\dd^3 k_2}{(2\pi)^3}\, \chi(\vk_1,\vk_2) \mathcal{M}(k_1)\mathcal{M}(k_2)\mathcal{M}(k_{12})\left[P_{\phi}(k_1)P_{\phi}(k_2)+\mbox{ 2 cyc. }\right]
\end{equation}

where $g$ and $\chi$ vary for each variable $X$ considered.

\subsection{Numerical approaches}\label{sec:onepointnumeric}

In parallel with analytic approaches, N-body simulations have been exploited both to test theoretical predictions and to find semi-analytic or fully fitted phenomenological formulas to be used in real data. 
The earliest fits to the non-Gaussian mass function from simulations with non-Gaussian initial conditions \cite{Dalal:2007cu,Pillepich:2008ka} focused on local-type non-Gaussianity with an amplitude of $\mathcal O(\fnl)\sim 100$s. In \cite{Pillepich:2008ka} they fitted the multiplicity function as

\begin{equation}
f(\sigma) = \left[D+B\left(\frac{1}{\sigma}\right)^A\right] \exp\left(-\frac{C}{\sigma^2}\right),
\end{equation}

where $A, B, C$ and $D$ each follow the functional form

\begin{equation}
P(z,\fnl) = p_1(1+p_2 z+p_3 z^2)(1+p_4 \fnl)\qquad P=A,B,C,D.
\end{equation}

This fit showed agreement within $\sim 10$\% of the measurements. 

In \cite{Achitouv:2011sq,Achitouv:2013oea}, the halo mass function from excursion set theory with a drifting diffusive barrier was tested against N-body simulations. Their prediction is an extension of Eq. \eqref{eq:mr} to include a stochastic barrier with linearly drifting average up to
next-to-leading order. The idea is to allow for stochastic deviations from the usual spherical collapse threshold value $\delta_{sc}$, parametrizing the deviation with two quantities: the rate $\beta$ at which the collapse threshold  deviates on average  from the spherical collapse $\delta_{sc}$ and  the amplitude $D_B$ of a constant scatter around the average collapse threshold at a given mass scale. They compare their model with measurements of the halo mass function in simulations with local-type non-Gaussianity in the initial conditions, first fitting $\beta$ and $D_B$ and then calibrating them on the Gaussian simulations. Using the first fit, their prediction agree within $5$\% with simulations. When calibrating on the Gaussian simulations, they find dependence of the parameters on non-Gaussianity, indicating that the stochastic nature of the barrier is correlated with it.

An important study  of the large-scale density field was performed in \cite{Mao:2014caa} using N-body simulations in order to understand whether measurements of the moments of large-scale structure can yield constraints on primordial non-Gaussianity. They found that most of the information is included in the variance of the galaxy density field, because of the effect of the scale-dependent bias, which we will discuss in the next Sec. \S\ref{sec:two-point}. The result on the variance was recently further refined in \cite{Nusser:2018vym}, where they use maxima and minima counts of the halo density field of N-body simulations to show that primordial non-Gaussianity of the local-type could be constrained at the level of $\fnl \approx 10$ with the Euclid survey using this method. Although these figures are not competitive with direct constraints from the power spectrum of galaxies (see for instance \cite{Amendola:2012ys}), the estimate of \cite{Nusser:2018vym} is still preliminary and should be refined.

In \cite{Adhikari:2014xua}, N-body simulations with primordial non-Gaussianity generated by a range of two-field models of inflation were run. The measurements were compared to semi-analytic estimations of the non-Gaussian mass function based on the Edgeworth expansion. The type of primordial non-Gaussianity they consider is such that cumulants of higher order than the skewness are more important than in the single-field local case. Their comparison with data shows that different prescriptions for the scaling of higher moments give appreciably different results. The analysis of multi-field inflation in N-body simulations proves an important point on being careful when truncating the Edgeworth expansion. Given that recent theoretical progress has demonstrated that the local-type non-Gaussianity for single-field models is not observable \cite{Tanaka:2011aj, Baldauf:2011bh,  Creminelli:2011rh, Pajer:2013ana, Dai:2013kra, Camera:2014sba, Bartolo:2015qva, dePutter:2015vga} (see discussion in the next Sec. \S\ref{sec:single-field}), this type of searches should be investigated further. 

\subsection{Final remarks of this section}\label{sec:onepointremarks}

In this section, we have focused on the effect of primordial non-Gaussianity on the probability distribution of dark matter halos. We have showed how the strongest effects are in the tails of the probability distribution. For this reason, studies on the very high mass tail of what observed in a galaxy survey, such as extreme-mass clusters \cite{Chongchitnan:2011eq,Chongchitnan:2015kwa} and x-ray detected clusters \cite{Roncarelli:2009pp,Sartoris:2010cr,Shandera:2013mha}, have been investigated as well. Similarly, very large voids probe the low-density tail of the PDF. Early studies of N-body simulations showed that this could be a promising venue for large values of $\fnl$ of the local type \cite{Grossi:2008fm}. On the theoretical side, calculations within the context of the extended Press-Schechter mass function were perfomed in  \cite{Kamionkowski:2008sr,Song:2008gb,Lam:2009nd} and later using the excursion set formalism \cite{DAmico:2010dwy}. It is also possible to constrain directly the dark matter density field using weak lensing measurements \cite{Jimenez:2009us,Lam:2009rz,Pace:2010sr,Marian:2010mh, Jeong:2011rh,Hilbert:2012gr}.

Despite almost 40 years of research into the possibility of constraining primordial non-Gaussianity using the probability distribution of dark matter, and dark matter tracers, in the Universe, it remains a great challenge to obtain competitive figures as compared to other probes, such as the CMB and correlation functions of dark matter tracers (see the following section \S\ref{sec:two-point}). The main reason for this is that most of the relevant information is in the high-mass tail of the halo (or possibly other tracers) distribution, where we lack the statistics to have high signal-to-noise ratios. Moreover, current limits on non-Gaussianity constrain the correction to the probability distribution to be very small: for instance, a local-type non-Gaussianity of order unity implies a correction at the subpercent level in the halo mass function.

\section{Imprints of primordial interactions on two-point halo statistics}\label{sec:two-point}

The interactions taking place during inflation that we introduced in Sec. \S\ref{sec:interactions} manifest themselves as non-trivial correlation functions of the primordial curvature perturbation, which provides the seed for structure formation. We have focused in particular on three-point correlation functions: the natural choice for looking for these primordial correlations in the large-scale structures of the late universe would be the three-point correlation function of galaxies, which is sourced by the primordial correlator at tree-level in a similar way as for the dark matter case of Eq. \eqref{eq:bistreedm}. 

It came as a breakthrough the finding  that the two-point function of halos on large scales contains a great deal of information about the class of primordial interactions whose bispectrum shape peak in the squeezed limit. The discovery of a characteristic scale dependence at large-scales in the linear bias of halos came as a surprise since it was a generic prediction of clustering models with Gaussian initial conditions that the bias at large scales was scale-independent. As we will show in this section, there are two main features that make this observable so promising to constrain primordial interactions: firstly, the presence of primordial non-Gaussianity manifests the strongest on the largest scales. In this regime, non-linear biasing, non-linear gravitational interaction and red-shift space distortions have negligible effects. Secondly, it manifests as a peculiar scale dependence, which is not produced by any other phenomenon on those scales. The particular scaling depends on the primordial bispectrum shape that generates it, hence a good constraint on the scale dependence could discriminate among different models of inflation.

This finding opened an entire new stream of research on the effects of PNG on the power spectrum of galaxies and it provides one of the most promising observables in LSS as of today. This section is devoted in reviewing recent efforts both on the theoretical side and on the numerical side.

{\bf Plan of the Section.} We start by introducing the imprint of local-type non-Gaussianity on the halo power spectrum (almost) as it was discovered, with a derivation using the peak-background split ansatz in Sec. \S\ref{sec:break}. We then proceed by generalizing to any model of primordial non-Gaussianity in Sec. \S\ref{sec:analytic}, describing different methods for modeling structure formation and halo clustering in the presence of such an effect. We then overview the current status of numerical analyses in Sec\S\ref{sec:numeric}, complementing the analytic part. We conclude with final remarks in Sec.\S\ref{sec:remarks}.



\subsection{The breakthrough: scale-dependent bias from local PNG}\label{sec:break}

In \cite{Dalal:2007cu}, the effect of primordial non-Gaussianity in the halo power spectra was measured using N-body simulations with non-Gaussian initial conditions of the local type. The measurements exhibit a strong $1/k^2$ scaling at large scales.  First physical interpretations and predictions were given in the context of the high peaks approximation \cite{Matarrese:2008nc,Valageas:2009vn,Shandera:2010ei}, multivariate bias expansions 
\cite{McDonald:2008sc,Giannantonio:2009ak}, and using a peak-background split (PBS) ansatz \cite{Slosar:2008hx,Afshordi:2008ru,Schmidt:2010gw,Desjacques:2011mq,Smith:2011ub,Scoccimarro:2011pz}.  As a first step into the study of this topic, in this section we provide a pedagogical derivation for the case of local-type primordial non-Gaussianity in the context of the peak-background split ansatz, loosely following \cite{Slosar:2008hx}.

\subsubsection{Derivation with the peak-background split ansatz}\label{sec:pbsderiv}
The peak-background split approach \cite{Kaiser:1984sw, Bardeen:1985tr} gives a very clean physical interpretation of the effect and it provides a rather model independent prediction. The starting point of a peak-background split ansatz is separating long- and short-wavelengths modes of the density field $\delta$. A typical short mode can be thought of as the Lagrangian radius of a proto-halo, i.e. the scale relevant for the formation of virialized objects. The long modes act as a local rescaling of the amplitude of short ones. The basic assumption for being able to perform such a separation is that the two regimes are decoupled. This is the case for Gaussian distributed fields. Here, we want to consider the case of non-Gaussian initial conditions, which by definition introduce coupling between long and short modes already at the primordial level. This coupling is manifest when considering the local ansatz for the primordial perturbations,

\begin{equation}\label{eq:localtypephi}
\zeta(\bx) = \zeta_{\rm G}(\bx)+\frac 65\fnl(\zeta_{\rm G}^2(\bx) - \langle \zeta_{\rm G}^2\rangle).
\end{equation}

If the initial conditions are non-Gaussian, as in Eq. \eqref{eq:localtypephi}, the separation of scales has to be performed not at the level of the density field $\delta$, but on Gaussian primordial fluctuations

\begin{equation}
\zeta_{\rm G}=\zeta_{\ell} + \zeta_{s},
\end{equation}

where $\ell$ indicates long modes and $s$ short ones. Applying the split to Eq. \eqref{eq:localtypephi} we get

\begin{equation}
\zeta = \zeta_{\ell} + \frac 65 \fnl \zeta^2_{\ell} + \left(1 + \frac {12}{5}\fnl \zeta_{\ell}\right) \zeta_{s} + \frac 65 \fnl \zeta^2_{s} + \mbox{const.}\,.
\end{equation}

On sufficiently large scales, the relation

\begin{equation}
\delta_{\ell}( k) = \mathcal{M} ( k ) \zeta_{\ell}(k)
\end{equation}

holds, while the short-wavelengths modes of the density field are affected, at first order, as

\begin{equation}\label{eq:moddelta}
\delta_{s} \approx \mathcal{M}\,  \left(1+ \frac {12}{5}\fnl \zeta_{\ell}\right)\,\zeta_{s},
\end{equation}

which can be interpreted as a local modulation of the amplitude of matter fluctuations
by the long modes proportional to $\fnl$,  

\begin{equation}\label{eq:sigmamod}
\sigma_{s} \longrightarrow \sigma_{s} \left(1+\frac {12}{5}\fnl \zeta_{\ell}\right) \equiv \hat\sigma_{s}.
\end{equation} 

As a result, the local number density of halos of mass $M$ is not a function of $\delta_{\ell}$ only, but also of this local modulation $\hat\sigma_{s}$ and, more in general, on all cumulants of the density field. The expansion in Lagrangian space for the halo overdensity reads, at first order in $\delta_{\ell}$, as

\begin{align}\label{eq:ngexp}
\delta_{\rm h} ( \vx)
&\approx -\frac{1}{\bar n_{\rm h}}\,\frac{\partial \bar n_{\rm h}}{\partial \delta_{\ell}}\,\delta_{\ell}(\vx) + \frac{\partial \ln \bar n_{\rm h}}{\partial \ln \hat\sigma_{s}} \frac{\partial \ln \hat\sigma_{s}}{\partial \delta_{\ell}}   \,\delta_{\ell}(\vx) + ...\,,\nonumber\\
&\approx -\frac{1}{\bar n_{\rm h}}\,\frac{\d \bar n_{\rm h}}{\d \delta_{\ell}}\,\delta_{\ell}(\vx) + \frac{12}{5} \fnl\, \frac{\partial \zeta_{\ell}}{\partial \delta_{\ell}} \,\frac{\partial \ln \bar n_{\rm h}}{\partial \ln \hat\sigma_{s}} \,\delta_{\ell}(\vx) + ...\,.
\end{align}

From this expression we can define the first order bias in Fourier space in the presence of primordial non-Gaussianity

\begin{equation}\label{eq:scaledepbias}
b_1 (M,k) \equiv b_1(M) + \Delta b^{\rm NG}_1(M,k) =  b_1^{\rm G}(M) + \frac{12}{5}\frac{\fnl}{\mathcal{M}(k)} b_1^{\rm NG}(M)
\end{equation} 

being

\begin{equation}\label{eq:biasPBS}
b_1 \equiv -\frac{1}{\bar n_{\rm h}}\,\frac{\partial \bar n_{\rm h}}{\partial \delta_{\ell}}
\end{equation}

the usual Gaussian Lagrangian bias and 

\begin{equation}\label{eq:ngbiasamp}
b_1^{\rm NG} \equiv \frac{\partial \ln \bar n_{\rm h}}{\partial \ln \sigma_8},
\end{equation}

the amplitude of the so called \emph{scale-dependent} bias. Here $\sigma_8$ is the variance of the density field for a smoothing at $R= 8$ Mpc/h, the typical reference amplitude of matter fluctuations. This result is general, at first order, as we did not make assumptions other than the scale separation in $\zeta_{\rm G}$. 

The characteristic feature of this additional bias contribution at first order is that it scales as $\Delta b_1^{\rm NG} \propto 1/k^2$ at large scales, since $\mathcal M (k) \propto k^2$ in this limit. It implies that the two-point correlation function of halos at large separations should show a clear enhancement, or suppression if $\fnl^{\rm loc}$ is negative, with respect to the Gaussian case. In Fourier space, we therefore have

\begin{equation}
P_{\rm hh}(k) = (b_1+\Delta b^{\rm NG}_1(k))^2 P_{\rm mm}(k),
\end{equation}

where here we should remember that also the matter auto-power spectrum has non-Gaussian corrections, according to Eq. \eqref{eq:mmng}. As first argued in \cite{Slosar:2008hx}, the amplitude of this non-Gaussian correction depends on the halo formation history, suppressing the amplitude of recently formed halos and enhancing it if they are formed early. We will not discuss the effect here and defer the reader to the available literature \cite{Slosar:2008hx, Reid:2010vc}.

\subsubsection{Universality of the mass function}\label{sec:universality}

The amplitude $b_1^{\rm NG}$ of Eq. \eqref{eq:ngbiasamp} cannot be directly measured from galaxy surveys, as we do not have multiple realizations of the Universe with varying $\sigma_8$. The simplest way around this problem is to assume universal mass functions of the form of Eq. \eqref{eq:fnu}, as in the case of the Press \& Schechter mass function, Eq. \eqref{eq:psmf}. In this case, the logarithmic derivative of the mass function with respect to the normalization amplitude $\sigma_8$ takes the simple form

\begin{equation}\label{eq:universal}
 \frac{\partial \ln \bar n_{\rm h}}{\partial \ln \sigma_8} =  \frac{\partial \ln \bar n_{\rm h}}{\partial \ln \nu} \frac{\partial \ln \nu}{\partial \ln \sigma_8}= -\frac{\nu}{\bar n_{\rm h}} \frac{\partial  \bar n_{\rm h}}{\partial \nu}= \delta_{sc} b_1.
\end{equation}

The above expression also assumes spherical collapse. In this way, the linear bias in the presence of local-type PNG can be characterized via a single bias parameter $b^{\rm G}_1$. 

The assumption of universality of the mass function has been thoroughly investigated for Gaussian initial conditions in the last decade  (see for instance \cite{Tinker:2008ff, Reed:2012ih, Despali:2015yla}), but not as precisely for non-Gaussian inital conditions. Indeed, studies of N-body simulations have shown that the assumption is not as reliable as needed by the increasing precision required by data. We expand on this studies in the next Sec. \S\ref{sec:lpngnum}.


\subsubsection{The single-field consistency relation }\label{sec:single-field}

Single-field models of inflation generate a minimal amount of non-Gaussianity, as explained in \S \ref{sec:singleint}. The minimal interaction of a single scalar field with gravity produces a minimal amount of non-Gaussianity, Eq. \eqref{eq:malda}, of order $\mathcal O(\fnl) \sim \mathcal O(\epsilon,\eta)$ . For squeezed configurations of the bispectrum, the inflationary consistency relation Eq. \eqref{eq:maldasqueezed} states that all single-field models of inflation are constrained to have small non-Gaussianity, $\mathcal O(\fnl)\propto n_s-1$. This non-Gaussianity can be connected to a local-type one by an appropriate matching \cite{Maldacena:2002vr}, so that one would be tempted to state that we are expected to see a minimal amount of non-Gaussianity in the scale dependent bias. However, following an interesting debate \cite{Tanaka:2011aj, Baldauf:2011bh,  Creminelli:2011rh, Pajer:2013ana, Dai:2013kra, Camera:2014sba, Bartolo:2015qva, dePutter:2015vga,Bravo:2017gct,Bravo:2017wyw}, it has been shown that the single-field consistency relation is equivalent to the statement that long-wavelength perturbations are not physically coupled to small-scale ones and therefore there is no scale-dependent bias for single-field models of inflation. The argument follows from studying how to connect the result of \cite{Maldacena:2002vr}, which was computed in comoving gauge at the epoch of inflation, with actual observations of galaxy number counts at the present day. The positive side of this finding is that any detection of a scale-dependent bias on large scales of the type of Eq. \eqref{eq:scaledepbias} would rule out the possibility that primordial perturbations were generated at early times by a single field. 

In fact, the statement is not only restricted to local-type non-Gaussianity, but to all squeezed limits of inflationary bispectra: one way to show it is to remove the unphysical contributions from any bispectrum by calculating it in Conformal Fermi Coordinates \cite{Dai:2015rda,Dai:2015jaa}. An example in which this subtraction is done for the resonant running model of Eq. \eqref{eq:resrun} is found in \cite{Cabass:2018roz}.




\subsection{Analytic approaches}\label{sec:analytic}

After warming up with the effect of local PNG on the clustering of halos in the previous section, here we discuss how to extend it to any type of primordial interaction. We will begin by introducing a framework which allows for minimal assumptions on the physics of halo formation and collapse using an effective field theory approach to LSS. This treatment is extremely powerful in characterizing the most promising signatures from the early Universe. Subsequently, we proceed with applying a few well-motivated ansatz on the physics of structure formation at late times thereby increasing the predictive power of the theory, but losing control of theoretical errors. The trade-off between model-independence and predictivity plays a major role in constraining PNG using the power spectrum of tracers from galaxy surveys.

\subsubsection{The squeezed limit of the primordial bispectrum }\label{sec:squeezedbispe}

The finding that the presence of a coupling between long and short modes in the initial conditions modulates the local number density of halos and therefore their clustering at large scales has great importance: it implies that there is a specific kinematic regime of the initial conditions in which halo clustering at large scales is particularly sensitive to early Universe physics. This kinematic regime is the squeezed limit of the bispectrum, which is precisely the triangle shape by which the power spectrum on short modes is modulated by long ones. 

By defining long and short modes for a bispectrum triangle as 

\begin{equation}
\bk_1=\bk_s-\frac{\bk_\ell}{2},\qquad \bk_2=-\bk_s-\frac{\bk_\ell}{2},\qquad \bk_3=\bk_\ell,
\end{equation}

we can expand the primordial bispectrum of $\zeta$ for $k_\ell \ll k_s$ and get

\begin{equation}\label{eq:bissqueezeexp}
B_\zeta(\bk_1,\bk_2,\bk_3) = \sum_{J=0}^{\infty} A_{2J}(k_s,k_\ell)\, \mathcal L_{2J}(\hat\bk_s\cdot\hat\bk_\ell)\, P_\zeta(k_s) P_\zeta(k_\ell) + \mathcal O\left(\frac{k_\ell^2}{k^2_s}\right),
\end{equation}

where $\mathcal L_J$ are Legendre polynomials which appear only with even orders since in the squeezed limit $\bk_1 \approx -\bk_2$, removing the degree of freedom for odd number of exchanges of $\bk_s$ into $-\bk_s$. The factor $A_{J}(k_s,k_\ell)$ encodes information about the primordial interaction we want to describe. Tables \ref{tab:single} and \ref{tab:multi} gathers the values of $A_{2J}$ for the models introduced in Sec. \S\ref{sec:interactions}, excluding interactions involving features: these cases generically break scale-invariance and need to be handled with extra care. We expand on this in the next paragraph.

\begin{table}[H]
\caption{List of values of $A_{2J}$ from the squeezed limit expansion of Eq. \eqref{eq:bissqueezeexp} for interactions produced in single-field models of inflation.}\label{tab:single}
\centering
\renewcommand{\arraystretch}{2.5} 
\begin{tabular}{| l | c | c | c |}
\hline
\textbf{Interactions in single-field models}	& $A_0$	& $A_2$ \\
\hline
Gravitational Floor (Eq. \eqref{eq:malda})		&  $2\epsilon+\eta + \left(\frac{29}{6}\epsilon+\frac 14\eta\right) \left(\frac{k_\ell}{k_s}\right)^2$    & $\left(\frac 16 \epsilon +\frac 54\eta\right)\left(\frac{k_\ell}{k_s}\right)^2$ \\
DBI (Eq.	\eqref{eq:shapepx})	& $\left[\frac{19}{12}\left(1-\frac{1}{c_s^2}\right)+\frac{3}{4}\left(\frac{1}{c_s^2}-1-\frac{2\lambda}{\Sigma}\right)\right] \,\left(\frac{k_\ell}{k_s}\right)^2$    & $\frac{5}{6}\,\left(1-\frac{1}{c_s^2}\right)\,\left(\frac{k_\ell}{k_s}\right)^2$ \\
Solid Inflation (Eq. \eqref{eq:solidsqueezed}) & $0$ & $-\frac{40}{9}\, \frac{F_Y}{F}\,\frac{1}{\epsilon}\, \frac{1}{c_L^2}$\\
\hline
\end{tabular}
\end{table}

\begin{table}[H]
\caption{List of values of $A_{2J}$ from the squeezed limit expansion of Eq. \eqref{eq:bissqueezeexp} for interactions produced in multi-field models of inflation. We have parametrized the amplitude of interactions involving the exchange of massive particles using $\alpha_{s}$ and $r_s$, where $s$ indicates the spin of the particle, for principal and complementary series respectively. Full expressions for these amplitudes can be found in \cite{Lee:2016vti}.}\label{tab:multi}
\centering
\renewcommand{\arraystretch}{2.5} 
\begin{tabular}{| l | c | c | }
\hline
\textbf{Interactions in multi-field models}	& $A_0$	& $A_{2J}$ \\
\hline
Local (Eq. \eqref{eq:slocal}) & $\frac{12}{5} \fnl^{\rm loc} + \frac 35 \fnl^{\rm loc} \left(\frac{k_\ell}{k_s}\right)^2 $ & $3 \fnl^{\rm loc} \left(\frac{k_\ell}{k_s}\right)^2$ \\
\hline
{\bf Massive particles} & & \\
Even Spin (Principal Series)  & $\alpha_0 \left(\frac{k_\ell}{k_s}\right)^{3/2} \cos\left[\mu\, \ln \left(\frac{k_\ell}{k_s}\right)\right]$ & $\alpha_{2J} \left(\frac{k_\ell}{k_s}\right)^{3/2} \cos\left[\mu_{2J}\, \ln \left(\frac{k_\ell}{k_s}\right)+\phi_{2J}\right]$\\
Even Spin (Complementary Series)  & $r_0 \left(\frac{k_\ell}{k_s}\right)^{3/2-\nu} $ & $r_{2J} \left(\frac{k_\ell}{k_s}\right)^{3/2-\nu_{2J}}$ \\
Odd Spin (Principal Series)  & $\alpha_{s} \left(\frac{k_\ell}{k_s}\right)^{2}$ & $\mathcal O\left[\left(\frac{k_\ell}{k_s}\right)^{5/2}\right]$\\
Odd Spin (Complementary Series)  & $\alpha_s \left(\frac{k_\ell}{k_s}\right)^{2} + r_{s} \left(\frac{k_\ell}{k_s}\right)^{5/2-\nu_{s}} $ & $\mathcal O\left[\left(\frac{k_\ell}{k_s}\right)^{5/2}\right]$\\
\hline
\end{tabular}
\end{table}

Notice that, as long as we are constraining non-Gaussianity using the halo power spectrum only, any information on the angle between $\bk_s$ and $\bk_\ell$ is lost. One needs to look at the bispectrum \cite{Baldauf:2010vn,Pollack:2011xp,Sefusatti:2011gt,Yokoyama:2013mta,Tasinato:2013vna,Dizgah:2015kqi,Hashimoto:2015tnv,Tellarini:2016sgp,Hashimoto:2016lmh,Yamauchi:2016wuc,DiDio:2016gpd,Chiang:2017vsq,An:2017rwo,MoradinezhadDizgah:2018ssw,dePutter:2018jqk,MoradinezhadDizgah:2018pfo} or, alternatively, to galaxy shapes \cite{Chisari:2013dda,Chisari:2016xki} and intrinsic alignments \cite{Kogai:2018nse} to be sensitive to it.

\paragraph{{\bf Scale-invariant primordial bispectra}}

The coefficients $A_J$ of the expansion contain information on the specific inflationary model from which the interaction giving rise to the bispectrum generates. It is instructive to start from the sub-class of models we defined in Section \S\ref{sec:interactions} which are scale-invariant. It is straightforward to show that 

\begin{equation}\label{eq:scaleinv}
A_{J}(k_s,k_\ell) =  a_J (k_\ell/k_s)^\Delta
\end{equation}

for scale-invariant bispectra. We can then explicitly write the modulation of a specific long mode $k_\ell$ on the local power spectrum around a small Lagrangian patch centered in $\bq$ as

\begin{equation}\label{eq:pmodu}
P_\zeta(\bk_s | \bq) = \left\{1+ \left[a_0 + \frac 32\left(\frac{k_\ell^i k_\ell^j}{k_\ell^2} -\frac 13\delta^{ij}\right)\, \frac{k_s^i k_s^j}{k_s^2}\, a_2 + ...\right]\, \zeta(\bk_\ell) \left(\frac{k_\ell}{k_s}\right)^\Delta e^{i\bk_\ell\cdot\bq}\right\} P_\zeta(k_s),
\end{equation}

where in the second term in the square brackets we have wrote explicitly the second order Legendre polynomial $\mathcal L_2(l_\ell, k_s)$, factorizing out short modes, and the ellipses indicate higher order terms in $J$.  Here it is important to stress that the modulation is imprinted in the initial conditions for structure formation, at the \emph{Lagrangian} position $\bq$, as compared to the Eulerian position at which we usually make observations of the (galaxy) density field, $\bx$ \cite{Giannantonio:2009ak,Baldauf:2010vn,Assassi:2015fma}. Eq. \eqref{eq:pmodu} suggests a generalization of Eq. \eqref{eq:sigmamod} by defining operators such as 


\begin{equation}\label{eq:psijzero}
\psi_{(\Delta)}(\bq)=\int \frac{\d^3 k}{(2\pi)^3} \,k^\Delta\, \zeta(\bk)\,e^{i\bk\cdot\bq}
\end{equation}

for $J=0$ and

\begin{equation}\label{eq:psitwo}
\psi_{(\Delta)}^{ij}(\bq)=\int \frac{\d^3 k}{(2\pi)^3} \, \frac 32\left(\frac{k^i k^j}{k^2} -\frac 13\delta^{ij}\right)  \, k^\Delta\, \zeta(\bk)\,e^{i\bk\cdot\bq}
\end{equation}

for $J=2$. The integration allows to account for the collective contribution from all long modes and has generically support for $k_\ell<\Lambda$, where $\Lambda$ is an arbitrary scale splitting long from short modes. Higher orders can be calculated by using $J-$th order trace-free projection operators with respect to $\bk_\ell$. These fields can be evaluated at the Eulerian position $\bx$ by expanding around $\bx(\bq,\tau) = \bq + {\bf s}(\bq,\tau)$, being ${\bf s}$ the Lagrangian displacement, in perturbation theory

\begin{equation}\label{eq:lagpsi}
\psi(\bx(\bq,\tau),\tau) = \psi(\bq) + {\bf s}^i(\bq,\tau)\partial_i \psi(\bq) + ... .
\end{equation} 

Using a similar reasoning as in the local case, we then have a scale-dependent bias term in the halo auto-power spectrum of the form

\begin{equation}\label{eq:a0ngbias}
P_{\rm hh}(k)|_{J=0} = \left( b_1 + \frac{a_0}{\mathcal M(k)} b_{\psi_{(\Delta)}}\, k^\Delta\right)^2 P_{mm}(k)
\end{equation}

for the case of $J=0$ at leading order. Here $b_{\psi_{(\Delta)}}$ is the bias parameter corresponding to the operator of Eq.  \eqref{eq:psijzero}, which is a generalization of Eq.  \eqref{eq:ngbiasamp}. Using a similar argument on the separation of scales leading to the expression of Eq. \eqref{eq:ngbiasamp}, the rescaling of the short-wavelengths modes by the long ones in this case reads

\begin{equation}
\delta_s\longrightarrow \left[1+\epsilon k^{-\Delta}\right] \delta_s,
\end{equation} 

where we have defined $\epsilon = a_0 \psi_{(\Delta)}$. Consequently, we get

\begin{equation}\label{eq:pbsgeneric}
b_{\psi_{(\Delta)}}=\left. \frac{\partial \ln \bar n_h}{\partial \epsilon}\right|_{\epsilon=0}.
\end{equation}

The peak-background split is extremely powerful to derive expressions for bias parameters based on the response of the local number density of halos to the variation of a particular physical quantity. These relations are inherently renormalized, physical quantities that can be directly used to compute n-point correlation functions of halos and tracers in general. In order to use them as predicted paramaters, a specific modelling of halo formation needs to be implemented, as we discuss in the next section. 

The local-type non-Gaussianity corresponds to $a_0=12/5\fnl$ and $\Delta=0$. For any $\Delta>0$ the enhancement on large scales is reduced. For instance, non-Gaussianity from DBI inflation has $\Delta=2$, as indicated in Tab. \ref{tab:single}.  In this case, the non-Gaussian contribution to the halo power spectrum is scale-independent and thus degenerate with the linear bias $b_1$. This can be understood by looking at Eq. \eqref{eq:psijzero}: $\psi_{(2)} \propto \nabla^2 \zeta$ is directly proportional to the matter density. Similarly, higher values of $\Delta$ give contributions which are degenerate with  bias parameters related to higher-derivative operators of the form $\nabla^{\Delta-2} \delta$. The intermediate range $0 \leq \Delta \leq 2$ is explored by models with massive particles of various spins produced during inflation, as discussed in Sec. \S\ref{sec:massivep}.  

The second term in Eq. \eqref{eq:a0ngbias} is the leading order correction to the linear bias in the presence of generic scale-invariant primordial bispectra. For $J>0$, the correction comes at higher orders: for instance for $J=2$, the operator $\psi_{(\Delta)}^{ij}$ needs to be multiplied by some operator $O^{ij}$ in order to appear in the bias expansion of Eq. \eqref{eq:ngexp}, simply because the halo overdensity $\delta_h$ is a scalar quantity. We make this statement more precise in a few paragraphs. 

Let us first discuss how to extend the squeezed limit expansion of the bispectrum of Eq. \eqref{eq:bissqueezeexp}. The extension is motivated by the fact that the limit is strictly valid until the modes $k$ we observe do not affect the local formation of halo. This can be diagnosed by checking how the variance $\sigma^2$ smoothed on the small-scale $R_s$ is affected by $k$, and it approximately coincides with modes at the peak of the matter power spectrum, $k\sim 0.02$ h/Mpc. Given that observational data hardly constrain modes much larger than $k\sim 10^{-3}$ h/Mpc, it is important to understand how much can we exploit from analytic methods before resorting to N-body simulations.

If we restrict to the scale-invariant case, we can perturbatively add terms in higher powers of $k_\ell/k_s$ and get 

\begin{equation}\label{eq:beyondbissqueezeexp}
B_\zeta(\bk_1,\bk_2,\bk_3) = \sum_{J, N=0}^{\infty} a_{2J,2N}\left(\frac{k_\ell}{k_s}\right)^{\Delta+2N}\, \mathcal L_{2J}(\hat\bk_s\cdot\hat\bk_\ell)\, P_\zeta(k_s) P_\zeta(k_\ell) ,
\end{equation}

where $N$ runs also only on even powers. Since the correction comes in the same form as the leading term at each order in $J$, we then just have to add operators of order $\delta^{2N}$ to the bias expansion. For instance, for $J=0$ we get for the halo auto-power spectrum

\begin{equation}\label{eq:beyonda0ngbias}
P_{\rm hh}(k)|_{J=0} = \left[ b_1 + \left(a_{0,0}\, b_{\psi_{(\Delta)}}+a_{0,2} \,b_{\psi_{(\Delta+2)}} k^2 + ...\right) \, \frac{k^\Delta}{\mathcal M(k)}\right]^2 P_{mm}(k).
\end{equation}

As mentioned in the squeezed limit case, these higher order terms are potentially degenerate with other bias terms: in addition with bias operators from higher-derivative terms present even with Gaussian initial conditions, in this case there is degeneracy also when combining lower order scale-dependent bias from non-Gaussian initial condition with Gaussian ones. For instance, if we combine $\psi_{(\Delta)}$ with a bias generated by the operator $\nabla^2\delta$, we get a contribution approximately of the same order as $b_{\psi_{(\Delta+2)}}$. Moreover, the transfer function $\mathcal M(k)$ also contributes in this expansion when considering the full matter bispectrum in the squeezed limit \cite{Schmidt:2013nsa} with terms which become relevant at the matter-radiation equality scale, $k_{\rm eq}$.

\paragraph{{\bf Beyond scale-invariance} }

In the previous paragraph we have made the approximation of a scale-invariant primordial bispectrum, which excludes non-Gaussianities generated by features during inflation, or any case in which a particular scale is involved in the primordial interaction which generates the bispectrum (or higher-order correlators). The effect of models with features on the scale-dependent bias was first investigated in \cite{CyrRacine:2011rx}. The results above can be generalized to the case of features, here we consider a simple example to illustrate the calculation. Let us consider a simplified form of the sharp feature shape of Eq. \eqref{eq:sharp} by defining

\begin{equation}
\mathcal S(k_1,k_2,k_3) \approx A \sin\left(\frac{K}{k_*}\right),
\end{equation} 

being $A$ some overall amplitude and $k_*$ the scale of the feature. Here we are neglecting the constant phase of the oscillations, the power law and damping factor of Eq. \eqref{eq:sharp} as they are not relevant for the argument. 
Let us also suppose that this bispectrum is generated by a single-field model of inflation, e.g. a sharp feature on the potential (see \cite{Bartolo:2013exa} for a treatment in the context of the EFT of inflation). According to what discussed in \S\ref{sec:single-field}, when expanding in the squeezed limit, terms of order $(k_\ell/k_s)^0$ and $(k_\ell/k_s)$ are not observable. Let us suppose therefore that we have made the calculation in CFC and removed such terms. The squeezed limit at leading order will then look like

\begin{equation}\label{eq:squeesha}
\mathcal S(k_s,k_\ell) \approx A'\left[\frac 13 \frac{k_s}{k_*}\cos\left(\frac{2 k_s}{k_*}\right) - \frac{k^2_s}{k^2_*}\sin\left(\frac{2 k_s}{k_*}\right)\right] \frac{k^2_\ell}{k^2_s}  
\end{equation}

where we have also integrated over the angle between $k_s$ and $k_\ell$. We have now to distinguish two regimes: the ultra-squeezed limit, in which $k_\ell\ll k_*$, can be described by the scale-invariant ansatz of Eq. \eqref{eq:scaleinv} with $\Delta=2$. On the other hand, as $k_\ell$ approaches $k_*$, the product of $k^2_s/k^2_* \times k^2_\ell / k^2_s$ from the second term of Eq. \eqref{eq:squeesha} becomes order unity. It is straightforward to show that going to higher orders does not help, and indeed each order becomes equally important. The method therefore fails unless we are able to resum the expansion. In \cite{Cabass:2018roz} it was shown in the case of resonant running corresponding to the shape of Eq. \eqref{eq:resrun} that this can be done under certain assumptions on the dependence of halo abundance on the statistics of short modes. We consider such methods in the Sec. \S\ref{sec:haloclustering}.

\subsubsection{The bias expansion}

The extension of the scale-dependent bias signature from local-type non-Gaussianity to a wider range of models provides the basic information to write a generic bias expansion with which we can systematically compute correlators of the halo density field. Here we highlight the main steps required to include PNG to the bias expansion, while we defer details on the Gaussian terms to the recent review \cite{Desjacques:2016bnm}.

A perturbative bias expansion connecting the distribution of halos (or tracers in general) to the dark matter field relies on the assumption that the formation of halos takes place over a very long period of cosmic time, but it is affected by a relatively short range of spatial scales. In other words, if we assume that the halo formation process develops within some scale $R_*$, which is typically the Lagrangian radius of the halo, and we study the halo distribution on much larger scales, then we can effectively write this process as local in space (but non-local in time). In this picture, the equivalence principle applied to the ``free-falling'' small region of size $\sim R_*^3$ imposes that only second derivatives of the gravitational potential appear in the expansion. Therefore, we can write the expansion   

\begin{equation}
\delta_h(\bx,\tau) = \sum_O b_O(\tau) [O](\bx,\tau),
\end{equation}

where $b_O$ are bias parameters related to the operators $O$ which are  all the scalar combinations that can be made, order by order, out of the matter density field $\delta$ and the tidal field\footnote{In the context of structure formation, it is conventional to work with the gravitational potential $\Phi$, which in matter domination is simply related to the curvature perturbation by $\Phi=3/5 \zeta$. } $\partial_i\partial_j\Phi$. Because of non-locality in time, that is, the fact that the halo formation process happens over a long range of time, the terms of the expansion will in general include convective time derivatives \cite{Mirbabayi:2014zca,Assassi:2015fma}. The square brackets indicate that the operators need to be renormalized, as typically, excepting the simple linear term, they involve products of fields evaluated at the same point, generating ultraviolate divergencies \cite{McDonald:2006mx,McDonald:2009dh,Assassi:2014fva,Schmidt:2012ys,Senatore:2014eva,Angulo:2015eqa}. 
Moreover, small deviations from locality can be accounted for perturbatively in terms of higher-derivative terms such as $\nabla^2 \delta$ to be added to the expansion. 

The final ingredient to be added to this expansion is stochasticity, i.e. the impact of small-scale perturbations on the formation of galaxies \cite{Dekel:1998eq,Taruya:1998hf,Matsubara:1999qq}. We can include these effects using techniques from the effective field theory approach, introducing stochastic fields in the expansion which do not correlate with long-wavelength mode fields. Stochasticity in the presence of PNG becomes especially relevant in the case in which non-Gaussianity is sourced in the initial conditions by a superposition of two or more fields (see Secs. \S\ref{sec:massivep} and \S\ref{sec:after}) . These models are characterized by large trispectra, which are not discussed here, therefore we will neglect stochasticity. A complete treatment, including the case of primordial non-Gaussian initial conditions, can be found in \cite{Desjacques:2016bnm}, while original papers date back a few years earlier \cite{Smith:2010gx,Baumann:2012bc}\footnote{See also \cite{McAneny:2017bbv} for a recent treatment in the context of quasi-single field inflation.}.

The assumption of separation of long- and short-scales is no longer valid in the case of primordial non-Gaussianity: indeed, we have seen above that more operators derived from $\zeta$, or equivalently $\Phi$, itself are allowed in the expansion. The coupling in this case takes place at the \emph{Lagrangian} position $\bq$, since it is imprinted in the initial conditions of structure formation. Following \cite{Desjacques:2016bnm} we quote here the full bias expansion  with PNG up to order $J=2$ in the scale-invariant approximation of Eq. \eqref{eq:scaleinv}

\begin{align}\label{eq:biasexp}
\delta_h (\bx) &= b_1 \delta(\bx) + b_{\nabla^2\delta}\nabla^2\delta(\bx) + a_0\, b_\psi \psi(\bq)+ a_0\, b_{\nabla^2\psi} \nabla^2_{\bq} \psi(\bq) \nonumber\\
&+b_2 \delta^2(\bx) + b_{\mathcal K} \left(\mathcal K_{ij}\right)^2 + a_0\, b_{\psi\delta} \psi(\bq)\, \delta(\bx)+ b_{\psi\mathcal K} \psi^{ij}(\bq) \mathcal K_{ij}(\bx) + \mbox{ stochastic terms } + \mathcal O( a_4, \delta^3,\nabla^4 \delta),
\end{align}

where we defined the tidal field 

\begin{equation}\label{eq:tidal}
\mathcal K_{ij} = \frac{2}{3\Omega_m \mathcal H^2} \,\partial_i\partial_j \Phi - \frac 13 \delta_{ij} \delta.
\end{equation}

The first line is leading order in $\delta$ and the non-Gaussian correction $\psi$, where we have considered the leading order in Eq. \eqref{eq:lagpsi}, while the second line is second order.  We do not include terms which are second order in $\psi$ given that from current CMB observations we expect PNG corrections to be small \cite{Akrami:2019izv}. We have only implicitly listed  stochastic terms: at first order one needs to introduce an additional field, $\epsilon_\psi$ with respect to the Gaussian case, which has only $\epsilon$ \cite{Assassi:2015fma}. We have truncated the expansion on three different levels:

\begin{itemize}[leftmargin=*,labelsep=5.8mm]
\item The squeezed limit expansion  $a_J$ with $J>2$. This corresponds to contributions from higher-spin fields. As highlighted in Sec. \S\ref{sec:massivep} a particular feature of this interaction is that the primordial bispectrum has dependence on the angle between the long and short mode. This information though is integrated over in the halo power spectrum. Moreover, we have suppressed powers of $a_J$ as we expect $a_J$s to be small from current constraints.
\item At third order in the density field. Since most of the effect of PNG is at large scales, where $\delta\ll 1$, the series expansion can be safely truncated at third order. As smaller and smaller scales are probed, higher orders need to be introduced.
\item Higher-derivatives terms. At large enough scales $k\ll 1/R_*$, higher-derivatives terms such as $b_{\nabla^4\delta}\,\nabla^4\delta$ can be neglected, as they typically scale as $R_*^4 k^4 \delta$ \cite{Desjacques:2016bnm}.
\end{itemize}

Up to the order considered, Eq. \eqref{eq:biasexp} is the most generic bias expansion for the halo density field. The unknowns in this expansion are the bias parameters $b_O$: similarly to other effective field theory expansions, these need to be measured from data, in this case either via direct observations of the galaxy statistical distribution or by analysing N-body simulations of the Universe. A series of recent efforts have focused on extracting these parameters from high-resolution simulations for Gaussian simulations \cite{Fujita:2016dne,Perko:2016puo,Abidi:2018eyd,delaBella:2018fdb}, while analyses on universes with primordial non-Gaussianity are still missing.

\subsubsection{Models of halo clustering }\label{sec:haloclustering}

The results of the previous paragraphs do not rely on a specific modelling of halo formation and clustering. On the assumption that on large enough scales, the formation of structures is well described by the sole action of gravity, it is possible to describe the statistics of tracers with a finite set of operators, with related bias parameters. Given a set of initial conditions, the perturbative treatment of the clustering of halos (or tracers in general) allows to be rather agnostic about the small scale details of the late universe evolution, and predictions have a calculable theoretical error at each order in the expansion. The drawback of this approach is that it has to rely on observations: the parameters of the expansion are free to vary and Eq. \eqref{eq:biasexp} clearly shows that, especially in the case of non-Gaussianity in the initial conditions, there are many of them to fit to observations or N-body simulations. 

For this reason, a certain degree of modelling can be combined to the perturbative expansion in order to provide reliable priors on the bias parameters. Chronologically, models of halo and galaxy formation and clustering date back even before the first perturbative methods were applied, with the models of spherical collapse 
first introduced in the $70$s. Since then, progress has developed into two main directions: the excursion-set approach and the peak model, which we briefly introduced in the previous section \S\ref{sec:onepoint}. The common underlying idea of these models is to build a correspondence between low redshift halos and their progenitors at early times based on a few simple, but well motivated, assumptions on the statistics and physical processes of the initial conditions of structure formation. They are therefore inherently \emph{Lagrangian} models, as opposed to the \emph{Eulerian} expansion of Eq. \eqref{eq:biasexp}. The advantage of the Lagrangian description, in the context of primordial non-Gaussianity, is that no additional operators have to be introduced with respect to the Gaussian case. There are several efforts in the literature to model the two-point correlation function of tracers including primordial non-Gaussianity, using thresholded regions \cite{Matarrese:2008nc,Desjacques:2011mq,Desjacques:2011jb}, the excursion set approach \cite{Desjacques:2011mq,Desjacques:2011jb,D'Aloisio:2012hr,Adshead:2012hs,Musso:2012ch,Desjacques:2013qx}, the peak model \cite{Desjacques:2013qx} (see also \cite{Desjacques:2016bnm} for an overview).  As in the previous section \S\ref{sec:onepoint}, we choose the example of the excursion set peaks model to derive the non-Gaussian contribution to the halo power spectrum and conclude by making the connection with the generic bias expansion of Eq. \eqref{eq:biasexp}.

\paragraph{{\bf Excursion Set Peaks}}

In order to compute the non-Gaussian correction to the power spectrum of peaks, we make use of the effective ESP bias expansion developed in a number of recent papers \cite{Desjacques:2012eb,Desjacques:2013qx,Biagetti:2013hfa,Lazeyras:2015giz,Desjacques:2017msa}. The basic idea is to write the \emph{peak overdensity field} $\delta_{\rm ESP}^{\rm L}$ as an effective perturbative expansion constructed from the rotational invariants of the system, namely the components of the vector $\boldsymbol{\omega}$ defined in Eq. \eqref{eq:peakvec}

\begin{equation}
\delta^{\rm L}_{\rm ESP} (\bq) = c_\nu \nu(\bq) + c_{J_1} J_1(\bq) + c_{\mu} \mu(\bq) + c_{\eta} \eta^2 + c_{J_2} J_2(\bq) + c_{J_3} J_3 + c_{\nu \, J_1} \nu(\bq) J_1(\bq) + ...,
\end{equation}

where $c_i$ are bias parameters and we have not respected any particular order in the expansion, but rather just shown that any combination of the invariants enter the expansion. In order to refine the expansion and make it useful to calculate correlation functions, we have to i) find a way to predict bias parameters and ii) ensure to remove all zero-lag terms such as for example $\langle \delta^{\rm L}_{\rm ESP}(\bq) \delta^{\rm L}_{\rm ESP}(\bq')\rangle \supset c^2_\nu$ that are generated when using the expansion to compute correlators of $\delta^{\rm L}_{\rm ESP}$. The way to go is to write the expansion in terms of an entire set of orthogonal polynomials $O_{\bf n} [\boldsymbol{\omega}(\bq)]$ 

\begin{equation}
\delta^{\rm L}_{\rm ESP} (\bq) = \sum_{{\bf n}\neq 0} \sigma^{[{\bf n}]} b^{\rm L}_{\bf n}O^*_{{\bf n}}(\boldsymbol{\omega}(\bq))
\end{equation}



where the indices ${\bf n}=\{i,j,k,\ell,m,n\}$ correspond to the six variables of $\boldsymbol{\omega}$, $\sigma^{[{\bf n}]}$ is an abbreviation for $\sigma^{[{\bf n}]}=\sigma_0^{i} \sigma_1^{2\ell} \sigma_2^{j+2m+3n}\varsigma_0^{k}$, where we have defined

\begin{equation}
\varsigma_j^2(R)\equiv \frac{1}{2\pi^2}\int_0^\infty\!\!dk\,k^{2(j+1)} P_L(k) \Big(\frac{\d{W(kR)}}{\d R}\Big)^2
\end{equation}

as the variance of the up-crossing variable $\mu=-\d\delta/\d R$ and $O^*$ is the dual of $O$. The polynomials $O_{\bf n} [\boldsymbol{\omega}(\bq)]$ are found by looking at the probability distributions of the rotational invariants, namely 

\begin{itemize}[leftmargin=*,labelsep=5.8mm]
\item The trivariate Hermite polynomials $H_{ijk}(\nu,J_1,\mu)$ are associated to the trivariate normal distribution $\mathcal N(\nu,J_1,\mu)$
\item The  Laguerre polynomial $L_i^{(1/2)}\left(\frac 32 \eta^2\right)$ is associated to the $\chi^2$-distribution with $3$ degrees of freedom of $\eta^2$ and $J_2$
\item The polynomials of the form

\begin{equation}
F_{ij}(5 J_2, J_3) = (-1)^i\sqrt{\frac{\Gamma\left(\frac 52\right)}{2^{3j}\Gamma\left(3j+\frac 52\right)}} L_i^{(3j+3/2)}\left(\frac 52 J_2\right) \mathcal L_j (x_3),
\end{equation}

where $\mathcal L$ are Legendre Polynomials and $x_3=J_3^2/J_2^3$.

\end{itemize}

It is straightforward to verify that indeed these polynomials satisfy the orthogonality conditions and that they are a complete basis for the variables of the system \cite{Lazeyras:2015giz}. The coefficients $b^{\rm L}_{\bf n}$ are renormalized bias parameters which can be measured through 1-point ensemble averages \cite{Lazeyras:2015giz} 

\begin{equation}
\sigma_0^{i}\, \sigma_1^{2\ell}\, \sigma_2^{j+2m+3n}\, \varsigma_0^k\, b^{\rm L}_{ijk\ell mn} = \frac{1}{\bar n_{\rm ESP}}\left\langle \bar n_{\rm ESP}\, H_{ijk}(\nu,J_1,\mu)\, (-1)^\ell L_\ell^{(1/2)}\left(\frac{3\eta^2}{2}\right) F_{mn}(5J_2, J_3) \right\rangle,
\end{equation}

where the presence of the moments $\sigma_{0}$, $ \sigma_1$, $\sigma_{2}$ and $\varsigma_0$ ensures that the expansion is written in terms of the physical variables $\delta, \eta$ and $\zeta$ rather than the normalized ones (cfr. Eqs. \eqref{eq:normpeak1} - \eqref{eq:normpeak3}). Putting everything together, we can write down the effective ESP expansion up to second order in $\delta$

\begin{align}
\label{eq:loc_exp}
\delta_\esp^L(\vq) &= b_{100}\delta_R(\vq) - b_{010} \nabla^2\delta_R(\vq) - b_{001} \frac{d\delta_R}{dR}(\vq)\nonumber \\
&+ \frac{1}{2} b_{200} (\delta_R^2(\vq)- \sigma_0^2) + \frac{1}{2} (b_{020} \big[\nabla^2\delta_R(\vq)\big]^2 -\sigma_2^2)
+ \frac{1}{2}b_{002} \left[\left(\frac{d \delta_R}{dR}\right)^2\!\!(\vq)-\varsigma_0^2\right] \nonumber \\
&-  b_{110}(\delta_R(\vq)\nabla^2\delta_R(\vq)+\sigma_2^2) - b_{101} \left(\delta_R(\vq) \frac{d\delta_R}{dR}(\vq)+\gamma_{\nu\mu}\right)
+ b_{011}\left( \nabla^2\delta(\vq)\frac{d\delta_R}{dR}(\vq)-\gamma_{J_1\mu}\right) \nonumber \\
&+ \chi_1 \left[\big(\nabla\delta_R)^2\!(\vq) - \sigma^2_1\right]+ \frac{3}{2}\omega_{10}\left\{
\left[\partial_{ij}\delta_R-\frac{1}{3}\delta_{ij}\nabla^2\delta_R\right]^2\!\!\!\!(\vq)-\sigma^2_2\right\} + \mathcal O(\delta_R^3)
\end{align}

where we have adopted the notation

\begin{align}
b^{\rm L}_{ijk000} &= b_{ijk}\\
b^{\rm L}_{000\ell00}&=\chi_\ell\\
b^{\rm L}_{0000mn}&=\omega_{mn}
\end{align}

 following previous literature. Let us make a few comments on this expansion. As compared to the Eulerian expansion of Eq. \eqref{eq:biasexp}, there are two crucial differences: first of all, this effective ESP expansion is done in the initial conditions in Lagrangian space, therefore it does not account for any non-linearity generated by gravitational evolution. The perturbative treatment in the context of Lagrangian space, as well as the connection to Eulerian space can be done using the formalism of Integrated Perturbation Theory \cite{Matsubara:2011ck,Matsubara:2012nc,Matsubara:2013ofa,Yokoyama:2013mta,Matsubara:2016wth}. This particular formulation of perturbation theory is specifically suited to embed the ESP framework.
 For the purpose of this review, working  with Eq. \eqref{eq:loc_exp} is enough to get the  intuition of the physics at play. Secondly, Eq. \eqref{eq:loc_exp} does not include any tidal shear, that is, scalar combinations of the field $\mathcal K_{ij}$ defined in Eq.  \eqref{eq:tidal}. In the presence of non-Gaussianity, including the tidal shear in the modeling of the ESP is tightly connected with the universality of the mass function and the way one models the scatter in the collapse barrier, so we dedicate the following paragraph to discuss it properly. For the time being, let us assume that the collapse is spherical and that we can neglect the tidal shear in the expansion of Eq. \eqref{eq:loc_exp}.
 
We want now to compute the non-Gaussian correction to the power spectrum of peaks as predicted by the ESP model. In order to get a contribution from the primordial bispectrum of $\zeta$ in the power spectrum, we need at least terms proportional to $\langle \delta_R^3\rangle$. At leading order, this is possible only by combining a first order bias term with a second order one, so that, in Fourier space, we get

\begin{equation}
\Delta P^{\rm NG}_{\rm ESP}(k) = \frac{c^{\rm L}_1(k)}{\mathcal M_R(k)} P_R(k) \, \int \frac{\d^3 q}{(2\pi)^3} c_2^{\rm L}(\bq, -\bq-\bk) \, \frac{\mathcal M_R(|\bk+\bq|)}{\mathcal M_R(q)}\, P_R(q)\, \frac{k\, q}{|\bk+\bq|^2}\, \mathcal S(k,q,|\bk+\bq|),
\end{equation}

where 

\begin{equation}
c_1^{\rm L}(k)=b_{100}+b_{010} k^2-b_{001} \frac{\d\ln W_R(k)}{\d R}
\end{equation}

 and 

\begin{align}\label{eq:espngbias}
c_2^{\rm L}(\bk_1,\bk_2) &= b_{200} + b_{020} \,k_1^2\, k_2^2 + b_{002} \,\frac{\d\,\ln W_R(k_1)}{\d R}\,\frac{\d\,\ln W_R(k_2)}{\d R}\nonumber\\
&+b_{110}(k_1^2+k_2^2)-b_{101} \left[ \frac{\d\,\ln W_R(k_1)}{\d R}+\frac{\d\,\ln W_R(k_2)}{\d R}\right]+b_{011} \left[k_1^2\, \frac{\d\,\ln W_R(k_2)}{\d R}+k_2^2\, \frac{\d\,\ln W_R(k_1)}{\d R}\right]\nonumber\\
&+2\chi_1(\bk_1\cdot\bk_2)+\omega_{10}\left[3(\bk_1\cdot\bk_2)^2-k_1^2k_2^2\right]
\end{align}  

are the first and second order ESP bias parameters in Fourier space
. Note that from now on we will neglect all the zero-lag correction terms from the calculation, leaving implicit that we subtract them whenever needed. The expression of Eq. \eqref{eq:espngbias} allows to compute the leading order non-Gaussian contribution for any given primordial bispectrum. 

It is instructive to take the squeezed limit $k\rightarrow 0$ of Eq. \eqref{eq:espngbias} in order to make the connection with Eq. \eqref{eq:a0ngbias} of the previous section \S\ref{sec:squeezedbispe}. For scale-invariant primordial bispectra of the type Eq. \eqref{eq:scaleinv}, we get

\begin{equation}
\Delta b^{\rm NG}_{\rm ESP}(k) \stackrel{k \rightarrow 0}{=} \frac{a_0}{\mathcal M_R(k)} b^{\rm NG}_{\rm ESP} \, k^\Delta,
\end{equation}

where we defined

\begin{align}
b^{\rm NG}_{\rm ESP} =& \,\,  \int \frac{\d^3 q}{(2\pi)^3}\, q^{-\Delta} c_2^{\rm L}(\bq, -\bq) P_R(q)\nonumber\\
=&\,\, \sigma^2_{-\Delta} b_{200} + \sigma^2_{2-\Delta} b_{020} + \varsigma^2_{-\Delta} b_{002} + 2\sigma^2_{1-\Delta} b_{110}\nonumber\\
&-2(\sigma^2_{-\Delta})' b_{101} + 2(\sigma^2_{1-\Delta})' b_{011}-2\sigma^2_{1-\Delta} \chi_1 + 2 \sigma^2_{2-\Delta} \omega_{10},
\end{align}

where primes denote derivation with respect to $R$. At first sight, this result seems at odds with the model-independent prediction from the peak-background split ansatz of Eq. \eqref{eq:pbsgeneric}. Nevertheless, in \cite{Gong:2011gx} it was proven that the two expressions are equivalent

\begin{equation}\label{eq:masterequi}
b_{\psi_{(\Delta)}}\equiv\left. \frac{\partial \ln \bar n_h}{\partial \epsilon}\right|_{\epsilon=0} = \sum_{i=0}^2\frac{\partial \ln \bar n_h}{\partial \ln \sigma_i} \frac{\sigma^2_{i-\Delta}}{\sigma^2_i} + \frac{\partial \ln \bar n_h}{\partial \ln \gamma_{\nu\mu}} \frac{(\sigma^2_{-\Delta})'}{\gamma_{\nu\mu}}+ \frac{\partial \ln \bar n_h}{\partial \ln \gamma_{J_1\mu}} \frac{(\sigma^2_{1-\Delta})'}{\gamma_{J_1\mu}}+ \frac{\partial \ln \bar n_h}{\partial \ln \varsigma_0} \frac{\varsigma^2_{-\Delta}}{\varsigma_0^2} \equiv b^{\rm NG}_{\rm ESP}
\end{equation}

 for any deterministic barrier for collapse. In the case of local PNG and in the approximation of spherical collapse, one recovers the well known results, $b^{\rm NG}_{\rm ESP}= \delta_{sc}\, b_{100} $, where now the linear bias is a direct prediction of the ESP model. Although this result holds in general for universal mass functions, in this case it is true also for the ESP mass function, Eq. \eqref{eq:fespnesp}, which is clearly not universal, since it depends non-trivially not only on $\nu_c$ but also the spectral moments $\sigma_i$. As argued in \cite{Desjacques:2013qx,Biagetti:2015exa}, the ESP mass function recovers the result for universal mass functions due to the fact that spectral moments appear only in ratios such as $\nu_c/\sigma_0$ or $\gamma_1=\sigma^2_1/\sigma_0 \sigma_2$.

  
  \paragraph{Tidal shear, the collapse barrier and primordial non-Gaussianity}

The results above can be used to predict the halo power spectrum at late redshift under the (strong) approximation that all halos form around peaks of the early matter density field that overcome a flat spherical collapse barrier. Several analyses \cite{Robertson:2008jr,Grossi:2009an,Tinker:2010my,Corasaniti:2011dr,Ludlow:2011jx,Achitouv:2011sq,Elia:2011ds,Achitouv:2012mk,Despali:2012ig,DelPopolo:2012jm,Borzyszkowski:2014xua,Castorina:2016tuc,Reischke:2016eza} of N-body simulations show that the collapse barrier is not really flat, but mass-dependent, and has significant scatter around the mean. The ESP model was therefore extended to allow for a non-spherical barrier with log-normally distributed scatter,

\begin{equation}\label{eq:phenobarrier}
B(\sigma_0)= \delta_{sc} + \beta\, \sigma_0
\end{equation}

 being $\beta$ the log-normally distributed variable and the parameters of the distribution are fitted to what found in the N-body analyses. This phenomenological barrier was shown to fit rather well simulations with Gaussian initial conditions. In parallel, it was realized that non-Gaussian bias amplitude prediction for universal mass functions, $b^{\rm NG}_{\rm ESP}= \delta_{sc}\, b_{1} $, did not reproduce well some N-body simulation measurements, systematically underestimating the signal (see discussion in the next paragraph \S\ref{sec:numeric}). Attempts at improving the agreement by considering ellipsoidal collapse barriers, so as to have moving barriers, but without scatter, have been proposed \cite{Afshordi:2008ru}. In the context of the ESP model, in \cite{Biagetti:2015exa} it was argued that the phenomenological barrier of Eq. \eqref{eq:phenobarrier} might improve the fitting to simulations. However, for this barrier, the equivalence of Eq. \eqref{eq:masterequi} does not hold. Since the PBS result was derived in full generality, the ESP prediction of the non-Gaussian bias using the barrier Eq. \eqref{eq:phenobarrier} is not consistent. The validity of the PBS prediction was also directly verified in N-body simulations \cite{Biagetti:2016ywx}, as we will show in  the next section. 
 
These considerations bring us to the tidal shear. It has been long known that the local tidal shear has an impact on the collapse threshold \cite{Sheth:1999su}. As remarked above, tidal shear is completely missing from the ESP model as we presented it up to now. As shown in \cite{Castorina:2016tuc}, a significant part of the scatter in the collapse barrier is related to the local value of the tidal shear at halo formation. In \cite{Desjacques:2016bnm}, it was demonstrated that if tidal shear is properly included in the ESP model, then the prediction of the non-Gaussian bias amplitude does agree with the PBS result. The crucial difference with respect to the phenomenological barrier of Eq. \eqref{eq:phenobarrier} is that  a physically motivated barrier should be determined by the mean density and tidal shear within the collapse region, and therefore it should not explicitly depend on $\sigma_0$. The reason why this is particularly important in the non-Gaussian bias prediction is that, according to the PBS derivation, the effect precisely comes from the modulation of the halo mass function to local changes in $\sigma_0$. 
While \cite{Desjacques:2016bnm} clarifies this issue and lays down all the necessary ingredients to model tidal shear in the context of the ESP model, a full implementation is still to be done. If tidal shear constitutes most of the observed scatter of the collapse barrier, one should expect that the ESP model successfully predicts the non-Gaussian bias amplitude.

\subsection{Numerical approaches}\label{sec:numeric}

Having discussed in detail various predictions from analytical approaches, we turn now to numerical studies. Calibrating the amplitude of the non-Gaussian bias from local PNG has been one of the main goals in the last decade \cite{Dalal:2007cu,Desjacques:2008vf,Pillepich:2008ka,Grossi:2009an,Wagner:2010me,Hamaus:2011dq,Scoccimarro:2011pz,Wagner:2011wx,Baldauf:2015vio,Biagetti:2016ywx}. Extensions to other shapes than the local case have also been investigated \cite{Shandera:2010ei,Desjacques:2011jb,Wagner:2010me,Wagner:2011wx,Desjacques:2011jb}. The importance of this task is clear from Eq. \eqref{eq:scaledepbias}: the non-Gaussian bias amplitude $b_1^{\rm NG}$ is degenerate in scale with $\fnl$, so any theoretical or numerical uncertainty in its prediction affects the measurement of $\fnl$. Generalizations to different types of PNG, such as Eq. \eqref{eq:a0ngbias}, show that we expect the degeneracy to remain, or even increase, since the non-Gaussian term becomes less and less distinguishable from the Gaussian one as the scaling in $k$ is weaker for $\Delta >0$. The degeneracy can be alleviated by measuring the $1/k^\Delta$ scaling in the power spectrum of tracer populations with different mass and redshift using the multi-tracers technique originally proposed in \cite{Seljak:2008xr}, optimal weighting \cite{Hamaus:2011dq,Mueller:2017pop} and shot-noise suppression \cite{Seljak:2009af}, but it is nonetheless important to have a handle on the uncertainty of $b_1^{\rm NG}$. In this sense, the major feature of Lagrangian models presented in the previous section, that is, being able to predict the value of $b_1^{\rm NG}$, loses part of its advantage since the theoretical error is unknown and therefore it cannot be reliably used in data analyses. 
In the next paragraphs, we briefly summarize the most recent efforts in understanding PNG using N-body simulations.

\subsubsection{Local-type PNG}\label{sec:lpngnum}
For N-body simulations with local-type primordial non-Gaussian initial conditions, measurements of the power spectrum at large scales should show the characteristic $1/k^2$ scaling which is absent in Gaussian ones. The measurement of the non-Gaussian power spectrum has several contributions, other than the non-Gaussian bias $\Delta b^{\rm NG}_1$ we are after, as first pointed out in \cite{Desjacques:2008vf, Pillepich:2008ka}. At leading order, it reads

\begin{equation}
\label{eq:deltabkhm}
\frac{P^{\rm NG}_{\rm hm}(k)}{P^{\rm G}_{\rm mm}(k)}  = b^{\rm G}_{\rm hm} + \Delta b^{\rm NG}_I(\fnl) + b^{\rm G}_{\rm hm}\, \beta^{\rm NG}_m(k) + \Delta b^{\rm NG}_1(k) + \mathcal{O}(b^{\rm G}_2, \fnl^2),
\end{equation}

where $P^{\rm NG}$ and $P^{\rm G}$ are the measured power spectra from simulations with non-Gaussian and Gaussian initial conditions respectively and the subscripts 'hm' and 'mm' indicate the halo-matter and matter-matter correlations, respectively. Let us go through each term separately: 

\begin{itemize}[leftmargin=*,labelsep=5.8mm]
    \item $b^{\rm G}_{\rm hm}$ is the linear Gaussian bias, which can be measured from simulations by measuring the ratio $P^{\rm G}_{\rm hm}(k)/P^{\rm G}_{\rm mm}(k)$ at large scales, 
    
    \vskip2.5mm
    
    \item $\Delta b^{\rm NG}_I(\fnl)$ is the scale independent correction which arises as a consequence of the change of the halo mass function $\bar n_h$ in the presence of PNG, which we extensively discussed in Sec. \S\ref{sec:onepoint}. This effect grows  with increasing halo mass, since the presence of $\fnl$ affects the high mass tale of $\bar n_h$. For the same reason, the correction has opposite sign with respect to $\fnl$, since the bias decreases (increases) whenever the halo mass function is enhanced (suppressed) with positive (negative) $\fnl$,
    
        \vskip2.5mm
        
    \item $\beta^{\rm NG}_m(k)$ is the correction due to the change of the matter power spectrum in the presence of PNG, which can be computed using perturbation theory methods (see \cite{Welling:2016dng} for a recent computation) or directly measured from simulations,

    \vskip2.5mm

    \item $\Delta b^{\rm NG}_1(k)$ is the term we want to measure which is proportional to $1/k^2$,

    \vskip2.5mm
    
    \item $\mathcal{O}(b^{\rm G}_2, \fnl^2)$ indicates that the leading order has corrections from two different directions: non-linear biasing and higher-order primordial correlators. Next-to-leading order corrections can be canceled out combining non-Gaussian simulations with opposite sign $\fnl$ \cite{Biagetti:2016ywx}.
\end{itemize}

Most studies of the non-Gaussian bias from local-type PNG in N-body simulations have used the assumption that the mass function maintains universality even for mildly non-Gaussian initial conditions. Universality simplifies considerably the measurement of the components of $\Delta b^{\rm NG}_1(k)$, because the non-Gaussian bias amplitude is proportional to the linear Gaussian bias $b^{\rm NG}_{\rm univ} = \delta_{sc} b_1$. Since the earliest measurements \cite{Desjacques:2008vf,Pillepich:2008ka}, it was noticed though that this prescription did not match the measurements with the expected precision, and moreover this discrepancy often changed depending on the halo finding algorithm used. In \cite{Pillepich:2008ka}, it was argued that fitted formulas based on the universality assumption deviate from the measured halo mass function from $10$\% up to $30$\% for low mass halos found using a FoF (Friends-of-Friends) algorithm. Moreover, the fit also degraded as a function of decreasing redshift, deviating $10$\% even for more massive halos. In \cite{Grossi:2009an}, the amplitude of the measured scale-dependent bias deviated about $25$\% from the prediction of Eq. \eqref{eq:universal}. Later analyses on both SO (Spherical Overdensity)  and FoF halos confirmed similar discrepancies \cite{Desjacques:2008vf, Hamaus:2011dq,Scoccimarro:2011pz,Wagner:2011wx,Baldauf:2015vio}.

\begin{figure}[!h]
\centering 
\resizebox{0.43\textwidth}{!}{\includegraphics{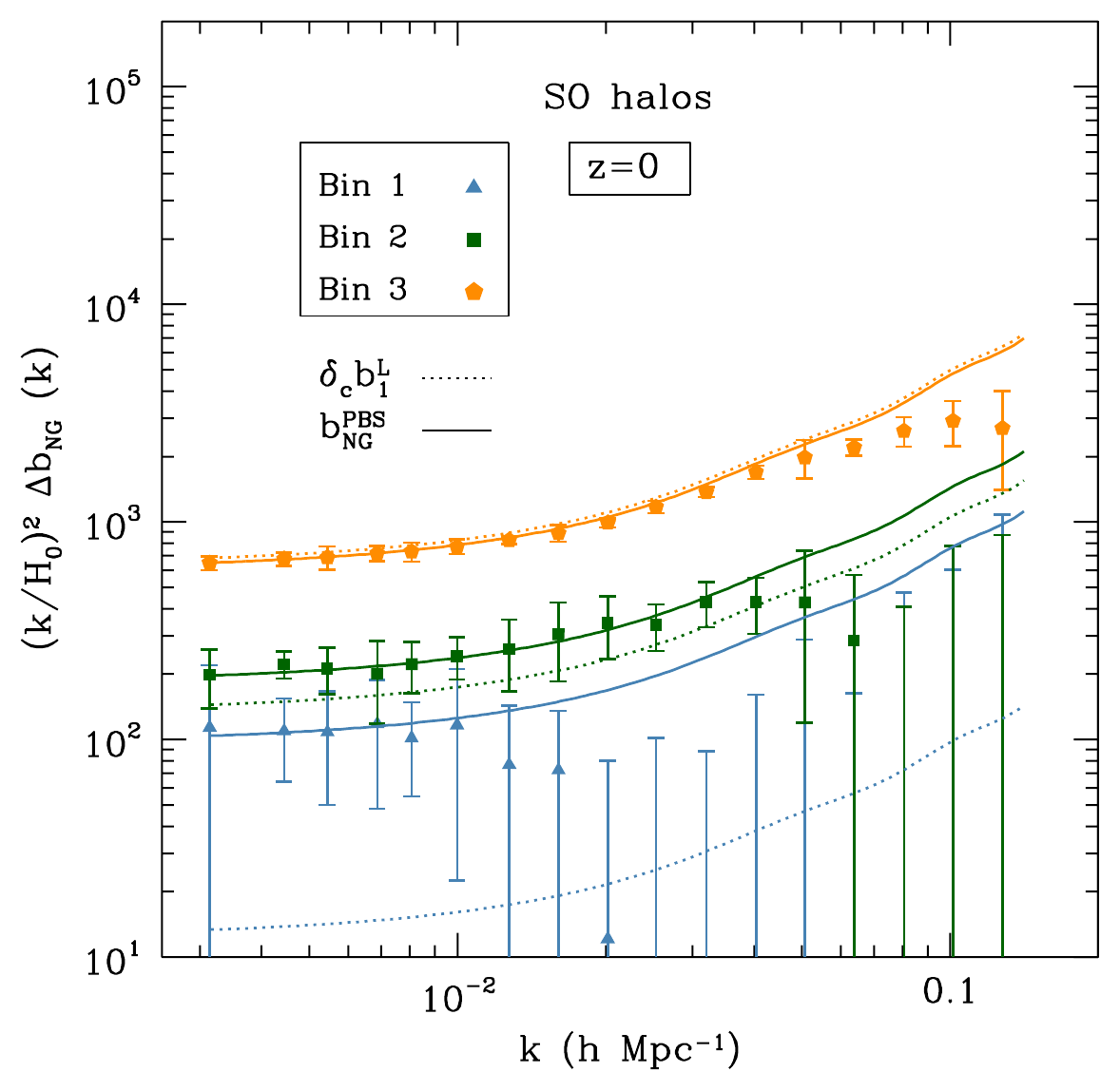}}
\resizebox{0.43\textwidth}{!}{\includegraphics{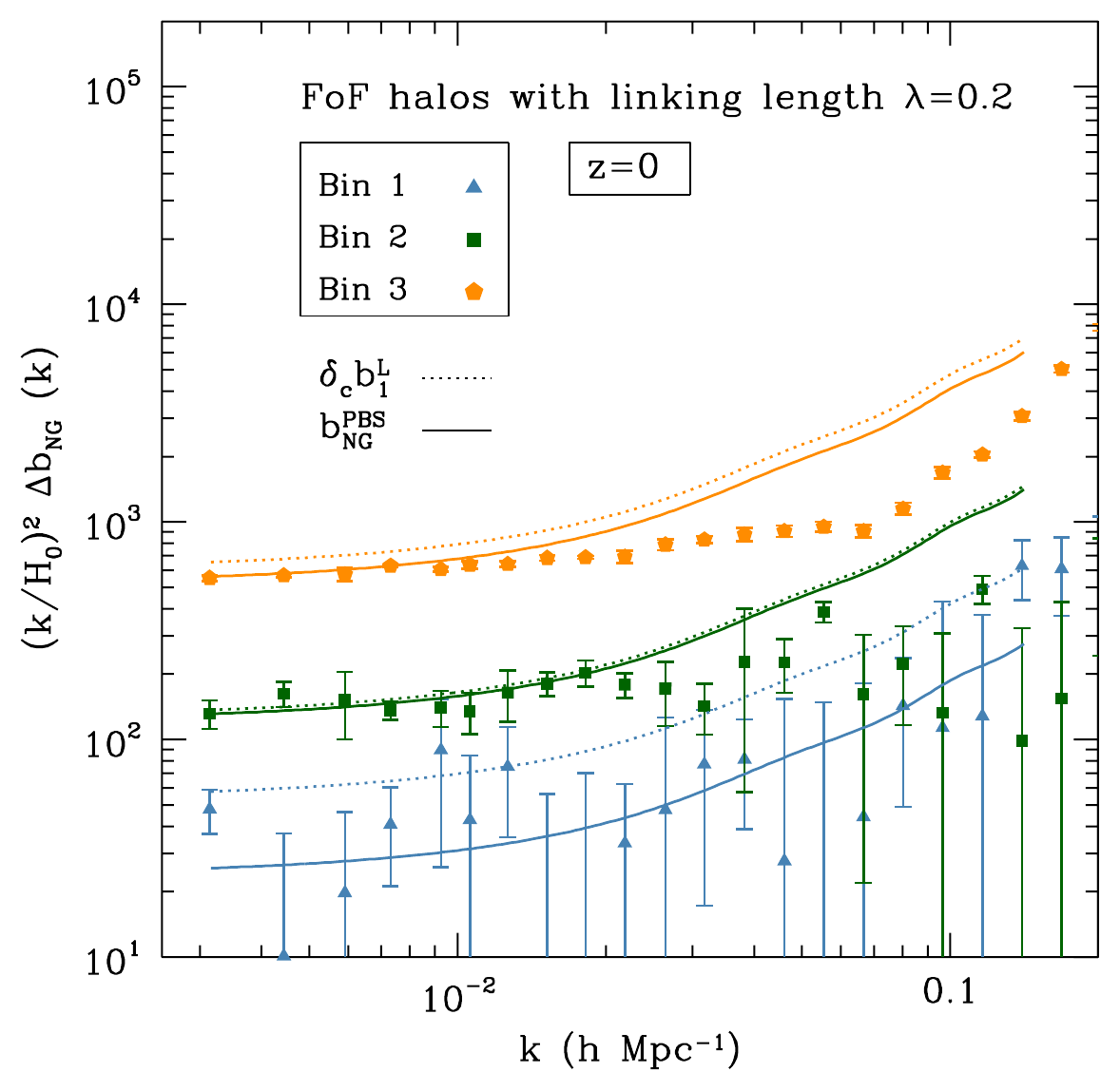}}
\caption{Non gaussian bias for the three different halo mass bins, rescaled by $k/H_0)^2$, as a function of wavenumber $k$ at redshift $z=0$. Left panel shows results from a Spherical Overdensity (SO) halo finder algorithm (AHF code \cite{Gill:2004km}) and right panel for a Friends-of-Friends algorithm with linking length $\lambda=0.2$. Credits to \cite{Biagetti:2016ywx}.}
\label{fig:ngbiasf}
\end{figure}

To settle this problem, \cite{Biagetti:2016ywx} has compared the measurements of the non-Gaussian bias amplitude both with the universal prescription of $b^{\rm NG}_{\rm univ} = \delta_{sc} b_1$ and the peak-background split model-independent amplitude $b^{\rm NG}_{\rm PBS} = \partial \ln \bar n_h / \partial \ln \sigma_8$, see Fig. \ref{fig:ngbiasf}. 

\begin{figure}[!h]
\centering 
\resizebox{0.32\textwidth}{!}{\includegraphics{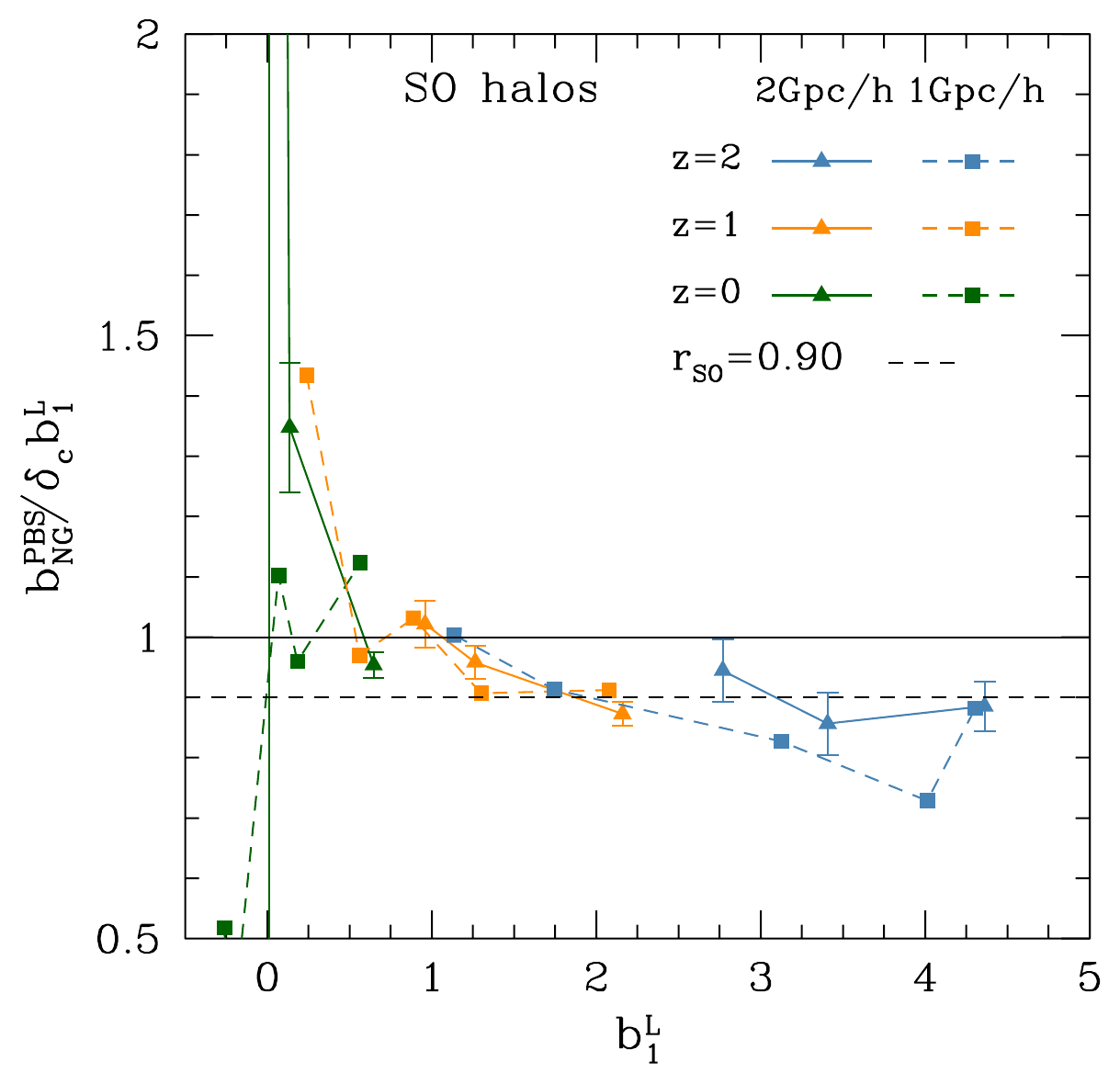}}
\resizebox{0.32\textwidth}{!}{\includegraphics{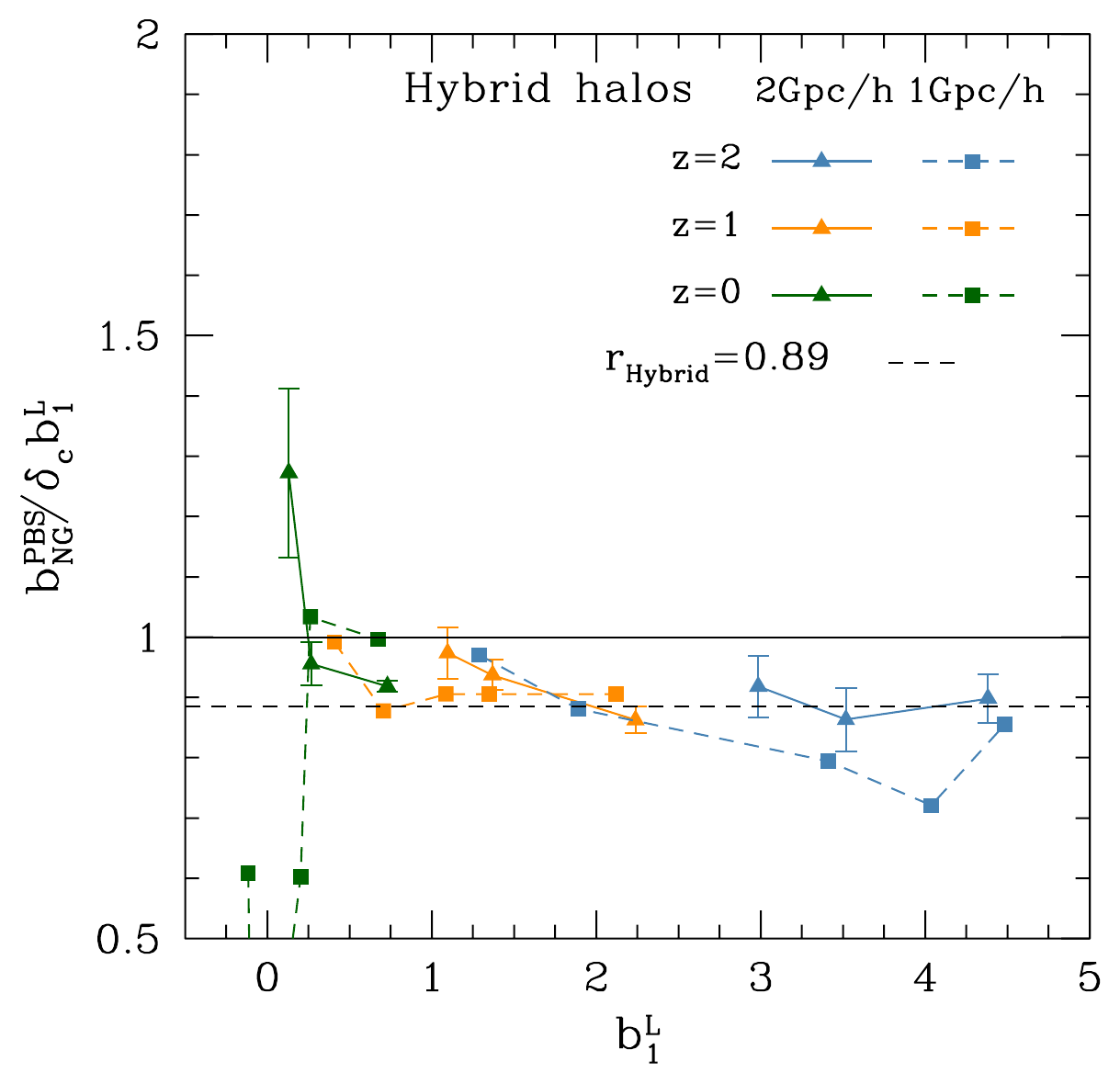}}
\resizebox{0.32\textwidth}{!}{\includegraphics{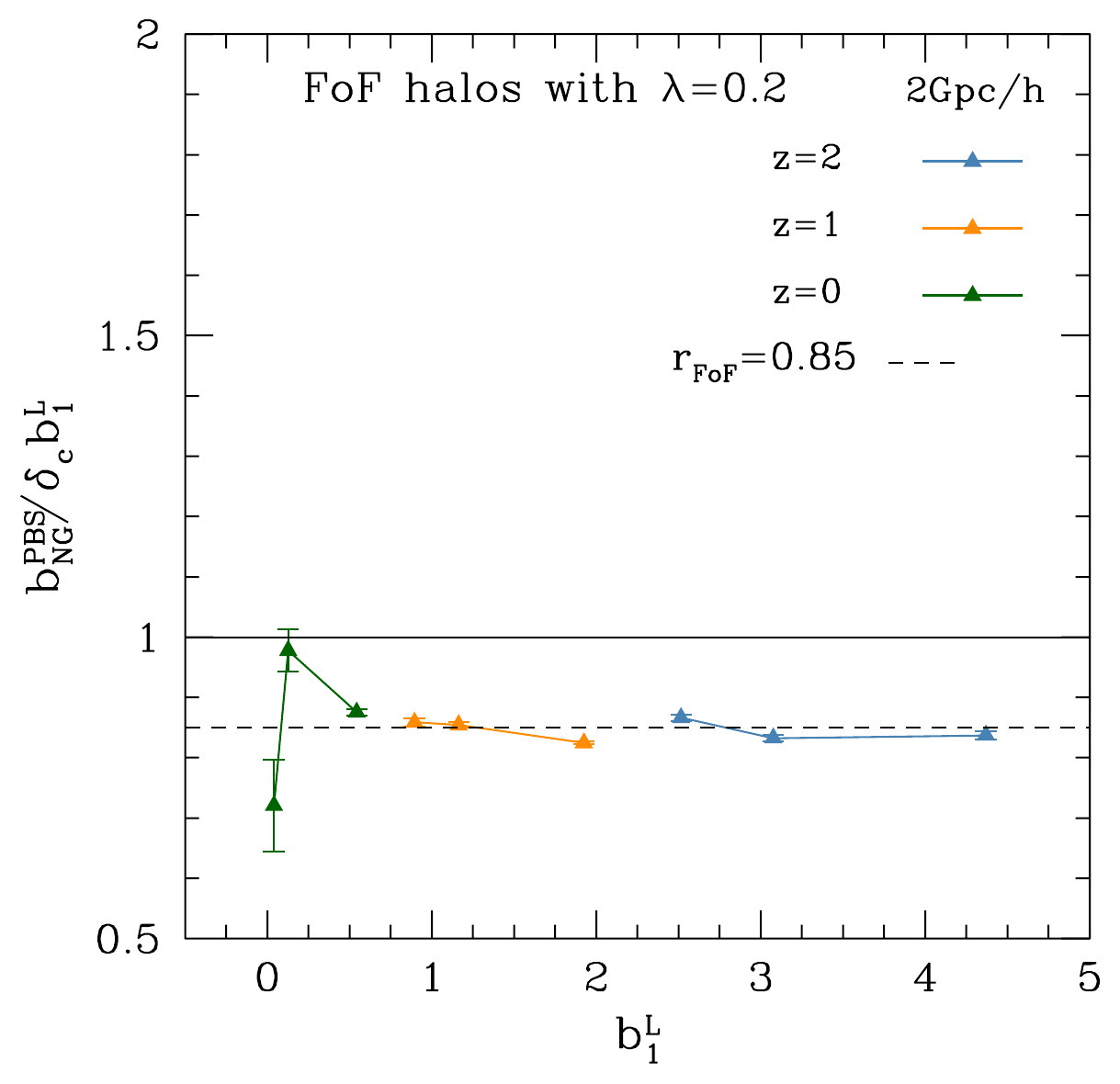}}
\caption{Comparison between PBS and universal predictions for three mass bins and at redshifts $z=0,1$ and $2$.  The different panels show results for different halo finders: SO , Hybrid (combination of FoF and SO, Rockstar code \cite{Behroozi:2011ju}) and a pure FoF with linking length $\lambda=0.2$. 
Black dashed lines indicate the fitted constant value of $b_{\rm NG}^{\rm PBS}/\delta_{sc} b^{\rm G}_1$ at $b^{\rm G}_1 \gtrsim1$ for each finder. Credits to \cite{Biagetti:2016ywx}.}
\label{fig:bngr}
\end{figure}

The latter prescription was directly measured from simulations by running a set of Gaussian simulations with varying matter amplitude $\sigma_8$ and calculating the numerical derivative of the mass function $\bar n_h$. The measurements clearly show that the PBS prediction works extremely well, while the universal mass function approximation systematically underestimates the amplitude. In order to clarify this mismatch, they also measured the ratio of the non-Gaussian bias amplitude as predicted by the PBS split ansatz, $b_{\rm NG}^{\rm PBS}$, to the standard universal prediction $\delta_{sc} b^{\rm G}_1$ as a function of $b^{\rm G}_1$ 
for a combination of three mass bins and at redshifts $z=0,1$ and $2$, see Fig. \ref{fig:bngr}. The discrepancy between the two prescriptions is evident at the level of $10\%$ or more for all the halo finder algorithms used. 

These results call again for a more accurate modeling of the collapse process, which we already discussed in the context of the ESP model in the previous section. Without invoking a specific prescription for the collapse barrier, several attempts in the literature have been made to partially explain these discrepancies. For instance, it is known that FoF halos with linking length $0.2$ tend to have an effective spherical collapse threshold $\delta_{sc}<1.687$ at the high mass end of the halo mass function. This would possibly explain why $b_{\rm NG}^{\rm univ}$ with $\delta_{sc}=1.687$ overestimates $b_{\rm NG}$ at high mass \cite{Grossi:2009an}. However, a change in $\delta_{sc}$ cannot explain the fact that the amplitude $b_1^{\rm NG}$ changes sign, as a function of mass, at a different value for the two different prescriptions. The positive outcome of this analysis is that there is indeed a way to calibrate in a model independent way the non-Gaussian bias amplitude, by measuring the response of the halo mass function to a change in the local value of $\sigma_0$. Besides running simulations with different $\sigma_8$ as done in \cite{Biagetti:2016ywx}, one could also perform this measurement using separate universe simulations \cite{Baldauf:2015vio}.

\subsubsection{N-body simulations with generic non-Gaussian initial conditions}\label{sec:nbodymulti}

While the effects of local-type PNG in the power spectrum of halos have been extensively investigated in N-body simulations, we know that a plethora of other possible interactions may arise from inflation, see Sec. \S\ref{sec:interactions}.  In \cite{Scoccimarro:2011pz}, a code to generate initial conditions to run simulations with equilateral and orthogonal templates was developed and used to analyse power spectrum and bispectrum measurements. Multi-field inflation was also implemented in N-body simulations, allowing to make precise considerations about the stochasticity generated by the presence of more than one field in the initial conditions \cite{Smith:2010gx,Adhikari:2014xua}. The implementation of a generic primordial bispectrum in the initial conditions of an N-body simulation is not trivial. The main challenge is the fact that physical bispectrum shapes from inflation are not separable, i.e. they cannot be factorized as a product of functions of $k_1$, $k_2$ and $k_3$. Separability is a crucial feature for efficient computational algorithm for simulations not only in LSS \cite{Wagner:2010me, Fergusson:2010ia}, but also in CMB applications \cite{Creminelli:2003iq,Fergusson:2009nv}. This is the reason why templates are often used in practical applications. In \cite{Wagner:2011wx}, it was argued though that, since the non-Gaussian contribution on the power spectrum is peaked in the squeezed limit, one should check that the templates reproduce correctly the physical shape in this limit, even in the case in which the physical shape does not peak in that limit. For instance, the orthogonal template \cite{Senatore:2009gt} generates a non-Gaussian bias which scales as $1/k$ in the squeezed limit, while typical orthogonal physical shapes from the effective field theory of inflation \cite{Bartolo:2010bj} generate a scale independent correction. 

As a consequence of these considerations, \cite{Fergusson:2010ia,Wagner:2011wx} have introduced methods to implement a generic inflationary bispectrum in the initial conditions of N-body simulations. For instance, \cite{Wagner:2011wx} has introduced the following ansatz

\begin{equation}
\Phi_{\bk}^{\rm NG} = \frac{1}{2}\int \frac{\d^3q}{(2\pi)^3} \frac{B_\Phi(k,q,|\bk+\bq|)\, \Phi^{*\rm G}_{\bq}\, \Phi^{*\rm G}_{\bk+\bq}}{\left[P_\Phi(k) P_\Phi(q)+P_\Phi(q) P_\Phi(|\bk+\bq|)+P_\Phi(k) P_\Phi(|\bk+\bq|)\right]^2},
\end{equation}

which, however, is not separable and therefore is computationally expensive. Note that the integration is bounded from both above and below, since the simulation has a finite box size and resolution. The method of \cite{Fergusson:2010ia} is separable, but the power spectrum of $\Phi$ receives spurious corrections at large scales \cite{Wagner:2011wx}.

\subsection{Final remarks of this section}\label{sec:remarks}
The last decade has seen an intense theoretical and numerical work in analysing the effect of primordial non-Gaussianity in the statistics of biased tracers. The appearance of a scale-dependent feature at large scales, sourced by certain types of interactions during inflation, has motivated observational efforts and even pushed the funding of a tailored experiment, SPHEREx \cite{Dore:2014cca}, which is scheduled to be launched in 2023. We have argued the importance of two main aspects of this search: on the inflationary side, a selection of theoretically motivated models with observable imprints; on the structure formation modeling side, the need for combining effective approaches with model-dependent ansatz in order to be able to accurately handle the astrophysical uncertainties which hamper our ability to extract the primordial signature. This procedure becomes more and more important when refining the prediction to include halo occupation distributions algorithms for the modeling of galaxy statistics, red-shift space distortions and survey-related uncertainties and when considering higher-order statistics such as the three-point correlation function of halos and galaxies. Another important ingredient, as for all LSS studies, is the use of N-body simulations. Not only the theoretical modeling has to be meticulously tested against numerical results, but also an accurate matching of the N-body outputs against real data needs to be performed. In the case of primordial non-Gaussianity, these two tasks are complicated by the fact that each inflationary prediction effectively realizes a different cosmology and therefore requires a different set of N-body simulations to run and test. For this reason, numerical efforts have concentrated mainly on simple templates for known non-Gaussianities, such as the local type. More work is surely needed in order to properly model the large amount of possible interactions, most of which we have summarized in Sec. \S\ref{sec:interactions}. Indeed, as remarked in the previous section, the risk of using simplified templates is to lose constraining power on more complicated, yet well motivated, inflationary signatures.
\section{Observational prospects}\label{sec:final}

While the focus of this review has been restricted to the theoretical study of the imprint of inflationary interactions on the statistics of dark matter halos, the discussion can, and has to, be extended towards several directions. First of all, the imprint of primordial interactions which we reviewed in Secs. \S\ref{sec:onepoint} and \ref{sec:two-point} has effects on a number of observables other than halos and galaxies number counts, which we summarize in the following list

\begin{itemize}
    \item Galaxy shapes contain this imprint in the shear and convergence field probed through weak-lensing \cite{Marian:2010mh,Maturi:2011am,Jeong:2011rh,Hilbert:2012gr} and galaxy intrinsic alignments \cite{Chisari:2013dda,Chisari:2016xki,Kogai:2018nse}.
    \item The Sunyaev-Zeldovich (SZ) effect \cite{Sunyaev:1970eu} can be exploited to observe dense clusters of galaxies which constrain the high tail of the density distribution function of galaxies which is sensitive to primordial non-Gaussianity \cite{Mak:2012yb}. The thermal SZ power spectrum \cite{Hill:2013baa} and kinetic SZ tomography \cite{Munchmeyer:2018eey} have also been used to put constraints on local-type non-Gaussianity.
    \item The pairwise velocity distribution of galaxies is an additional probe to galaxy density statistics, as primordial non-Gaussianities induce a non-zero skewness and higher-order momenta in the distribution \cite{Catelan:1994hb,Schmidt:2010pf,Lam:2010yz}.
    \item The scale-dependent bias outlined in Sec. \S\ref{sec:two-point} can be constrained using extrema counts as predictedy by peak theory \cite{Nusser:2018vym}
    \item The two-point statistics of voids, when combined with halos, give an order $O(1)$ improvement in local-type non-Gaussianity constraints \cite{Chan:2018piq}
    \item The covariance of galaxy number counts also helps in combination with number counts and variance \cite{Cunha:2010zz}
    \item Higher-redshift probes have been shown to be promising in constraining non-Gaussianity in the future: Lyman-alpha forest \cite{Seljak:2012tp,Chongchitnan:2014jaa}, 21-cm power spectra \cite{Joudaki:2011sv,Chongchitnan:2012we,Chongchitnan:2013oxa,Camera:2013kpa,Sekiguchi:2018kqe,Witzemann:2018cdx}, CO and CII lines \cite{MoradinezhadDizgah:2018zrs,MoradinezhadDizgah:2018lac}, cosmic reionization  \cite{DAloisio:2013mgn,Lidz:2013tra,Mao:2013yaa} and cross-correlations with CMB measurements \cite{Raccanelli:2014kga,Schmittfull:2017ffw}.
    \item The response of the small-scale power spectrum to a squeezed bispectrum is particularly effective for the imprint of primordial non-Gaussianity \cite{Chiang:2017vsq,Yuan:2017pck,dePutter:2018jqk}.
\end{itemize}

This review has also neglected any general relativistic effects. These become important when probing the largest scales of the galaxy power spectrum, where the imprint of sizable primordial bispectra in the squeezed configuration is strongest. Several analyses have computed these corrections \cite{Wands:2009ex,Bruni:2011ta,Baldauf:2011bh,Jeong:2011as,Yoo:2011zc,LopezHonorez:2011cy,Yoo:2012se,Villa:2014foa,DiDio:2016gpd,Castiblanco:2018qsd} and the importance of these relativistic corrections has been quantified for forthcoming surveys \cite{Maartens:2012rh,Raccanelli:2013dza,Lorenz:2017iez}.

So far, we have been concerned only with one- and two-point statistics of LSS observables. Higher-order tracers statistics are able to directly trace primordial interactions. For instance, the bispectrum of galaxies is sourced at tree-level by all the non-Gaussian shapes in Sec. \S\ref{sec:interactions}. Differently than for the power spectrum, where only the squeezed configurations contribute to the signature, the full shape of non-Gaussianity can be probed in the bispectrum. Theoretical and numerical work has been also making progress in this direction in the last decade \cite{Baldauf:2010vn,Pollack:2011xp,Sefusatti:2011gt,Yokoyama:2013mta,Tasinato:2013vna,Dizgah:2015kqi,Hashimoto:2015tnv,Tellarini:2016sgp,Hashimoto:2016lmh,Yamauchi:2016wuc,DiDio:2016gpd,Chiang:2017vsq,An:2017rwo,MoradinezhadDizgah:2018ssw,dePutter:2018jqk,MoradinezhadDizgah:2018pfo,Eggemeier:2018qae}. The main challenges to overcome with the galaxy bispectrum is the fact that gravitational non-linearities developed at late times dominate over primordial ones. The accurate modeling of these non-linearities requires introducing a large number of new parameters, as compared to the case of the power spectrum. Nevertheless, combining power spectrum and bispectrum information can significantly improve constraints, see for instance the forecast for the SPHEREx mission \cite{Dore:2014cca}. 

Besides improving the modeling and combining different statistics, several optimisation techniques have been proposed. One of the major problems to overcome, common to both CMB and LSS, is sampling variance. The multi-tracer technique \cite{Seljak:2008xr,Slosar:2008ta,McDonald:2008sh,GilMarin:2010vf} is based on the assumption that on large scales halos are biased, but not stochastic, tracers of the dark matter density field. Recently it has been argued that similar improvements on the sampling variance can be achieved by selecting tracers with no bias with respect to the dark matter density field \cite{Castorina:2018zfk}. One can therefore eliminate the cosmic variance error by correlating a highly biased population of galaxies against an unbiased one. Another limitation is caused by the discrete nature of the galaxy distribution statistics. This is taken into account usually by adding a Poisson shot noise term to the galaxy power spectrum. This term is particularly relevant for populations of galaxies with a large mass, due to their low density. These are also the galaxies which are more sensitive to the imprint of primordial non-Gaussianity, as showed in Secs. \S\ref{sec:onepoint} and \ref{sec:two-point}. A mitigation of this problem is provided by optimally weighting populations of galaxies \cite{Seljak:2009af,Hamaus:2010im,Hamaus:2011dq}. 

The current best constraints on primordial non-Gaussianities as set by CMB observations by the Planck satellite. Constraints are given in terms of the local, equilateral and orthogonal templates to be $f^{\rm loc}_{\rm NL} = -0.9 \pm 5.1$, $f^{\rm equi}_{\rm NL} = -26 \pm 47$ and $f^{\rm orth}_{\rm NL} = -38 \pm 24$ at $65\%$ confidence level, respectively. LSS searches have been also putting constraints since more than a decade \cite{Slosar:2008hx,Tseliakhovich:2010kf,Xia:2010pe,Xia:2011hj,Benson:2011uta,Ross:2012sx,Mana:2013qba,Giannantonio:2013uqa,Shandera:2013mha,Ma:2013mpa,Karagiannis:2013xea,Agarwal:2013qta,Ho:2013lda,Giannantonio:2013kqa,Leistedt:2014zqa,Castorina:2019wmr}, but they do not give competitive constraints with these figures for any of the observables listed above. The latest constraint was put by the eBOSS collaboration \cite{Castorina:2019wmr} and gives $-51<\fnl^{\rm loc}<21$ at $95\%$ confidence level and the full data analysis of the experiment including quasars might reach the CMB sensitivity. A strong effort in forecasting the possible imprint of primordial non-Gaussianity in LSS has been pushing the limits to order $O(1)$ for local-type non-Gaussianities\footnote{See Table 12 of \cite{Desjacques:2016bnm} for a summary table of future prospects on $f_{\rm NL}^{\rm loc}$.} and $O(10)$ for equilateral and orthogonal ones \cite{Fedeli:2010ud,Namikawa:2011yr,Raccanelli:2011pu,Giannantonio:2011ya,Pillepich:2011zz,Takeuchi:2011ej,Sefusatti:2012ye,Becker:2012yr,Biagetti:2012xy,Biagetti:2013sr,Yamauchi:2014ioa,Ferraro:2014jba,Byun:2014cea,Camera:2014bwa,Xu:2014bya,dePutter:2014lna,Dore:2014cca,Amendola:2015pha,Sartoris:2015aga,Raccanelli:2015vla,Alonso:2015uua,Gariazzo:2015qea,Alonso:2015sfa,Fonseca:2015laa,Raccanelli:2015oma,Yamauchi:2015mja,Hashimoto:2016lmh,Amendola:2016saw,Kovetz:2016hgp,Chisari:2016xki,Adhikari:2016wpj,Fonseca:2016xvi,Yamauchi:2016wuc,dePutter:2016trg,Gleyzes:2016tdh,Li:2017jnt,MoradinezhadDizgah:2017szk,MoradinezhadDizgah:2018ssw,Bellomo:2018lew,MoradinezhadDizgah:2018lac,Bacon:2018dui,Chan:2018piq}. Several future surveys such as SPHEREx \cite{Dore:2014cca}, Euclid \cite{Amendola:2016saw}, MeerKAT+DES \cite{Fonseca:2016xvi} and SKA \cite{Bacon:2018dui} might achieve these limits. 
\vspace{6pt} 




\funding{This research was funded by the Nederlands Organisation for Scientific Research (NWO) under the VENI  grant  016.Veni.192.210.}

\acknowledgments{The author thanks Daniel Baumann, Guilherme Pimentel and Toni Riotto for useful discussions and Vincent Desjacques, Azadeh Moradinezhad Dizgah, Guilherme Pimentel and Toni Riotto for comments on a draft.}

\conflictsofinterest{The author declares no conflict of interest.} 

\abbreviations{The following abbreviations are used in this manuscript:\\

\noindent 
\begin{tabular}{@{}ll}
DS & De-Sitter\\
CMB & Cosmic Microwave Background\\
LSS & Large Scale Structures\\
\end{tabular}}

\appendixtitles{no} 
\appendixsections{multiple} 
\appendix
\section{ESP model}
\subsection{The curvature function of density peaks}
\label{app:bigg}

Without entering into the details of the full calculation (see \cite{Bardeen:1985tr}), the integral Eq.  \eqref{eq:barnpk} can be simplified into the only integration over $J_1$ 
\begin{equation}\label{eq:npkxnu}
\bar n_\pk(\nu, R_s, \vx) =  \int_0^{+\infty} \d J_1\, \mathcal{N}_\pk(J_1, \nu)
\end{equation}
where we define
\begin{equation}
\mathcal{N}_\pk(J_1, \nu)=\frac{e^{-\nu^2/2}}{\sqrt{2\pi}} \frac{1}{(2\pi R*^2)^{3/2}}\,F_1(J_1)\, P_{\rm G}( J_1 - \gamma_1 \nu ; 1-\gamma_1^2)
\end{equation}
and
\begin{eqnarray}
F_1(x) &=& \frac 12 (x^3 - 3x) \left[ {\rm erf} \left(x\sqrt{\frac 52}\right) + {\rm erf} \left(x\sqrt{\frac 58}\right)\right] +\nonumber\\ 
&&\sqrt{\frac{2}{5\pi}} \left[ \left(\frac{31 x^2}{4} + \frac 85 \right) e^{-5x^2/8} + \left(\frac{x^2}{2} - \frac 85 \right) e^{-5x^2/2}\right]
\end{eqnarray}
and we used Bayes' theorem to write $\mathcal{N}(\nu, J_1) = \mathcal{N}(\nu)\, \mathcal{N}(J_1 | \nu)$ and consequently written the conditional gaussian distribution $P_{\rm G}(x-\mu ; s)$ with shifted mean $\mu=\gamma_1\nu$ and variance $s=1-\gamma_1^2$. The integral can be computed analytically and gives the final result
\begin{equation}
\bar n_\pk (\nu)= \frac{e^{-\nu^2/2}}{\sqrt{2\pi}} \frac{1}{V_*}\mathcal{G}^{(1)}_0(\gamma_1, \gamma_1 \nu),
\end{equation}
where we define  the integrals
\begin{equation}
G^{(\alpha)}_\kappa(\gamma_1, \gamma_1\nu) \equiv \int_0^\infty \d x x^\kappa F_\alpha(x) P_{\rm G}(x-\gamma_1\nu; 1-\gamma_1^2)
\end{equation}
and the $0$-th order is the curvature function of density peaks \cite{Bardeen:1985tr}
\begin{align}
\mathcal{G}^{(\alpha)}_0(x) &= \frac{1}{\alpha^4}
\Biggl\{\frac{e^{-{5\alpha x^2\over 2}}}{\sqrt{10\pi}}
\left(\alpha x^2-\frac{16}{5}\right) \\ 
&\quad +\frac{e^{-{5\alpha x^2\over 8}}}
{\sqrt{10\pi}}\left(\frac{31}{2}\alpha x^2+\frac{16}{5}\right)
+\frac{\sqrt{\alpha}}{2}\left(\alpha x^3-3x\right) \nonumber \\
& \qquad \times
\left[{\rm Erf}\left(\sqrt{\frac{5\alpha}{2}}\frac{x}{2}\right)+
{\rm Erf}\left(\sqrt{\frac{5\alpha}{2}}x\right)\right]\Biggr\} 
\nonumber \;.
\end{align}
Note that \cite{Desjacques:2010gz} introduced the extra variable $\alpha$
in order to get a closed form expression for their 2-point peak correlation, while 
\cite{Desjacques:2012eb} showed that $\alpha\neq 1$ can be interpreted as a long-wavelength 
perturbation in $J_2(\vx)$. 

\subsection{The non-Gaussian ESP mass function}\label{app:dnesp}

Performing the transformations Eqs. \eqref{eq:htransf} for all the terms and defining

\begin{equation}
\bar n_{\rm ESP}^{\rm NG}= \bar n_{\rm ESP}^{\rm G}(1+ \sigma_0^4\delta \bar n_{\rm ESP}^{\rm NG})
\end{equation}

we finally get

\begin{align}
\delta \bar n_{\rm ESP}^{\rm NG}=&\, \frac{1}{\sqrt{6}} \Biggr\{b_{300} S_3^{(\nu^3)}  +b_{030} S_3^{(u^3)} + b_{003}\Biggr[  \sqrt{6} S_3^{(\mu^3)}\left(\frac{\ggam \ggamnm(\ggam\ggamnm-\ggamum)(\ggamnm-\ggam\ggamum)}{(1-\ggam^2)^2}\right)\nonumber\\
& + \sqrt{6} S_3^{(\nu u^2)} \left(\frac{\ggamnm(\ggam \ggamnm-\ggamum)^2}{(1-\ggam^2)^2}\right)- \sqrt{6} S_3^{(\nu^2 u)} \left(\frac{\ggamnm(\ggam \ggamnm-\ggamum)(\ggamnm(1+\ggam^2)-2\ggam \ggamum)}{(1-\ggam^2)^2}\right)\nonumber\\
& - \sqrt{6} S_3^{(\nu^2\mu)} \left(\frac{ \ggam \ggamnm(\ggam \ggamnm-\ggamum)}{1-\ggam^2}\right)+\sqrt{6} S_3^{(\nu u \mu)} \left(\frac{ \ggamnm(\ggam \ggamnm-\ggamum)}{1-\ggam^2}\right) + S_3^{(\mu^3)}\Biggr]\Biggr\}\nonumber\\
& +\frac{1}{\sqrt{2}}\Biggr\{b_{210} S_3^{(\nu^2 u)} + b_{201} S_3^{(\nu^2 \mu)} + b_{120} S_3^{(\nu u^2)} + b_{102} \Biggr[\sqrt{2}S_3^{(\nu^3)}\left(\frac{\ggam(\ggam\ggamnm-\ggamum)(\ggamnm-\ggam\ggamum)}{(1-\ggam^2)^2}\right)\nonumber\\
&-\sqrt{2}S_3^{(\nu^2 u)}\left(\frac{(\ggam\ggamnm-\ggamum)(\ggamnm(1+\ggam^2)-2\ggam\ggamum)}{(1-\ggam^2)^2}\right)+\sqrt{2}S_3^{(\nu u^2)}\left(\frac{(\ggam\ggamnm-\ggamum)^2}{(1-\ggam^2)^2}\right)\nonumber\\
&-\sqrt{2}S_3^{(\nu^2\mu)}\left(\frac{\ggam(\ggam\ggamnm-\ggamum)}{1-\ggam^2}\right)+S_3^{(\nu\mu^2)}+S_3^{(\nu u \mu)}\left(\frac{\ggam\ggamnm-\ggamum}{1-\ggam^2}\right)\Biggr]\nonumber\\
&+b_{012} \Biggr[ \sqrt{2}S_3^{(\nu^3)}\left(\frac{\ggam(\ggamnm-\ggam\ggamum)(\ggamnm(-1+2\ggam^2)-\ggam\ggamum)}{(1-\ggam^2)^2}\right) \nonumber\\
&- \sqrt{2}S_3^{(\nu^2 u)}\left(\frac{(\ggamnm(1+\ggam^2)-2\ggam\ggamum)(\ggamnm(-1+2\ggam^2)-\ggam\ggamum)}{(1-\ggam^2)^2}\right)\nonumber\\
&+ \sqrt{2}S_3^{(\nu u^2)}\left(\frac{(\ggam\ggamnm-\ggamum)(\ggamnm(-1+2\ggam^2)-\ggam\ggamum)}{(1-\ggam^2)^2}\right)- \sqrt{2}S_3^{(\nu^2 \mu)}\left(\frac{\ggam(\ggamnm(-1+2\ggam^2)-\ggam\ggamum)}{1-\ggam^2}\right)\nonumber\\
&+S_3^{(u \mu^2)}+\sqrt{2}S_3^{(\nu u\mu)}\left(\frac{(\ggamnm(-1+2\ggam^2)-\ggam\ggamum)}{1-\ggam^2}\right)\Biggr] + b_{021}\Biggr[\sqrt{2}S_3^{(\nu^3)} \left(\frac{\ggam^2(-\ggamnm+\ggam\ggamum)}{1-\ggam^2}\right) \nonumber\\
&+\sqrt{2}S_3^{(\nu^2 u)} \left(\frac{\ggam(\ggamnm(1+\ggam^2)-2\ggam\ggamum)}{1-\ggam^2}\right)-\sqrt{2}S_3^{(\nu u^2)} \left(\frac{\ggam(\ggam\ggamnm-\ggamum)}{1-\ggam^2}\right)\nonumber\\
&+\sqrt{2} S_3^{(\nu^2\mu)}\ggam^2+S_3^{(u^2\mu)}-\sqrt{2} S_3^{(\nu u\mu)}\ggam\Biggr]+\sqrt{2} b_{111}\Biggr[ S_3^{(\nu^3)}\left(\frac{\ggam(\ggam\ggamum-\ggamnm)}{1-\ggam^2}\right)\nonumber\\
&+S_3^{(\nu^2 u)}\left(\frac{(\ggamnm(1+\ggam^2)-2\ggam\ggamum)}{1-\ggam^2}\right)-S_3^{(\nu u^2)}\left(\frac{(\ggam\ggamnm-\ggamum)}{1-\ggam^2}\right)+S_3^{(\nu^2 \mu)}\ggam\Biggr]\Biggr\}\nonumber\\
&-\sqrt{\frac 32}\Biggr[\chi_{10}^{100} S_3^{(\nu\eta^2)}+\chi_{10}^{010} S_3^{(u\eta^2)}+\chi_{10}^{001} S_3^{(\mu\eta^2)}\Biggr]-\sqrt{\frac 52}\Biggr[\chi_{01}^{100} S_3^{(\nu\zeta^2)}+\sqrt{\frac 52}\chi_{01}^{010} S_3^{(u\zeta^2)}+\sqrt{\frac 52}\chi_{01}^{001} S_3^{(\mu\zeta^2)}\Biggr]\nonumber\\
&+\frac{5}{\sqrt{21}} J_3 S_3^{(J_3)}
\end{align}

\reftitle{References}


\newpage 

\externalbibliography{yes}
\bibliography{reference_thesis}


\end{document}